\documentclass[11pt]{article}
\usepackage{amsthm,amsmath,amssymb}
\usepackage{natbib}
\usepackage{multirow}
\usepackage[pdftex]{graphicx}
\usepackage{subfigure}
\usepackage{makecell}
\usepackage{booktabs}
\usepackage{array}
\usepackage{fullpage}
\usepackage{url}
\usepackage[ruled]{algorithm2e}
\usepackage{bm}
\usepackage{smile}
\usepackage{wrapfig}
\usepackage{lipsum}
\usepackage{mathrsfs}
\usepackage{dsfont}
\usepackage{float}
\usepackage{natbib}
\usepackage{multirow}

\usepackage{appendix}

\newcolumntype {Q}{>{$\displaystyle}l<{$}}
\newcolumntype {A}{>{$}c <{$}}

\def\tr{\mathop{\text{tr}}\kern.2ex}

\def\P{{\mathbb P}}
\def\E{{\mathbb E}}

\def\S{{\mathbb S}}

\def\sign{\mathrm{sign}}

\def\card{\mathop{\text{card}}}

\long\def\comment#1{}

\def\tr{\mathop{\text{Tr}}}

\def\largmax{\mbox{local-argmax}}
\def\largmin{\mbox{local-argmin}}

\def\be{{\mathbf e}}

\providecommand{\norm}[1]{\vvvert#1\vvvert}

\newcommand{\bel}{\begin{eqnarray}\label}
\newcommand{\eel}{\end{eqnarray}}
\newcommand{\bes}{\begin{eqnarray*}}
\newcommand{\ees}{\end{eqnarray*}}

\def\reals{{\mathbb{R}}}
\def\R{{\reals}}
\def\T{{\sf T}}

\def\cov{{\rm Cov}}
\def\ones{\mathds{1}}

\newcommand{\hf}[1]{{
#1}}

\usepackage[usenames,dvipsnames,svgnames,table]{xcolor}
\usepackage[colorlinks=true,
            linkcolor=blue,
            urlcolor=blue,
            citecolor=blue]{hyperref}
            
\numberwithin{equation}{section}
\numberwithin{theorem}{section}
\numberwithin{corollary}{section}
\numberwithin{asmp}{section}
\numberwithin{definition}{section}  

\graphicspath{{./figs/}}

\usepackage{xr}
\begin{document}

\setlength{\abovedisplayskip}{5pt}
\setlength{\belowdisplayskip}{5pt}
\setlength{\abovedisplayshortskip}{5pt}
\setlength{\belowdisplayshortskip}{5pt}

\title{\LARGE {A Provable Smoothing Approach for High Dimensional Generalized Regression with Applications in Genomics}}

\author{Fang Han\thanks{Department of Statistics, University of Washington, Seattle, WA 98195, USA; e-mail: {\tt fanghan@uw.edu}.}, ~Hongkai Ji\thanks{Department of Biostatistics, Johns Hopkins University, Baltimore, MD 21205, USA; e-mail: {\tt hji@jhu.edu}.}, ~Zhicheng Ji\thanks{Department of Biostatistics, Johns Hopkins University, Baltimore, MD 21205, USA; e-mail: {\tt zji4@jhu.edu}.}, ~and~ Honglang Wang\thanks{Department of Mathematical Sciences, Indiana University-Purdue University Indianapolis, Indianapolis, IN 46202, USA; e-mail: {\tt hlwang@iupui.edu}.} 
\thanks{The work of Fang Han is supported by NSF grant DMS-1712536. The work of Hongkai Ji and Zhicheng Ji is supported by NIH grants R01HG006841 and R01HG006282.} }

\date{}

\maketitle

\begin{abstract}

In many applications, linear models fit the data poorly.  This  article studies an appealing alternative, the generalized regression model. This model
only assumes that there exists an unknown monotonically increasing link function connecting the response $Y$ to a single index $\bX^\T\bbeta^*$ of explanatory variables $\bX\in\reals^d$. 
The generalized regression model is flexible and covers many widely used statistical models. It fits the data generating mechanisms well in many real problems, which makes it useful in a variety of applications where regression models are regularly employed. 
In low dimensions, rank-based M-estimators are recommended to deal with the generalized regression model, giving root-$n$ consistent estimators of $\bbeta^*$. Applications of these estimators to high dimensional data, however, are questionable. 
This article \hf{studies, both theoretically and practically,} a simple yet powerful smoothing approach to handle the high dimensional generalized regression model. \hf{Theoretically,} a family of smoothing functions is provided, and the amount of smoothing necessary for efficient inference is carefully calculated. 
\hf{Practically,} our study is motivated by an important and challenging scientific problem: decoding gene regulation by predicting transcription factors that bind to cis-regulatory elements. Applying our proposed method to this problem shows substantial improvement over the state-of-the-art alternative in real data. 
\end{abstract}



\section{Introduction}

Regression models play a fundamental role in characterizing the relation among variables. Nonparametric and semiparametric regression models are commonly used alternatives to linear regression when the latter fails to fit the data well.  Their advantages over simple linear regression models have been established in various fields \citep{huber2011robust,ruppert2003semiparametric}. 

In this article, we study a semiparametric generalization of linear regression as such an alternative. We assume 
\begin{align}\label{eq:model}
Y=D\circ \Lambda(\bX^\T\bbeta^*,\epsilon),~~{\rm where}~~Y, \epsilon\in\reals~{\rm and}~\bX, \bbeta^*\in\reals^d.
\end{align}
Here $Y$ is the scalar response variable, $D(\cdot)$ is an unknown increasing function, $\Lambda(\cdot,\cdot)$ is an unknown strictly increasing function regarding each of its arguments, $\bX$ represents the vector of explanatory variables, $\epsilon$ is an unspecified noise term independent of $\bX$, and $\bbeta^*$ is the regression coefficient characterizing the relation between $\bX$ and $Y$. The coefficient $\bbeta^*$ is assumed to be sparse. Model \eqref{eq:model} is referred to as the generalized regression model. It was first proposed in econometrics \citep{han1987non}, and is a very flexible semiparametric model, containing a parametric part, encoded in the linear term $\bX^\T\bbeta^*$, and a nonparametric part, encoded in link functions $D(\cdot), \Lambda(\cdot,\cdot)$, and the noise term $\epsilon$. In practice, we assume that $n$ independent realizations of $(Y,\bX)$, denoted as $\{(Y_i,\bX_i), i=1,\ldots,n\}$, are observed. These observations will be used to fit the model.  

\subsection{A motivating genomic challenge}
\begin{figure}[!htbp]
\centering
\subfigure[introfigure1a][Illustration of TF, motif, cis-element and DNase-seq]{
\includegraphics[width=0.9\textwidth]{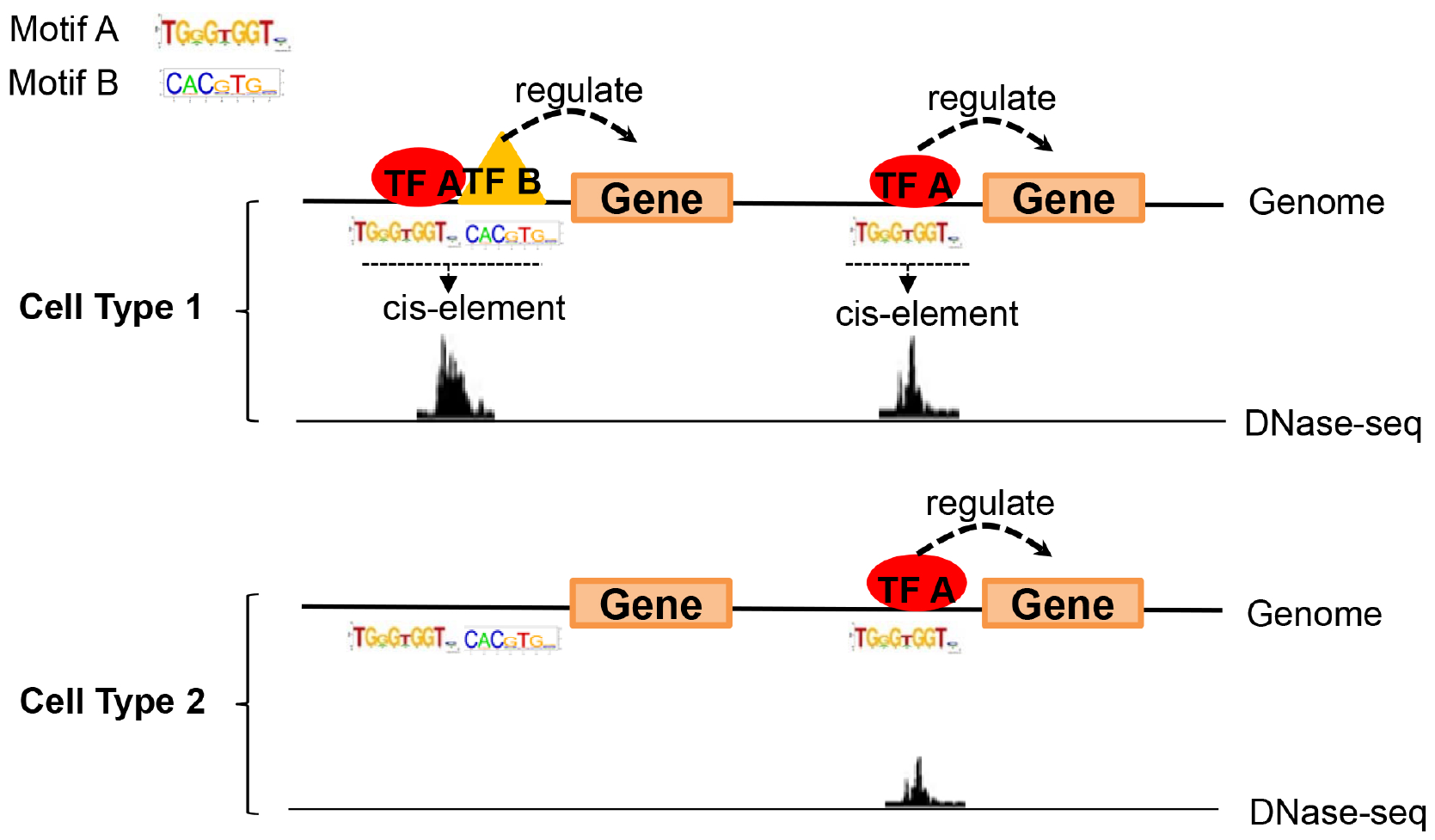}
\label{introfigure1a}}
\hspace{-0.1cm}
\subfigure[introfigure1b][Overview of data structure]{
\includegraphics[width=0.9\textwidth]{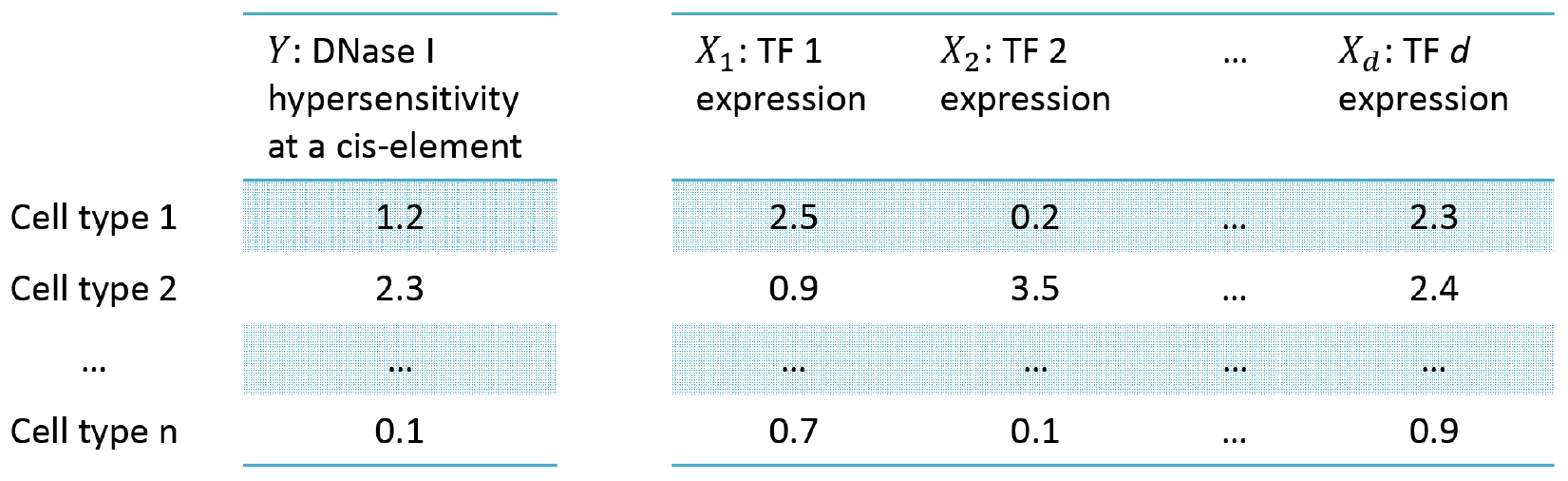}
\label{introfigure1b}}
\caption{\textbf{Background.} (a) An illustration of TF, motif, cis-element, and DNase I hypersensitivity measured by DNase-seq. 
For each cell type, two panels are displayed. In the top panel, the horizontal line represents the genome. TFs (ellipse and triangle) bind to cis-elements to activate or repress the transcription of nearby genes (rectangle). Each TF binds to a specific DNA motif located within the cis-element. In the bottom panel, DNase I hypersensitivity measured by DNase-seq at each cis-element correlates with the TF binding activity. Since TF binding activities are different in different cell types, the DNase-seq signals also vary across cell type.
(b) Data structure used for predicting a cis-element's binding TFs. For each cis-element, $Y$ is the DNase-seq signal measured in different cell types, and $\bX$ is gene expression levels for $d$ TFs in the same cell types.}
\label{introfigure} 
\end{figure}

Model \eqref{eq:model} naturally occurs in many applications. Below we elaborate a challenging scientific problem that motivates our study. To help put the model into context, some background introduction is necessary. One fundamental question in biology is how genes' transcriptional activities are controlled by transcription factors (TFs). TFs are an important class of proteins that can bind to DNA to induce or repress the transcription of genes nearby the binding sites (Figure \ref{introfigure1a}). Human has hundreds of different TFs. Each TF can bind to $10^2$-$10^5$ different genomic loci to control the expression of $10$-$10^3$ target genes. The genomic sites bound by TFs are also called cis-regulatory elements or cis-elements. Promoters and enhancers are typical examples of cis-elements. TF binding activities are context-dependent. 
The binding activity of a given TF at a given cis-element varies from cell type to cell type. It depends on the TF's expression level in each cell type as well as numerous other factors. One cis-element can be bound by multiple collaborating TFs that form a protein complex. Different TFs can bind to different cis-elements. 
In order to comprehensively understand the gene regulatory program, a crucial step is to identify all active cis-elements and their binding TFs in each cell type and biological condition. 

{The state-of-the-art technology for mapping genome-wide TF binding sites (TFBSs) is Chromatin Immunoprecipitation coupled with sequencing (ChIP-seq) \citep{park2009chip}. Unfortunately, each ChIP-seq experiment can only analyze one TF. Using this technology to map binding sites of all human TFs would require one to conduct hundreds of such experiments. Moreover, ChIP-seq requires high-quality antibodies which are not available for all TFs. Therefore, mapping binding sites for all TFs using ChIP-seq is both costly and technically infeasible. An alternative approach to mapping TFBSs is based on genome-wide sequencing of DNase I hypersensitive sites (DNase-seq) \citep{boyle2008high}.  TFBSs are often sensitive to the cleavage of DNase I enzyme. Thus, DNase I hypersensitivity (DH), which can be measured in a genome-wide fashion using DNase-seq, can be used to locate cis-elements actively bound by TFs (Figure \ref{introfigure1a}). This approach is capable of mapping binding sites of all TFs in a biological sample through a single experimental assay. However, a major limitation of DNase-seq is that it does not reveal the identities of TFs that bind to each cis-element. If one could solve this problem by correctly predicting which TFs bind to each element, one would then be able to combine DNase-seq with computational predictions to identify all active cis-elements and their binding TFs in a biological sample using a single experimental assay. This would help scientists to remove a major roadblock in the study of gene regulation.}

Many TFs recognize specific DNA sequence patterns called motifs (Figure \ref{introfigure1a}).  Different TFs recognize different motifs. A conventional way to infer the identities of TFs that bind to a cis-element is to examine which TFs' DNA motifs are present in the cis-element. Unfortunately, motifs for 2/3 of all human TFs are unknown.
Therefore, solely relying on DNA motifs is not sufficient to solve this problem. This motivates development of an alternative solution that leverages massive amounts of gene expression and DNase-seq data in public databases to circumvent the requirement for DNA motifs. The Encyclopedia of DNA Elements (ENCODE) project \citep{encode2004encode} has generated DNase-seq and gene expression data for a variety of different cell types. Using these data, one may examine how the protein-binding activity measured by DNase-seq at a cis-element varies across different cell types and how such variation is explained by variations in the expression of TFs (Figure \ref{introfigure1b}). Through this analysis, one may infer which TFs bind to each cis-element.


{Formally, let $Y$ be the activity of a cis-element in a particular cell type measured by DNase-seq, and let $\bX=(X_1,\ldots,X_d)^\T$ be the expression level of $d$ different TFs in the same cell type (Figure \ref{introfigure1b}). 
The relationship between cis-element activity and TF expression can be described using a generalized regression model, $Y=D\circ \Lambda(\bX^\T\bbeta^*,\epsilon)$, with a high dimensional sparse regression coefficient vector $\bbeta^*$. One expects the relationship between $Y$ and $\bX^\T\bbeta^*$ to be monotonic since a TF has to be expressed in order to be able to bind to cis-elements. Also, increased TF expression may lead to increased binding. The relationship may not be linear and the noise may not be normal since DNase-seq generates counts data. Although after normalization, the data may no longer be integers, they usually are still non-normal and may be zero-inflated. The model is high dimensional since there are hundreds of of TFs (i.e., $d=10^2-10^3$), whereas the sample size (i.e., the number of ENCODE cell types with both DNase-seq and gene expression data) is small ($n$=50-100).  Lastly, $\bbeta^*$ has to be sparse since the number of TFs that can bind to a cis-element is expected to be small. This is because cis-elements are typically short. Each element cannot have physical contacts with too many different proteins. In this model, the non-zero components of $\bbeta^*$ may be used to infer the identities of binding TFs.} 


\subsection{{Existing works on generalized regression} }

{In order to properly position our results in the literature, below we briefly review existing methodological works  that are most relevant to our study on deciphering the generalized regression model.} 
The generalized regression model contains many widely-used econometrical and statistical models, including important sub-classes such as the monotonic transformation model and monotonic single-index model, of the following forms:
\begin{align}
&Y=G(\bX^\T\bbeta^*+\epsilon)\quad \mbox{(monotonic~transformation~model)}, \label{eq:model1}\\
&Y=G(\bX^\T\bbeta^*)+\epsilon\quad \mbox{(monotonic~single-index~model)}, \label{eq:model2}
\end{align}
where the univariate link function $G(\cdot)$ is assumed to be strictly increasing. 


There has been research in estimating the generalized regression model or its variants in low dimensions. These works follow two tracks. 
In the first track, \cite{han1987non} and \cite{cavanagh1998rank} proposed rank-based M-estimators for directly estimating $\bbeta^*$, while treating link functions $D(\cdot)$ and $\Lambda(\cdot,\cdot)$ as nuisance parameters. The corresponding estimator $\hat\bbeta$ aims to maximize certain rank correlation measurement between $Y$ and $\bX^\T\bbeta$, and hence often involves a discontinuous loss function. Based on an estimate of $\bbeta^*$,  \cite{horowitz1996semiparametric}, \cite{ye1997nonparametric}, and \cite{chen2002rank} further proposed {methods} to estimate the link function $D\circ \Lambda(\cdot,\cdot)$ under different parametric assumptions on link functions. \hf{This method is also extended in \cite{dai2014variable} and \cite{shi2014penalized} to ultra-high dimensional settings via coupling it with a lasso-type penalty.}
In the second track, 
\cite{ahn1996simple}, \cite{tanaka2008semiparametric}, \cite{kakade2011efficient}, among many others, focused on studying more specific models in \eqref{eq:model1} and \eqref{eq:model2}, and suggested to approximate $D\circ \Lambda(\cdot,\cdot)$ for estimating $\bbeta^*$ via exploiting the kernel regression and sieve approximation. 
These approaches therefore naturally require geometric assumptions and smoothness conditions on $D\circ \Lambda(\cdot,\cdot)$. 

In high dimensions when $d$ could be much larger than the sample size $n$, serious drawbacks are associated with methods in both tracks.  

For  the second track,  first, simultaneous estimation of $D\circ \Lambda(\cdot,\cdot)$ and $\bbeta^*$ requires extra prior assumptions on $D\circ \Lambda(\cdot,\cdot)$, which may not hold in practice. Secondly, 
nonparametric estimation is well known to be difficult in high dimensions. 
This could hurt the estimation of $\bbeta^*$. 
Thirdly, 
these estimation procedures usually are very sensitive to outliers, which are common in real applications. 

For the first track, rank-based M-estimators treat $D\circ \Lambda(\cdot,\cdot)$ as nuisance and hence could potentially gain efficiency and modeling flexibility in estimating $\bbeta^*$. However, these rank-based methods also have serious computational and theoretical drawbacks. Computationally, the loss functions of rank-based M-estimators are commonly discontinuous. This violates the regularity conditions in most optimization algorithms \citep{luo1993error, nesterov2012efficiency} and makes the optimization problem intractable. It could create serious computational issues, especially in high dimensions. Theoretically, the discontinuity of loss functions adds substantial difficulty for analysis, especially in high dimensions. This further makes the statistical performance of corresponding estimators intractable.

\subsection{Overview of key results}

This article \hf{studies} a simple smooth alternative to the above rank-based methods. This results in estimators that are computationally efficient to calculate, while keeping the modeling flexibility in estimating $\bbeta^*$. 
The core idea is to replace the non-smooth rank-based loss function $\hat\cL(\cdot)$ by  a smooth loss function $\hat\cL_n(\cdot)$, indexed by $n$. $\hat\cL_n(\cdot)$ is designed to {become} closer to $\hat\cL(\cdot)$ when $n$ increases. A family of smoothing functions is accordingly studied. 

\hf{Of note, the idea to approximate a non-smooth loss function using a smooth one has proven its successfulness in literature: see, for example, \cite{horowitz1992smoothed}, \cite{ma2005regularized}, \cite{horowitz1998bootstrap}, \cite{zhang2013statistical}, and \cite{shi2014penalized} for smoothing Manski's maximum score estimator, an estimator targeting at the area under the receiver operating characteristic curve (AUC), quantile regression estimator, and Han's rank-based M-estimator in low and high dimensions. However, our results add new observations to this track of works both theoretically and in some new biological applications, which will be outlined below.}

\hf{Theoretically,} 
although a fairly studied approach in low dimensions, smoothing {approximation} to non-smooth and non-convex loss functions has received little attention in high dimensions. This is mainly due to the extreme irregularity of original loss functions, which form  discontinuous U-processes \citep{sherman1993limiting}.  Theoretically speaking, smoothing renders two types of errors: (i) the bias which characterizes the error introduced by employing $\hat\cL_n(\cdot)$ to approximate $\hat\cL(\cdot)$; (ii) the variance which characterizes the error introduced by the randomness of $\hat\cL_n(\cdot)$.  Our study characterizes behaviors of both types of errors, {based on which one can} calculate the amount of smoothing necessary to balance the bias and variance. 
Our theory holds without any assumption on $D\circ \Lambda(\cdot,\cdot)$ other than monotonically increasing. Additionally, the noise term $\epsilon$ is allowed to be non-centered, arbitrarily heavy-tailed, {including these Cauchy distributed ones with median possibly non-zero}, and contain a substantial amount of outliers. \hf{Our theory hence confirms, mathematically rigorously, several advantages of smoothed Han's maximum rank estimator.}

\hf{Practically,} the aforementioned \hf{advantages of the studied procedure} are important for problems where a large number of different regression models need to be fitted. Consider our motivating problem of predicting binding TFs of cis-elements. Since different cis-elements behave differently, the relationship between $Y$ and $\bX$ may have different functional forms for different cis-elements even though all these functions are monotonic. Additionally, different cis-elements may have different noise distributions. Despite this heterogeneity, different cis-elements can be conveniently handled in a unified fashion using our generalized regression procedure regardless of the form of $D\circ \Lambda(\cdot,\cdot)$ and $\epsilon$. By contrast, manually constructing parametric models of different forms for differet cis-elements would be difficult due to the large number of models that need to be constructed. The advantages of our approach over existing high dimensional linear and robust regression couterparts (e.g., those in \cite{loh2015statistical} and \cite{fan2014robust}) are hence clear. 


Applying our method to the binding TF prediction problem, we find that our approach is substantially more accurate than {{the competing lasso method}} for predicting binding TFs. This demonstrates the practical utility of the smoothing approach and shows that it can provide a solution to a long-standing open problem in biology.

\subsection{Other related works}

Monotonic single-index and transformation models are important subclasses of the generalized regression model. In contrast to the monotonic transformation model, there exists an increasing amount of research studying the monotonic single-index model. These include \cite{kalai2009isotron}, \cite{kakade2011efficient}, and \cite{foster2013variable}, to just name a few. However, they require much stronger modeling assumptions, and are sensitive to different types of data contamination.

There are two more related semiparametric approaches to generalize the linear regression model. The first is the general single-index model, with no explicit geometric constraint on $G(\cdot)$ in \eqref{eq:model2}. In high dimensions, \cite{plan2014high} and \cite{yi2015optimal} studied such a relaxed model. For this, {besides being sensitive to data contamination}, they need to first sphericalize the data, which is extremely difficult in high dimensions. Recently, \cite{radchenko2015high} proposed an alternative approach that does not require data sphericity. However, boundedness assumption on $\bX$, subgaussianity of $\epsilon$, and certain smoothness conditions on $G(\cdot)$ are still required. 

The second is sufficient dimension reduction. Related literature in this direction includes \cite{li1991sliced}, \cite{li1992principal}, \cite{cook1991comment}, \cite{cook1998principal}, \hf{and \cite{yin2015sequential}}. Sufficient dimension reduction approaches only assume $Y$ is independent of $\bX$ conditional on some linear projections of $\bX$. However, a data sphericalization step is also crucial in all these approaches, and in each step of derivation, we need $d/n\rightarrow 0$ to proceed. These make the sufficient dimension reduction approaches vulnerable to high dimensionality, \hf{which will be further illustrated in simulations and real data experiments.}

\subsection{Paper organization}

The rest of the paper is organized as follows. The next section presents our smoothing approach to {estimating} the generalized regression model.  
In Section \ref{sec:real}, we use this method to solve our motivating scientific problem of decoding transcription factor binding. We demonstrate that this semiparametric regression approach is capable of substantially improving the accuracy over the state-of-the-art alternatives in real data. Section \ref{sec:theory} gives theoretical results for understanding the proposed approach and calculates the appropriate smoothing amount. \hf{Section \ref{sec:discussion} provides discussions. 
Finite-sample simulation results and proofs are relegated to an appendix. }

\subsection{Notation}

Let $\bv=(v_1,\ldots,v_d)^{\T}$ and $\Mb=[\Mb_{jk}]\in\reals^{d\times d}$ be a $d$ dimensional real vector and a $d$ by $d$ real matrix. For sets $I,J \subset \{1,\ldots,d\}$, let $\bv_I$ be the subvector of $\bv$ with entries indexed by $I$, and $\Mb_{I,J}$ be the submatrix of $\Mb$ with rows and columns indexed by $I$ and $J$. Let $\card(I)$ represent the cardinality of the set $I$. For $0<q<\infty$, we define the vector $\ell_0$, $\ell_q$, and $\ell_{\infty}$ (pseudo-)norms of $\bv$ to be $\norm{\bv}_0:=\card(\{j:v_j\ne 0\})$, $\norm{\bv}_q:=(\sum_{i=1}^d |v_i|^q)^{1/q}$, and $\norm{\bv}_{\infty}:=\max_{1\leq i\leq d}|v_i|$. For the symmetric matrix $\Mb$, let $\lambda_{\max}(\Mb)$ and $\lambda_{\min}(\Mb)$ represent the largest and smallest eigenvalues of $\Mb$. We write $\Mb\succeq 0$ if $\Mb$ is positive semi-definite. For any $x\in\reals$, we define the sign function $\sign(x):=x/|x|$, where by convention we write $0/0=0$. For any two random vectors $\bX,\bY\in \reals^d$, we write $\bX\stackrel{{\sf d}}{=}\bY$ if $\bX$ and $\bY$ are identically distributed. We let $c, C$ be two generic absolute positive constants, whose actual values may vary at different locations. We write $\mathds{1}(\cdot)$ to be the indicator function. We let $\S^{d-1}$ represent the ball $\{\bv\in\reals^d: \norm{\bv}_2=1\}$. For any two real sequences $\{a_n\}$ and $\{b_n\}$, we write $a_n \lesssim b_n$, or equivalently $b_n \gtrsim a_n$, if there exists a constant $C$ such that $|a_n|\leq C|b_n|$ for any large enough $n$. We write $a_n\asymp b_n$ if $a_n \lesssim b_n$ and $b_n\lesssim a_n$. The symbols $\stackrel{\P}{\lesssim}, \stackrel{\P}{\gtrsim},$ and $\stackrel{\P}{\asymp}$ are analogous to $\lesssim, \gtrsim$, and $\asymp$, but these relations hold stochastically.

\section{Problem setup and methods}\label{sec:method}

In this section, we provide some background on the generalized regression model and associated rank-based $M$-estimators. We further describe the class of nonconvex smoothness approximations that will be exploited and covered by our theory. 

Throughout the paper, we assume Model \eqref{eq:model} holds, and we have $n$ independent and identically distributed (i.i.d.) observations $\{(Y_i,\bX_i),i=1,\ldots,n\}$ generated from \eqref{eq:model}. We do not pose any parametric assumption on either the link functions $D(\cdot), \Lambda(\cdot,\cdot)$ or the noise term $\epsilon$, except for assuming , 
\begin{align}\label{eq:constraint}
D(\cdot) ~\text{is non-degenrate}, ~~D(a)\geq D(b), ~~\Lambda(a,\cdot)>\Lambda(b,\cdot),~~{\rm and}~~\Lambda(\cdot,a)>\Lambda(\cdot,b),
\end{align}
as long as $a>b$. For model identifiability, in the sequel, we assume we know at least one specific entry of $\bbeta^*$ that is nonzero, and fix it to be one. Without loss of generality, we assume $\beta_1^*=1$. 

The generalized regression model of form \eqref{eq:model} {represents  a large class of models}. These include the monotonic transformation and single-index models of the forms \eqref{eq:model1} and \eqref{eq:model2}. More specifically, the generalized regression model covers the linear regression model, with $D\circ \Lambda(u,v)=u+v$; the Box-Cox transformation model \citep{box1964analysis}, with $D\circ \Lambda(u,v)=(u+v)^{\lambda}$ for some $\lambda>0$;  the log-linear model and accelerated failure time model \citep{kalbfleisch2011statistical}, with $D\circ \Lambda(u,v)=\exp(u+v)$; the binary choice model \citep{maddala1986limited}, with $D\circ \Lambda(u,v)=\ones(u+v\geq 0)$; the censored regression model \citep{tobin1958estimation}, with $D\circ \Lambda(u,v)=(u+v)\ones(u+v\geq 0)$.

We focus on the following rank-based M-estimator, called the maximum rank correlation (MRC), which is first proposed in \cite{han1987non}:
\begin{align}\label{eq:loss-han}
\argmax_{\bbeta\in\reals^d,\beta_1=1}\Big\{\frac{2}{n(n-1)}\sum_{1\leq i<i'\leq n}\sign(Y_i-Y_{i'})\sign(\bX_i^\T\bbeta-\bX_{i'}^\T\bbeta) \Big\}.
\end{align}
The formulation of MRC is a U-process \citep{honore1997pairwise,zhang2013statistical}.

Intrinsically, MRC aims to maximize the Kendall's tau correlation coefficient \citep{kendall1938new} between $Y$ and $\bX^\T\bbeta$, while treating the link functions as nuisance. This is attainable only via assuming $D\circ\Lambda(\cdot,\cdot)$ is monotonic, and has its roots in transformation and copula models \citep{nelsen2013introduction}. For such models, rank-based approaches have been well-known to be of certain efficiency properties \citep{klaassen1997efficient, hallin2006semiparametrically}.

For fully appreciating the rationality of MRC, we provide a proposition that characterizes MRC's relation to the linear regression model, and further addresses the identifiability issue. Although the result in the first part is very straightforward, we do not find an explicit one in the literature that shows this relation, 

\begin{proposition}\label{prop:1}
Suppose $\bX$ is continuous and has a positive definite covariance matrix. We have, (i) when the link function $D\circ \Lambda(u,v)=u+v$ corresponds to the linear regression model (without requiring $\bX$ and $\epsilon$ to be centered), $\bbeta^*$ is the unique optimum to maximize the Pearson correlation between $Y$ and $\bX^\T\bbeta$ up to a scaling:
\[
\argmax_{\bbeta\in\reals^d}\Big\{\frac{\E [(Y_1-Y_2) (\bX_1^\T\bbeta-\bX_2^\T\bbeta)]}{\sqrt{\cov(Y_1-Y_2)}\sqrt{\cov(\bX_1^\T\bbeta-\bX_2^\T\bbeta)}}\Big\};
\]
(ii) As long as the link functions $D(\cdot)$ and $\Lambda(\cdot,\cdot)$ satisfy \eqref{eq:constraint}, $\bbeta^*$ is the unique optimum to maximize the Kendall's tau correlation coefficient between $Y$ and $\bX^\T\bbeta$ up to a scaling:
\[
\argmax_{\bbeta\in\reals^d}\Big\{\E\Big(\sign(Y_1-Y_2)\sign(\bX_1^\T\bbeta-\bX_2^\T\bbeta)\Big)\Big\}.
\]
\end{proposition}

Throughout, we are interested in the settings where the dimension $d$ could be much larger than the sample size $n$. In such settings, due to restrictive information obtainable, we have to further regularize the parameter space. In particular, sparsity on  the parametric space is commonly assumed. A seemingly natural regularized MRC estimator is as follows:
\begin{align}\label{eq:loss-han2}
\argmax_{\bbeta\in\reals^d,\beta_1=1}\Big\{\frac{2}{n(n-1)}\sum_{1\leq i<i'\leq n}\sign(Y_i-Y_{i'})\sign(\bX_i^\T\bbeta-\bX_{i'}^\T\bbeta) - \lambda_n\sum_{j=2}^d|\bbeta_j| \Big\},
\end{align}
or its equivalent formulation:
\begin{align*}
\argmax_{\bbeta\in\reals^d, \beta_1=1} \!\Big\{\frac{2}{n(n-1)}\sum_{1\leq i<i'\leq n} \Big(\mathds{1}(Y_i>Y_{i'})\mathds{1}(\bX_i^\T\bbeta\!>\!\bX_{i'}^\T\bbeta)+\\
\mathds{1}(Y_i<Y_{i'})\mathds{1}(\bX_i^\T\bbeta<\bX_{i'}^\T\bbeta) \Big)- \lambda_n\sum_{j=2}^d|\bbeta_j| \Big\}.
\end{align*}

However, \eqref{eq:loss-han2} involves a discontinuous loss function that has abrupt changes. As stated in the introduction, this incurs serious computational and statistical issues. For tackling such issues, we propose the following smoothing approximation to \eqref{eq:loss-han2}, using a set of cumulative distribution functions (CDFs) $\{F_{ii'}(\cdot),1\leq i<i'\leq n\}$ to approximate the indicator function $\ones(\cdot)$: 
\begin{align*}
\hat\bbeta_{\alpha_n} \in \underset{\bbeta\in\reals^d, \beta_1=1}{\largmax} \Big\{\frac{2}{n(n-1)}\sum_{1\leq i<i'\leq n}\Big(S_{ii'}F_{ii'}(\alpha_nZ_{ii'}(\bbeta))+(1-S_{ii'})(1-F_{ii'}(\alpha_nZ_{ii'}(\bbeta)))\Big) \\
- \lambda_n\sum_{j=2}^d|\bbeta_j| \Big\}.
\end{align*}
Here ``$\largmax\{\cdot\}$" and ``$\largmin\{\cdot\}$" represent the sets of local maxima and minima for a given function respectively. In addition, we write 
\[
Z_{ii'}(\bbeta):=(\bX_i-\bX_{i'})^\T\bbeta~~~{\rm and}~~~S_{ii'}:=\mathds{1}(Y_i>Y_{i'}). 
\]
Of note, $\alpha_n$ is an explicitly stated smoothness parameter controlling the approximation speed, presumably increasing to infinity with $n$. For any pair $(i,i')$, $F_{ii'}(\cdot)$ is a pre-determined fixed smooth continuous CDF, satisfying $F_{ii'}(-u)=1-F_{ii'}(u)$ for arbitrary $u\geq 0$. 

Note we can equivalently write 
\begin{align}\label{eq:opt1}
\hat\bbeta_{\alpha_n} \in  \underset{\bbeta\in\reals^d, \beta_1=1}{\largmin} \Big\{\underbrace{-\frac{2}{n(n-1)}\sum_{1\leq i<i'\leq n}F_{ii'}(\tilde S_{ii'}\alpha_nZ_{ii'}(\bbeta))}_{\hat\cL_n(\bbeta)}+\lambda_n\sum_{j=2}^d|\bbeta_j|\Big\},
\end{align}
where $\tilde S_{ii'}:=\sign(Y_i-Y_{i'})$ is the signed pairwise difference. 

On one hand, the smoothed loss function $\hat\cL_n(\bbeta)$ is close to the MRC loss function in \eqref{eq:loss-han2} when $\alpha_n$ is large enough. This guarantees ``the bias term" is small enough. On the other hand, $\hat\cL_n(\bbeta)$ is smooth, giving computational and statistical guarantees for convergence in optimizing the loss function \eqref{eq:opt1}. 

There are several notable remarks for the proposed smoothing approach. 

\begin{remark}\label{remark:example}
In practice, we can take the approximation function $F_{ii'}(u)$ of the following forms:
\begin{itemize}
\item sigmoid function: $F_s(u):=1/(1+\exp(-u))$; 
\item standard Gaussian CDF: $F_g(u):=\Phi(u)$, where $\Phi(\cdot)$ represents the standard Gaussian CDF; 
\item standard double exponential CDF: $F_e(u):=1/2+\sign(u)(1-\exp(-|u|))/2$.
\end{itemize}
As will be shown later, the above approximation functions all guarantee efficient inference. For the approximation parameter $\alpha_n$, theoretically, we recommend choosing it using a specific rate. Such a rate depends on $n,d$, and a sparseness parameter, and will be stated more  explicitly in Section \ref{sec:theory}. In practice, cross-validation is recommended \citep{stone1974cross}. 
\end{remark}

\begin{remark} \hf{We note the formulation \eqref{eq:opt1} is related to those smoothing approaches introduced in \cite{brown2005standard}, \cite{zhang2013statistical}, and \cite{shi2014penalized}. Actually, their approaches could be regarded as special cases of \eqref{eq:opt1}, by taking $F_{ii'}(\cdot)$ to be the Gaussian CDF or sigmoid function. However, as will be seen in Sections \ref{sec:real} and \ref{sec:theory}, we will add new contributions to literature in both theory and applications.} 
\end{remark}

\begin{remark} It is also worth comparing the formulation in \eqref{eq:opt1} to the other robust regression formulations introduced in the high dimensional statistics literature. The original lasso estimator is well-known to be vulnerable to non-linear link functions \citep{van2008high}, and heavy-tailed noise term $\epsilon$ \citep{bickel2009simultaneous}.  \cite{lozano2013minimum}, \cite{fan2014robust}, and \cite{loh2015statistical} proposed different robust (non)convex approaches to address the possible heavy-tailedness issue of $\epsilon$. In particular, \cite{loh2015statistical} provided a framework for investigating a group of (non)convex loss functions (e.g., Huber's loss and Tukey's biweight), and studied the corresponding estimators. However, these procedures all stick to the linear link function, and hence will lead to inconsistent estimation, invalid statistical inference, and erroneous predictions when the link function is non-linear. As is discussed in the introduction, non-linearity is common in complex biology systems.
\end{remark}

\section{Real data example}\label{sec:real}


In this section, we apply our approach to the motivating scientific problem -- predicting TFs that bind to individual cis-elements. As introduced before, solving this problem is crucial for studying gene regulation.   DNase-seq experiments can be used to map active cis-elements in a genome-wide fashion. If one can correctly predict which TFs bind to each cis-element, one would be able to couple DNase-seq with computational predictions to efficiently predict genome-wide binding sites of a large number of TFs simultaneously in a new biological sample. This cannot be achieved using any other existing experimental technology. 

The conventional approach that predicts binding TFs based on DNA motifs is contingent on the availability of known TF motifs. However, 2/3 of human TFs do not have known motifs. This motivates us to investigate an alternative solution that does not require DNA motif information. This new approach leverages large amounts of publicly available DNase-seq and gene expression data generated by the ENCODE project. Using data from multiple ENCODE cell types, this approach models the relationship between a cis-element's protein-binding activity $Y$ measured by DNase-seq and the gene expression levels of $d$ TFs, $\mathbf{X}$, measured by exon arrays. It predicts binding TFs by identifying important variables in $\mathbf{X}$ in the regression model. Below we present real data results from a small-scale pilot study as a way to illustrate our semiparametric regression approach and compare it with other methods. A comprehensive whole-genome analysis and investigation of biology are beyond the scope of this article and will be addressed elsewhere. 

In our pilot study, human DNase-seq and matching gene expression exon array data from 57 cell types were obtained from the ENCODE project. After data processing and normalization (see the appendix Section \ref{supp:pp} for details), 169 TFs whose gene expression varies across the 57 cell types were obtained. In parallel, 1,000 cis-elements were randomly sampled from a total of over $10^6$ cis-elements in the human genome for method evaluation. For each cis-element, the objective is to identify which of the 169 TFs may bind to it. Let $Y$ be a cis-element's DNase I hypersensitivity (a surrogate for protein-binding activity), measured by its normalized and log-transformed DNase-seq read count, in a cell type. Let $\bX$ be the normalized gene expression values of the 169 TFs in the same cell type. We want to use $Y$ and $\bX$ observed from 57 different cell types to learn their relationship and subsequently predict binding TFs. 

\hf{Six} competing methods are compared, listed below. \hf{For all methods except random prediction, the non-zero components of the estimate were used to predict which TFs can bind to a cis-element. The code that implements our method has been released online (\url{https://github.com/zji90/RMRCE}). }

\begin{itemize}
\item RMRCE: the generalized regression model $Y=D\circ \Lambda(\bX^\T\bbeta^*,\epsilon)$ was fitted using Regularized Maximum Rank Correlation Estimator (RMRCE) in (\ref{eq:opt1}). \hf{The tuning parameter $\alpha_n$ was selected using cross validation and the Gaussian smoothing approximation was used.}
\item \hf{Hinge: the indicator function in Han's proposal is replaced by a hinge loss approximation. Specifically, the loss function is changed to:
\[ \underset{\bbeta\in\reals^d, \beta_1=1}{\largmax} \Big \{\frac{1}{n(n-1)}\sum_{i\ne i'}\ind(Y_i>Y_{i'}) [\max\{0,(\bX_i-\bX_{i'})^\T\bbeta+1\}]-\lambda_n\sum_{j=2}^d|\bbeta_j|\Big\}.
\]}
\item Lasso: the lasso \citep{tibshirani1996regression} was used.
\item \hf{SIM: the method as proposed by \cite{radchenko2015high}.}
\item \hf{SDR: the sufficient dimension reduction method as proposed by \cite{yin2015sequential}.}
\item Random: TFs randomly sampled from the 169 TFs were treated as the predicted binding TFs.  
\end{itemize}

Among these methods, the lasso represents the state-of-the-art linear model for characterizing the relationship between a response and sparse predictors. \hf{SIM and SDR represent competing semiparametric regression models. Hinge is a simple convex relaxation of Han's proposal. We include this comparison to find out whether the smoothed rank correlation is better than convex relaxation.}
The random method serves as a negative control. For \hf{all methods except random prediction}, we tried different tuning parameters ($\alpha_n$ in RMRCE was \hf{selected using cross validation}) and calculated the overall accuracy under each parameter setting. Detailed implementation strategy for the methods is presented in Section \ref{sec:algorithm} in the appendix.

  \begin{figure}[!htbp]
      \centering
      \includegraphics[width=0.69\textwidth]{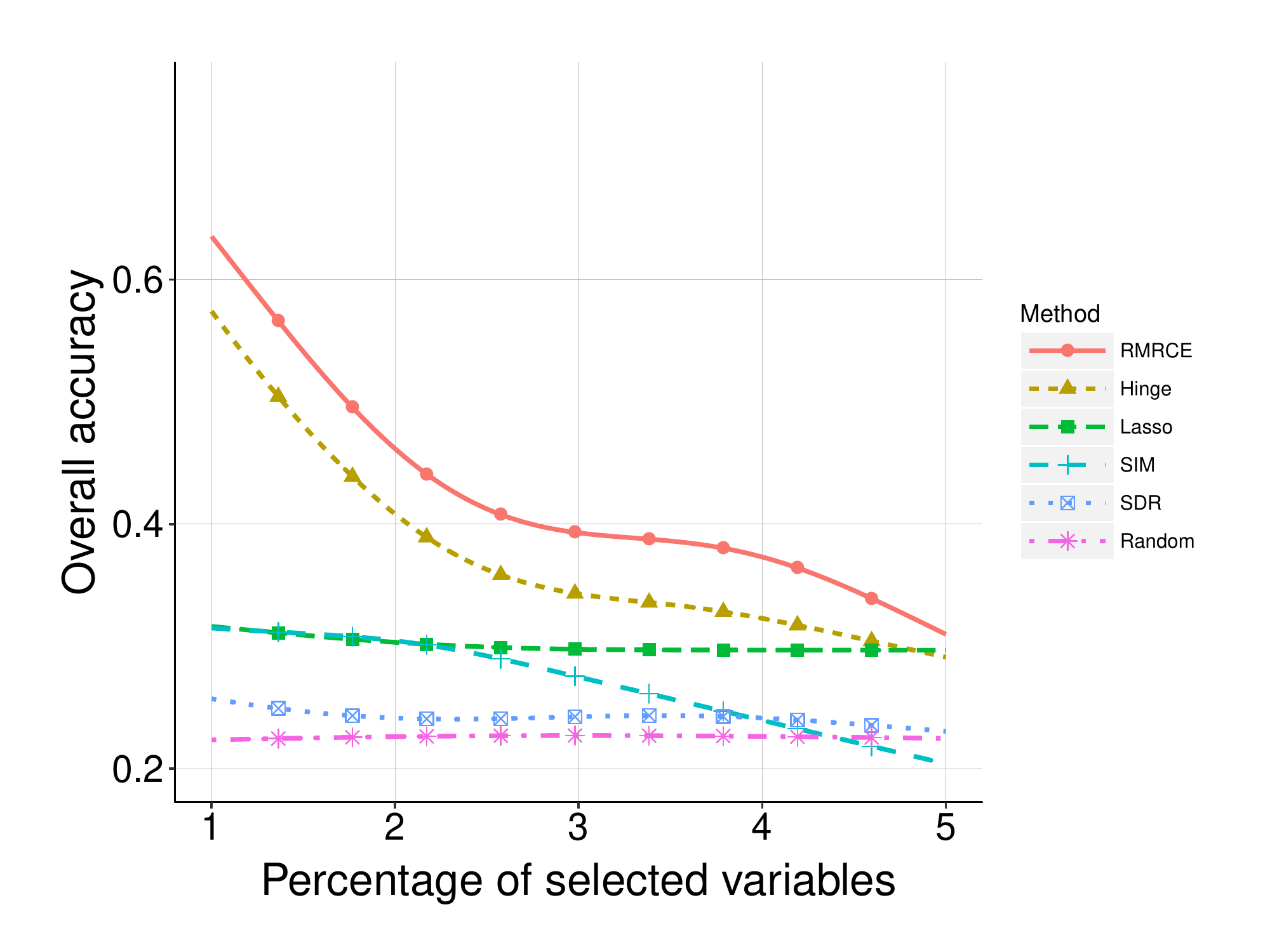}
      \caption{Overall accuracy of \hf{RMRCE ($\alpha_n$ selected by cross validation, Gaussian smoothing approximation), Hinge, the lasso, SIM, SDR, and the random prediction method}. X-axis shows the averaged percentage of selected TFs out of all 169 TFs. Y-axis shows the overall accuracy.}
      \label{TF_accuracy_allmethod}
      \end{figure}

{Prediction accuracy of different methods is compared using DNA motif information. The rationale is that DNA motifs were not used to make predictions, therefore they provide an independent source of information for validation. 
For a method with better prediction accuracy, one should expect that its predicted binding TFs for a cis-element are more likely to be supported by the presence of corresponding motifs in the cis-element. By contrast, it is less likely to find motifs in a cis-element for incorrectly predicted binding TFs. Based on this reasoning, we downloaded all known vertebrate motif matrices from the JASPAR database \citep{mathelier2015jaspar} and mapped them to human genome using the CisGenome software \citep{ji2008integrated} under its default setting. For a given TF, if its motif(s) occurred one or multiple times within 500 base pairs (bp's) of the center of a cis-element, then the TF's motif was called to be found in the cis-element. 
Let $i \in \{1,\ldots,1,000\}$ be the index of cis-elements. Let $M_i$ denote the set of TFs whose motifs were found in cis-element $i$. In order to characterize a method's prediction accuracy, we applied the method to predict binding TFs for each cis-element. If a predicted TF does not have any known DNA motif, we lack information for evaluating the correctness of the prediction. Therefore, for each cis-element, we only retained the predicted TFs that had known DNA motifs in the JASPAR database for estimating the prediction accuracy. Among all the 169 TFs, 63 TFs had known DNA motifs and were included in the evaluation. Let $A_i$ be the set of retained TFs for cis-element $i$, and let $|A_i|$ be the number of TFs in $A_i$. Let $B_i = A_i \bigcap M_i$ be the subset of TFs in $A_i$ whose motifs were found in cis-element $i$ (and hence validated), and let $|B_i|$ be the number of TFs in $B_i$.
The prediction accuracy of a method was characterized by the following ratio
\[
\sum_{i=1}^{1,000} |B_i|\Big/\sum_{i=1}^{1,000} |A_i|.
\]
This ratio is the percentage of all testable predictions that were validated by the presence of DNA motifs. The higher the ratio, the more accurate a method is.
}

{Figure \ref{TF_accuracy_allmethod} compares the accuracy of different methods. For each method, we gradually increased the number of reported TFs by relaxing the tuning parameter (e.g., setting a smaller tuning parameter $\lambda_n$ will result in more TFs being reported by RMRCE), and the accuracy was plotted as a function of increasing number of predicted binding TFs. This figure shows that RMRCE is significantly more accurate than \hf{all the other methods, and the random prediction method has the worst performance}. Of note, as the number of selected TFs increases, the accuracy of all methods drops (except for the random, which remains stable). This is as expected since the overall signal strength decreases as more TFs are reported. \hf{RMRCE performance with different choices of $\alpha_n$ is presented in the appendix (Section \ref{sec:realdataalpha}). A model diagnostic heuristic is developed to check the monotonicity assumption of the proposed model. The detailed descriptions and model diagnostic results in real data application can be found in the appendix (Section \ref{sec:realdatadiagnostic}).} 

{To shed light on why RMRCE substantially outperformed the lasso, Figure \ref{cislassoexample}  shows data from two cis-elements as examples. For each cis-element, we used the lasso to identify binding TFs from the 169 TFs. The observed response $\bY$ was then plotted against its fitted value $\bX^\T \widehat {\bm \beta}^{\rm lasso}$ in Figure \ref{cis1lassofit} and Figure \ref{cis2lassofit}. The blue curve in each plot represents a smooth curve fitted using {the generalized additive model with cubic splines and default parameters as implemented in the R package {\bf mgcv}}. {It treats} $\bY$ as response and $\bX^\T \widehat {\bm \beta}^{\rm lasso}$ as independent variable. Clearly, $\bY$ and $\bX^\T \widehat {\bm \beta}^{\rm lasso}$ do not have a linear relationship. Moreover, the figures show that the relationship between $\bY$ and $\bX^\T \widehat {\bm \beta}^{\rm lasso}$ for different cis-elements have different functional forms. This makes the use of parametric models complicated as one would need to build models with different functional forms for different cis-elements, which would be tedious if one wants to analyze millions of cis-elements in the whole genome.
Figure \ref{cis1lassoresidual} and Figure \ref{cis2lassoresidual} show the normal qqplots for the residuals that were obtained from the fitted smooth curves. These figures show that, even when a non-linear smoothed function was fitted to the data, the residuals are  still non-normal and may have a complicated distribution.} 

\begin{figure}[!htbp]
\centering
\subfigure[Subfigure 2 list of figures text][Fitted curve for cis-element 1 using the lasso]{
\includegraphics[width=0.42\textwidth]{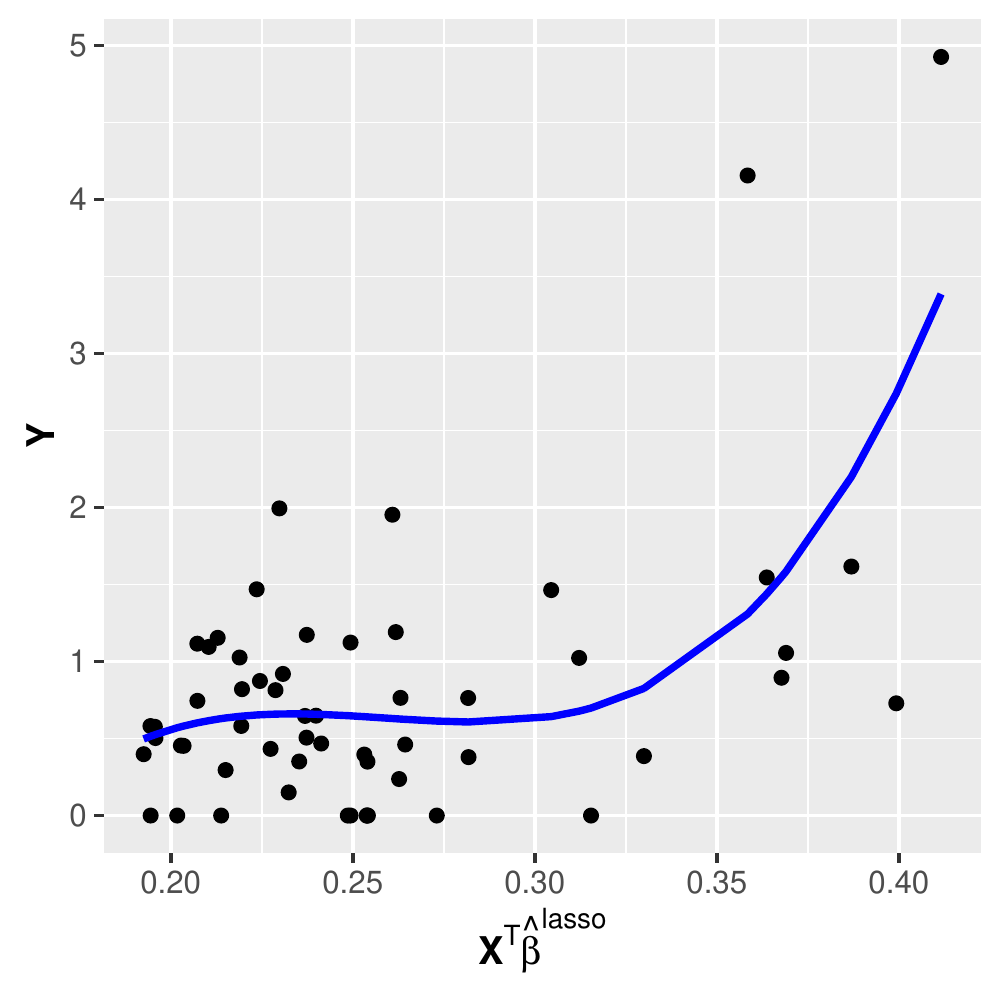}
\label{cis1lassofit}}
\subfigure[Subfigure 2 list of figures text][Residual qqplot for cis-element 1 using the lasso]{
\includegraphics[width=0.42\textwidth]{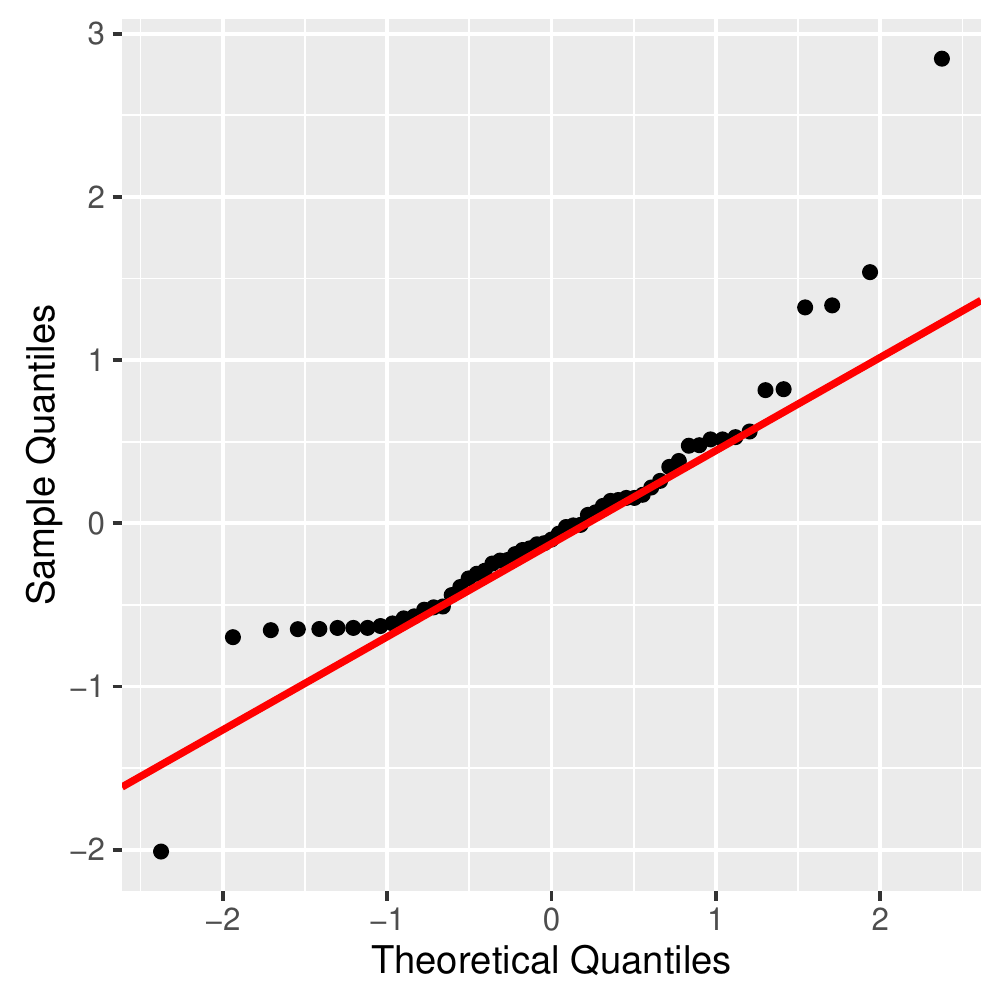}
\label{cis1lassoresidual}}

\subfigure[Subfigure 2 list of figures text][Fitted curve for cis-element 2 using the lasso]{
\includegraphics[width=0.42\textwidth]{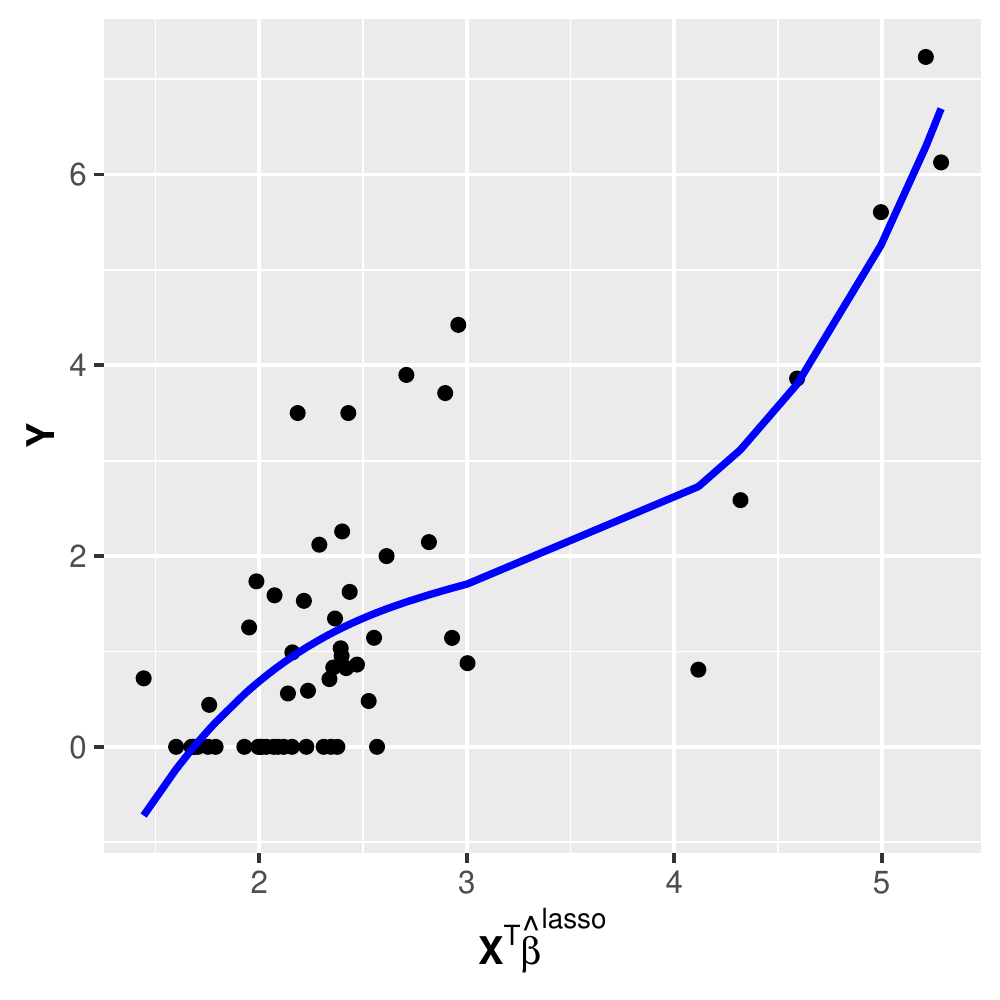}
\label{cis2lassofit}}
\subfigure[Subfigure 1 list of figures text][Residual qqplot for cis-element 2 using the lasso]{
\includegraphics[width=0.42\textwidth]{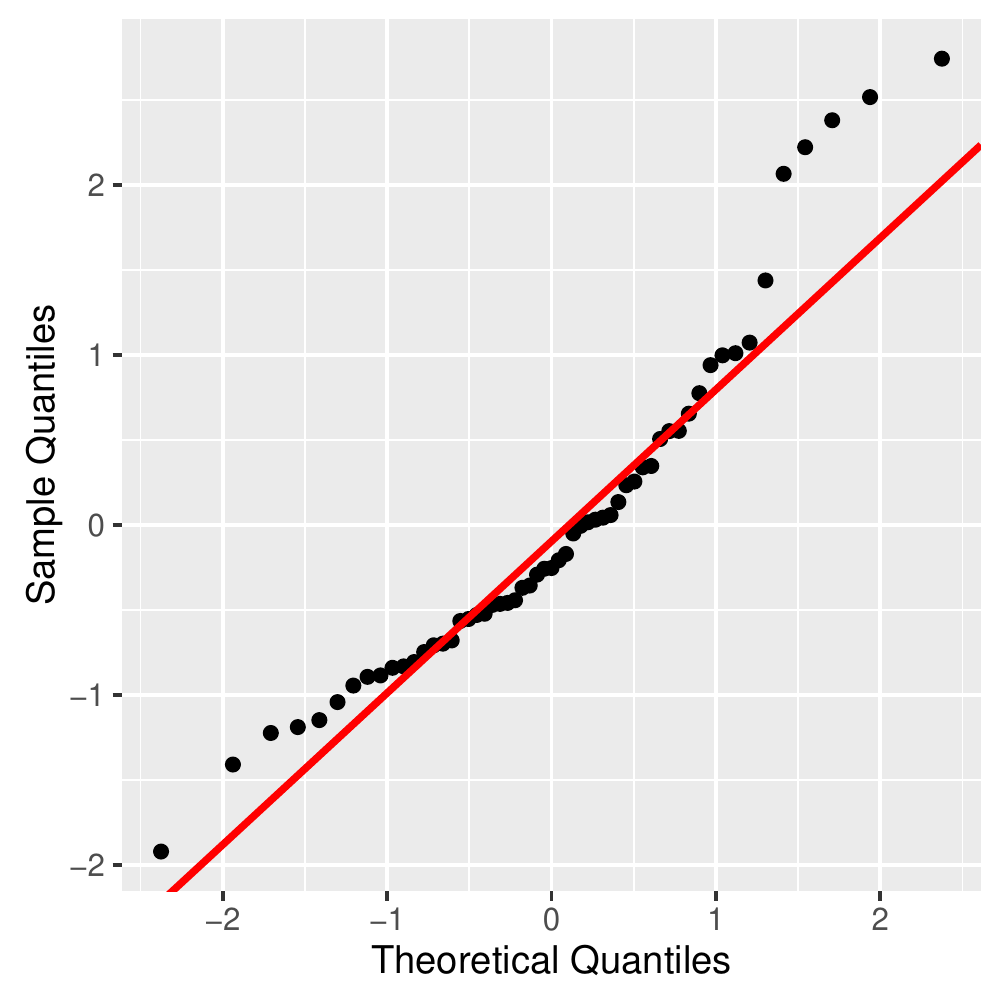}
\label{cis2lassoresidual}}
\caption{\textbf{Fitted curves and residual qqplots using the lasso for two cis-elements.} For each cis-element, the lasso was fitted to obtain the regression coefficients $\widehat {\bm \beta}^{\rm lasso}$. Blue lines in (a) and (c) are fitted smoothed curves treating $\bY$ as response and $\bX^\T \widehat {\bm \beta}^{\rm lasso}$ as independent variable. (b) and (d) are residuals qqplots for the smoothed curves. The quantiles of the residuals (sample quantiles) are compared to the quantiles of a normal distribution (theoretical quantiles). The residuals are calculated as $\bY-g(\bX^\T \widehat {\bm \beta}^{\rm lasso})$ where $g$ represents the smoothed functions fitted in (a) and (c).}
\label{cislassoexample}
\end{figure}

{Figure \ref{cisRMRCEexample} shows a similar analysis using RMRCE. In Figure \ref{cis1RMRCEfit} and Figure \ref{cis2RMRCEfit}, $\bY$ was plotted against the RMRCE fitted $\bX^\T \widehat {\bm \beta}^{\rm RMRCE}$. Clearly, the relationship between $\bY$ and $\bX^\T \widehat {\bm \beta}^{\rm RMRCE}$ is non-linear, but such a non-linear{, yet monotonically increasing,} relationship can be handled by our method in a unified fashion regardless of the specific functionnon-linearal forms. Figure \ref{cis1RMRCEresidual} and Figure \ref{cis2RMRCEresidual} are the residual qqplots where the residuals were obtained from the fitted smooth curves in Figure \ref{cis1RMRCEfit} and Figure \ref{cis2RMRCEfit}. These qqplots show that the distributions of the residuals are non-normal. The non-normal residuals, however, can be naturally handled by our generalized regression model.}

The above analyses demonstrate the value of our approach for handling noisy, monotonic, non-normal, and non-linear data. 
Whereas the simple linear regression and marginal screening cannot fully capture the delicateness of such a complex system, RMRCE handles these challenges very well. 
Thus, RMRCE is a more appealing method to tackle the studied problem than simple linear models based methods such as the lasso. \hf{Similar conclusion applies to the comparison with the other four competing methods. In particular, as will be shown in the Section of synthetic data analysis (Section \ref{sec:simulation}), Hinge is usually a convex approximation too crude to the studied method, and when the generalized regression model is reasonable in modeling the data (which is hinted by the experimental results in this section), SIM and SDR are much less efficient in handling the data than RMRCE.}

\begin{figure}[!htbp]
\centering
\subfigure[Subfigure 2 list of figures text][Fitted curve for cis-element 1 using RMRCE]{
\includegraphics[width=0.35\textwidth]{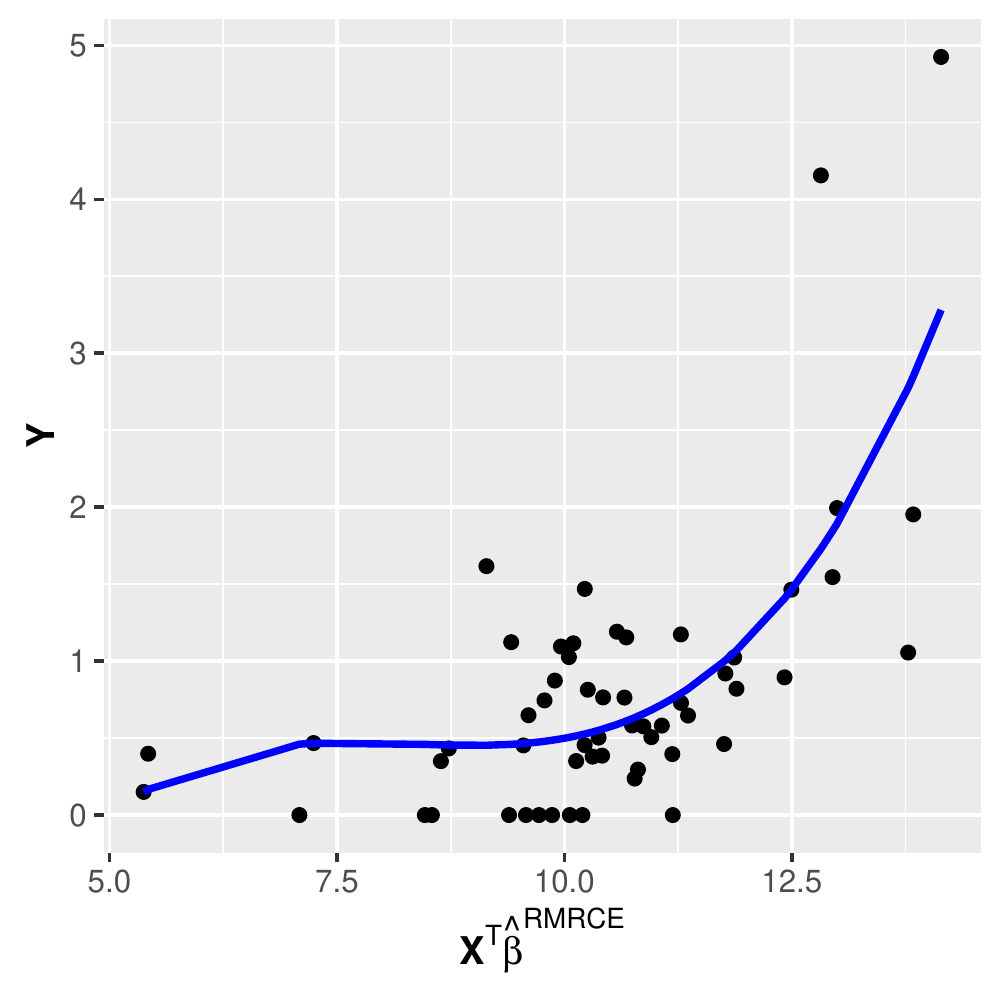}
\label{cis1RMRCEfit}}
\subfigure[Subfigure 2 list of figures text][Residual qqplot for cis-element 1 using RMRCE]{
\includegraphics[width=0.35\textwidth]{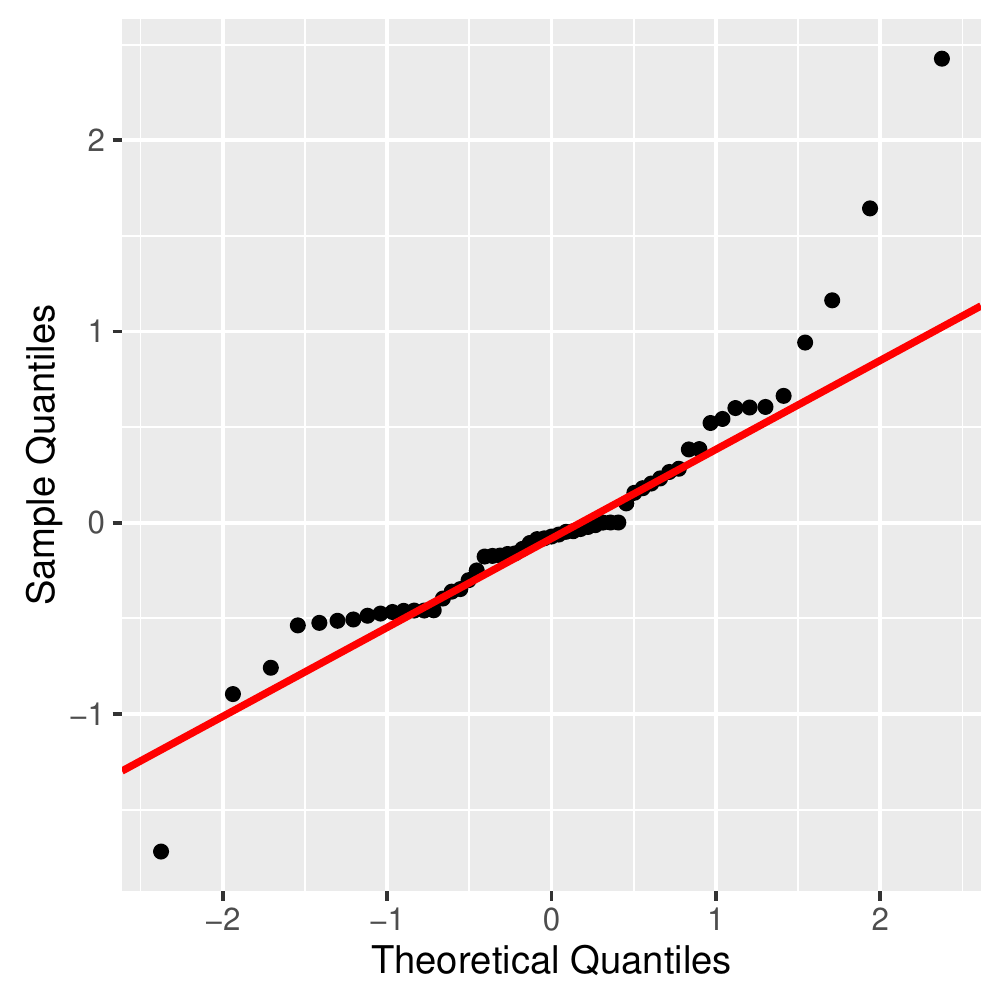}
\label{cis1RMRCEresidual}}
\qquad
\subfigure[Subfigure 2 list of figures text][Fitted curve for cis-element 2 using RMRCE]{
\includegraphics[width=0.35\textwidth]{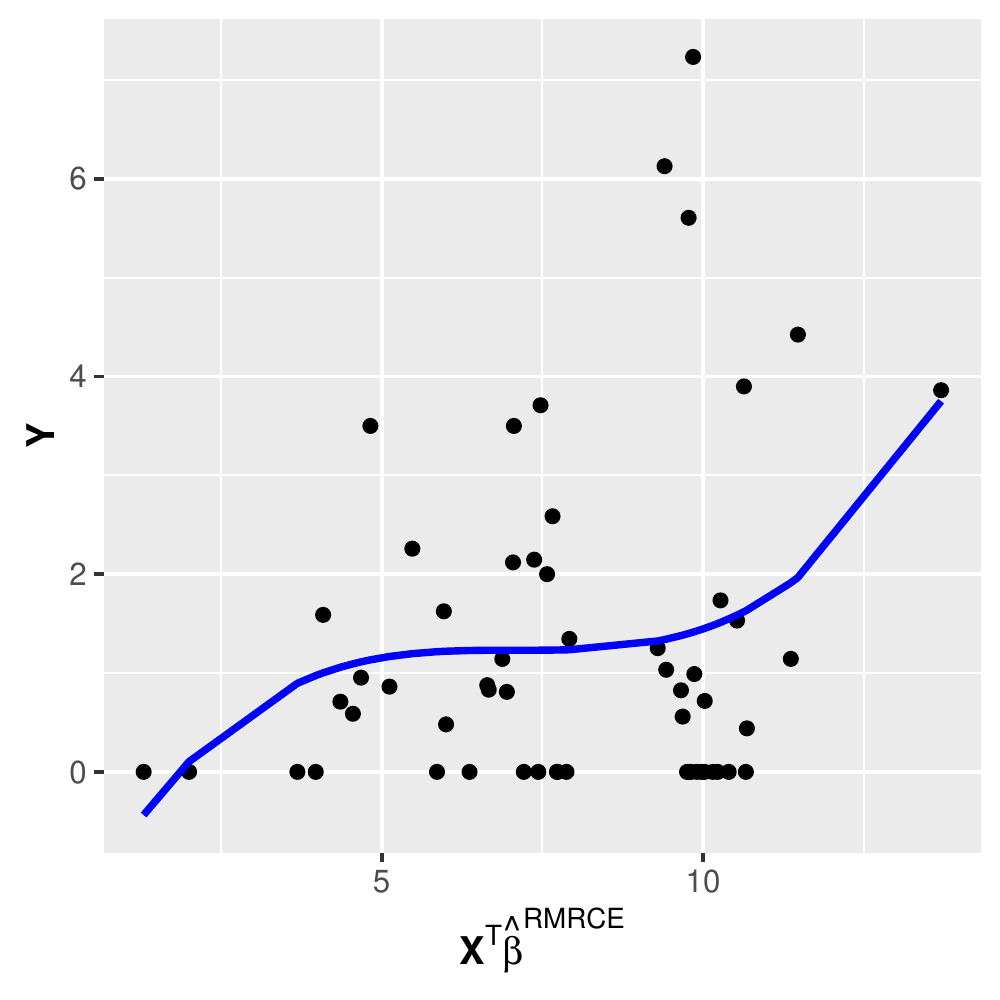}
\label{cis2RMRCEfit}}
\subfigure[Subfigure 1 list of figures text][Residual qqplot for cis-element 2 using RMRCE]{
\includegraphics[width=0.35\textwidth]{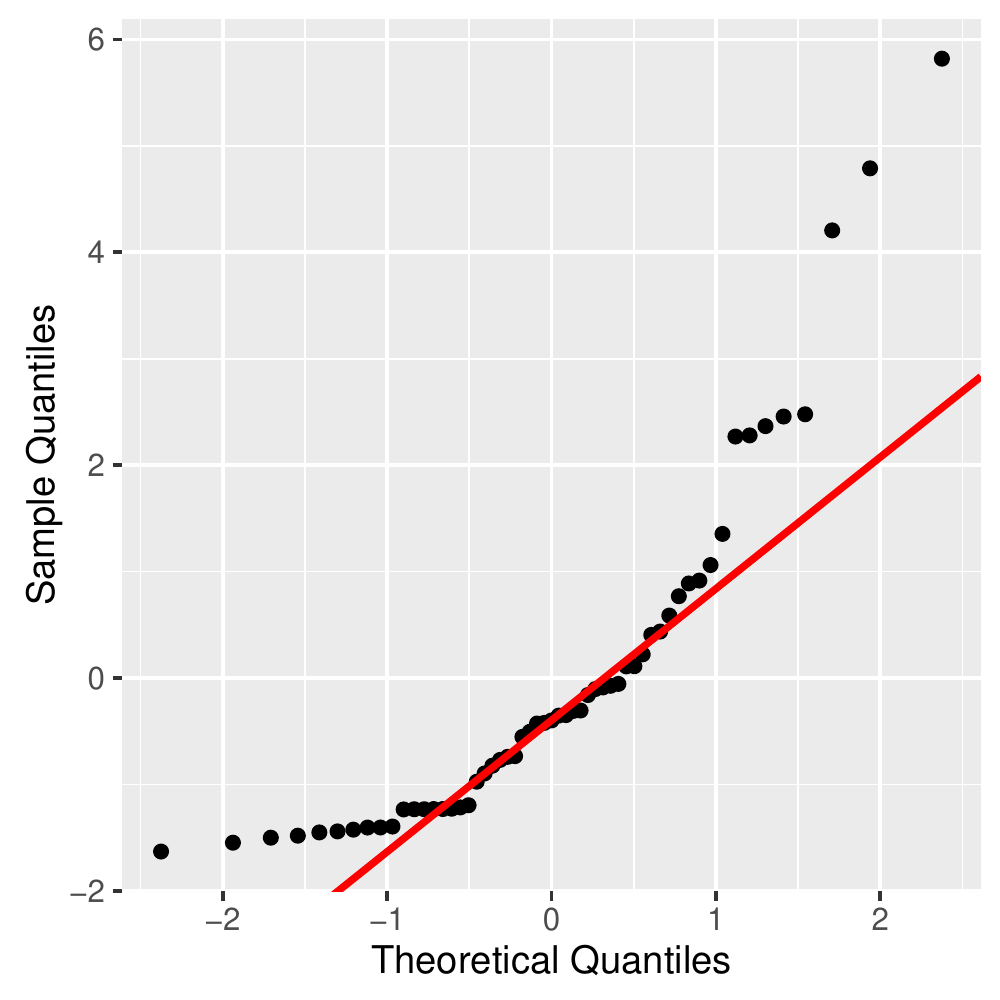}
\label{cis2RMRCEresidual}}
\caption{\textbf{Fitted curves and residual qqplots using RMRCE for two cis-elements.} For each cis-element, RMRCE was fitted to obtain the regression coefficients $\widehat {\bm \beta}^{\rm RMRCE}$. Blue lines in (a) and (c) are fitted smoothed curves treating $\bY$ as response and $\bX^\T \widehat {\bm \beta}^{\rm RMRCE}$ as independent variable. (b) and (d) are residuals qqplots for the smoothed curves. The quantiles of the residuals (sample quantiles) are compared to the quantiles of a normal distribution (theoretical quantiles). The residuals are calculated as $\bY-g(\bX^\T \widehat {\bm \beta}^{\rm RMRCE})$ where $g$ represents the smoothed functions fitted in (a) and (c).}
\label{cisRMRCEexample}
\end{figure}

\section{Theory}\label{sec:theory}

This section is devoted to investigating the statistical performance of the smoothing estimator $\hat\bbeta_{\alpha_n}$ in \eqref{eq:opt1}. For characterizing the estimation error of $\hat\bbeta_{\alpha_n}$ to $\bbeta^*$, we separately study the approximation error (bias part) and the stochastic error (variance part).

Before introducing main theorems, let's first introduce some additional notation. Let's first define the population approximation minimizer $\bbeta^*_{\alpha_n}$ as 
\[
\bbeta^*_{\alpha_n} \in \underset{\bbeta\in\reals^d, \beta_1=1}{\largmin } \cL_n(\bbeta), 
\]
where $\cL_n(\bbeta):=\E\hat\cL_n(\bbeta)$ can be easily derived as
\begin{align}\label{eq:Ln}
\cL_n(\bbeta)=-\frac{2}{n(n-1)}\sum_{1\leq i<i'\leq n}\E\Big\{S_{ii'}F_{ii'}(\alpha_nZ_{ii'}(\bbeta))+(1-S_{ii'})(1-F_{ii'}(\alpha_nZ_{ii'}(\bbeta))) \Big\} .
\end{align}
Note, analogously, Proposition \ref{prop:1} shows
\begin{align}\label{eq:L}
\bbeta^*=\argmin_{\bbeta\in\reals^d, \beta_1=1}\underbrace{-\E\big(S_{ii'}\mathds{1}(Z_{ii'}(\bbeta)>0)+(1-S_{ii'})\mathds{1}(Z_{ii'}(\bbeta)<0)\big)}_{\cL(\bbeta)}.
\end{align}
We decompose the statistical error $\norm{\hat\bbeta_{\alpha_n}-\bbeta^*}_2$ into two parts: 
\[
\norm{\hat\bbeta_{\alpha_n}-\bbeta^*}_2\leq \underbrace{\norm{\bbeta^*_{\alpha_n}-\bbeta^*}_2}_{\rm approximation~error} + \underbrace{\norm{\hat\bbeta_{\alpha_n}-\bbeta^*_{\alpha_n}}_2}_{\rm stochastic~error}.
\]
We will first characterize the approximation error  in the next section. Studies of the stochastic error are in Section \ref{sec:variance}. 

Because the loss functions $\cL(\bbeta)$, $\cL_n(\bbeta)$, and $\hat\cL_n(\bbeta)$ are all non-convex, adopting a similar argument as in \cite{loh2015statistical}, we first focus on the following specified minima:
\[
\hat\bbeta_{r,\alpha_n}:=\argmin_{\beta_1=1,\norm{\bbeta-\bbeta^*}_2\leq r}\Big\{\hat\cL_n(\bbeta)+\lambda_n\sum_{j=2}^d|\bbeta_j|\Big\}~~~{\rm and}~~~
\bbeta_{r,\alpha_n}^*:=\argmin_{\beta_1=1,\norm{\bbeta-\bbeta^*}_2\leq r} \cL_n(\bbeta).
\]
The estimators $\hat\bbeta_{r,\alpha_n}$ and $\bbeta_{r,\alpha_n}^*$ are constructed merely for theoretical purposes. In addition, the parameter $r$ there, controlling the convexity region around the truth $\bbeta^*$, is no need to be specified in practice. 

As will be shown in the next two subsections, $\hat\bbeta_{r,\alpha_n}$ and $\bbeta_{r,\alpha_n}^*$ are local optima to $\hat\cL_n(\cdot)$ and $\cL_n(\cdot)$ under some explicitly characterized regularity conditions. Thus, we will end this section with a general theorem for quantifying the behaviors of some local optimum $\hat\bbeta_{\alpha_n}$.

\subsection{Approximation error}

This section first shows  $\cL_n(\cdot)$ in \eqref{eq:Ln} could well approximate $\cL(\cdot)$ in \eqref{eq:L}. We will then study the behaviors of minima of $\cL_n(\cdot)$ and $\cL(\cdot)$.

\subsubsection{Generalized regression model}

For the generalized regression model of the form \eqref{eq:model}, to guarantee the fast approximation of $\cL_n(\cdot)$ to $\cL(\cdot)$, we require the following two assumptions on data generating schemes and approximation functions $\{F_{ii'}(\cdot), 1\leq i<i'\leq n\}$.
\begin{itemize}
\item[] {\bf (A1).} Model \eqref{eq:model} and the constraint \eqref{eq:constraint} hold with $\bbeta^*$ non-degenerate, $\{\bX_i,~i=1,\ldots,n\}\stackrel{i.i.d.}{\sim} N_d(\bmu,\bSigma)$ and independent of i.i.d. absolutely continuous noises $\{\epsilon_i,~i=1,\ldots,n\}$. Of note, the noise term $\epsilon_i$ is not necessarily of mean or median zero.
\item[] {\bf (A2).} Assume $F_{ii'}(u)$ satisfies $F_{ii'}(-u)=1-F_{ii'}(u)$ for arbitrary $u\geq 0$. Further assume there exist some absolute constants $C_1, C_2>0$ such that $1-F_{ii'}(u)\leq C_1\exp(-C_2u)$ for any $u>0$.
\end{itemize}

On one hand, Assumption {\bf (A1)} is a commonly adopted assumption in the high dimensional regression literature \citep{raskutti2010restricted, negahban2012unified}. Of note, Assumption {\bf (A1)} is by no means necessary. By checking the proof of Lemma \ref{lem:approx}, we could readily relax it to a multivariate subgaussian assumption \citep{cai2012optimal}. However, for presentation clearness, this paper is focused on the multivariate Gaussian case. On the other hand, Assumption {\bf (A2)} is mild. It is easy to check sigmoid function, Gaussian, and double exponential CDFs discussed in Remark \ref{remark:example} all satisfy {\bf (A2)}.

Under Assumptions {\bf (A1)} and {\bf (A2)}, the next lemma investigates the approximation error of $\cL_n(\cdot)$ to $\cL(\cdot)$. 
\begin{lemma}\label{lem:approx}
Assume Assumptions {\bf (A1)} and {\bf (A2)} hold. We then have
\[
\sup_{\bbeta:\beta_1=1}|\cL_n(\bbeta)-\cL(\bbeta)|\leq\frac{2C_1}{C_2\alpha_n}\sup_{\bbeta:\beta_1=1}\frac{1}{\sqrt{2\bbeta^\T\bSigma\bbeta}},
\]
where absolute constants $C_1$ and $C_2$ are given in Assumption {\bf (A2)}. 
\end{lemma}

Lemma \ref{lem:approx} shows $\cL_n(\cdot)$ uniformly approximates $\cL(\cdot)$ linearly with regard to $1/\alpha_n$. However, it is not enough to show convergence for $\bbeta^*_{\alpha_n}$ to $\bbeta^*$. For guaranteeing this, we require a ``local strong convexity" of $\cL(\cdot)$ around the truth $\bbeta^*$. To this end, we write
\[
\bGamma:=\nabla^2\cL(\bbeta^*) \in \reals^{d\times d}
\]
to be the second derivative of the population loss function $\cL(\cdot)$ in \eqref{eq:L}.  The next assumption regularizes the behavior of $\bGamma$'s spectrum. 
\begin{itemize}
\item[] {\bf (A3).} Assume $\lambda_{\min}(\bGamma)$ and $\lambda_{\max}(\bGamma)$ are respectively lower and upper bounded by two absolute positive constants. 
\end{itemize}
Assumption {\bf (A3)} is also posed inexplicitly in \cite{sherman1993limiting}. With Assumption {\bf (A3)}, the next proposition, essentially coming from \cite{sherman1993limiting}, shows the local strong convexity of $\cL(\cdot)$. 

\begin{proposition}[\cite{sherman1993limiting}]\label{prop:convexity} Assume Assumptions {\bf (A1)} and {\bf (A3)} hold. We can then pick positive absolute constant set $\gamma:=\{\gamma_1,\gamma_2\}$ with $\gamma_2/\gamma_1-1$ close to 0, such that for some small enough $r(\gamma)>0$ only depending on $\gamma$, as long as  $ \|\bbeta-\bbeta^*\|_2\leq r(\gamma)$, we have
\[
\gamma_1\lambda_{\min}(\bGamma)\norm{\bbeta-\bbeta^*}_2^2 \leq \cL(\bbeta)-\cL(\bbeta^*)\leq  \gamma_2\lambda_{\max}(\bGamma)\norm{\bbeta-\bbeta^*}_2^2.
\]
\end{proposition}

Combining Lemma \ref{lem:approx} and Proposition \ref{prop:convexity}, the next lemma characterizes the approximation error of $\bbeta^*_{r(\gamma),\alpha_n}$ to the truth $\bbeta^*$. 
\begin{lemma}\label{lem:app-error}
Assume Assumptions {\bf (A1), (A2),} and {\bf (A3)} hold. We then have
\[
\|\bbeta_{r(\gamma),\alpha_n}^*-\bbeta^*\|_2^2\lesssim \alpha_n^{-1}\cdot\sup_{\bbeta:\beta_1=1}\frac{1}{\sqrt{2\bbeta^\T\bSigma\bbeta}}.
\]
\end{lemma}
Lemma \ref{lem:app-error} verifies that, when we choose $\alpha_n$ to increase to infinity with $n$, the approximation error decays to zero at a rate $\alpha_n^{-1/2}$ while assuming the boundedness of the eigenvalues of $\bGamma$.  We are focused on such $\bbeta^*_{r(\gamma),\alpha_n}$ and the corresponding estimator $\hat\bbeta_{r(\gamma),\alpha_n}$. 

\subsubsection{Monotonic transformation model}

This section aims to study how sharp the results in the last subsection are. We focus on the monotonic transformation model \eqref{eq:model1}. We first show Lemma \ref{lem:approx} cannot be improved too much, even under a much more restrictive monotonic transformation model. 

\begin{lemma}\label{lem:approx2}
Under Model \eqref{eq:model1}, assume Assumption {\bf (A1)} and $F_{ii'}(u)$ satisfies $F_{ii'}(-u)=1-F_{ii'}(u)$ for arbitrary $u\geq 0$. For arbitrary vectors $\norm{\bbeta}_2=\norm{\bbeta^*}_2=M<\infty$, we then have the following three statements true.
\begin{itemize}
\item[] (1). Supposing $\bbeta^\T\bSigma\bbeta^*=0$, we have $\cL_n(\bbeta)=\cL(\bbeta)$ for any $\alpha_n>0$.
\item[] (2). With $\epsilon$ Gaussian distributed with bounded parameters, fixing $\alpha_n$ and supposing $\bSigma=\Ib_d$, $|\cL_n(\bbeta)-\cL(\bbeta)|$ is an increasing function of $|\bbeta^\T\bbeta^*|$ for all three examples of approximations given in Remark \ref{remark:example}. 
\item[] (3). If we further assume sigmoid approximation and $\epsilon$ Gaussian distributed with bounded parameters, we have 
\[
|\cL_n(\bbeta^*)-\cL(\bbeta^*)| \asymp \alpha_n^{-2}.
\]
Furthermore, Combined with the second fact yields, for any $\norm{\bbeta}_2= M$, 
\[
|\cL_n(\bbeta)-\cL(\bbeta)| \lesssim \alpha_n^{-2}.
\]
\end{itemize}
\end{lemma}
Lemma \ref{lem:approx2} shows, even under a very ideal parametric model, the approximation error in Lemma \ref{lem:approx} can only be improved slightly from linearly to quadratically decaying.

Secondly, we note Proposition \ref{prop:convexity} relies on an inexplicit assumption {\bf (A3)}. To fully appreciate this proposition, we provide an alternative way to clearly reveal its connection to the data generating parameter $\bSigma$. For this, we assume the monotonic transformation model \eqref{eq:model1} and one more assumption. 
\begin{itemize}
\item[] {\bf (A3').} Assume Model \eqref{eq:model1} holds with explanatory variables $\{\bX_i,i=1,\ldots,n\}$ and noises $\{\epsilon_i,i=1,\ldots,n\}$ satisfying Assumption {\bf (A1)}. We further assume the noises  are absolutely continuous, satisfying 
\[
\int_{-\infty}^{0}f_{\epsilon}(x)\exp(-x^2/(2b_n^2))dx=Cb_n(1+o(1))~~ \rm{as}~~b_n\rightarrow 0. 
\]
Here $f_{\epsilon}(\cdot)$ represents the probability density function (PDF) of $\epsilon_2-\epsilon_1$. 
\end{itemize}
We note the noise assumption in {\bf (A3')} is mild. It does not require any moment condition on the noises $\{\epsilon_i, i=1,\ldots,n\}$. In particular, the next proposition shows both Gaussian and Cauchy distributions satisfy Assumption {\bf (A3')}. 
\begin{proposition}\label{prop:gaussian_cauchy}
Suppose noises $\{\epsilon_i,~i=1,\ldots,n\}$ are arbitrarily Gaussian or Cauchy distributed with bounded parameters. Then they satisfy Assumption {\bf (A3')}. 
\end{proposition}
With Assumption {\bf (A3')}, we are now ready to prove the local strong convexity of $\cL(\cdot)$. 
\begin{lemma}\label{lem:convexity}
Assume Assumptions {\bf (A1)} and {\bf (A3')} hold. We can then pick positive absolute constant set $\gamma:=\{\gamma_1,\gamma_2\}$ with $\gamma_2/\gamma_1-1$ close to 0, such that for some small enough $r(\gamma)>0$ only depending on $\gamma$, as long as  $ \|\bbeta-\bbeta^*\|_2\leq r(\gamma)$, we have
\[
\gamma_1 \cdot \Big(1-\frac{\bbeta^\T\bSigma\bbeta^*}{\sqrt{\bbeta^\T\bSigma\bbeta}\sqrt{\bbeta^{*\T}\bSigma\bbeta^*}}\Big) \leq \cL(\bbeta)-\cL(\bbeta^*)\leq \gamma_2  \cdot \Big(1-\frac{\bbeta^\T\bSigma\bbeta^*}{\sqrt{\bbeta^\T\bSigma\bbeta}\sqrt{\bbeta^{*\T}\bSigma\bbeta^*}}\Big).
\]
As a simple consequence, for $\bSigma=\Ib_d$ and $\norm{\bbeta^*}_2=\norm{\bbeta}_2=M$, we have 
\[
\frac{\gamma_1}{2M^2}\norm{\bbeta-\bbeta^*}_2^2 \leq \cL(\bbeta)-\cL(\bbeta^*)\leq \frac{\gamma_2}{2M^2}\norm{\bbeta-\bbeta^*}_2^2.
\]
\end{lemma}
\begin{remark}
Compared to the result in Proposition \ref{prop:convexity}, Lemma \ref{lem:convexity} gives explicit inequalities based on $\bSigma$, and the proof techniques exploited are utterly different from these in \cite{sherman1993limiting} that are based on Taylor's expansion. 
\end{remark}

\subsection{Stochastic error} \label{sec:variance}

This section investigates the stochastic error term $\norm{\hat\bbeta_{\alpha_n}-\bbeta^*_{\alpha_n}}_2$. This falls in the application regime of the high dimensional M-estimators theory \citep{negahban2012unified} with some slight modifications due to the additional constraint on $\bbeta:\beta_1=1$ and $\norm{\bbeta-\bbeta^*}_2\leq r$. 

\subsubsection{A general framework for constrained M-estimators}

In this section we consider studying general M-estimators. In detail, let's be focused on the following constrained M-estimator:
\[
\hat\btheta:=\argmin_{\btheta\in\cA\subset\reals^d}\Big\{L_n(\btheta) + \lambda_nP(\btheta) \Big\},
\]
which aims to estimate the truth
\[
\btheta^*:=\argmin_{\btheta\in\cA\subset\reals^d}\E L_n(\btheta).
\]
Here $\cA\subset\reals^d$ is a subset of the $d$-dimensional real space, $L_n(\cdot)$ is the loss function, and $P(\cdot)$ is the penalty term. We pose the following five assumptions on $\cA$, $L_n(\cdot)$, and $P(\cdot)$. 
\begin{itemize}
\item {\bf (B1).} Assume $\cA-\btheta^*:=\{\bv\in\reals^d:\bv=\btheta-\btheta^*~{\rm for~some}~\btheta\in\cA\}$ is star-shaped. 
\item {\bf (B2).} Assume $L_n(\cdot)$ is convex differentiable in $\cA$, and $P(\cdot)$ is a semi-norm.
\item {\bf (B3).} (Decomposability). There exist subspaces $\cM\subset\overline\cM\subset \reals^d$ such that
\[
P(\btheta+\bgamma)=P(\btheta)+P(\bgamma)~~{\rm for~all}~\btheta\in \cM ~{\rm and}~\bgamma\in \overline\cM^{\perp}.
\]
\item {\bf (B4).} (Restricted strong convexity). Define the set
\[
\cC(\cM,\overline\cM^\perp;\btheta^*):=\{\bDelta\in\{\cA-\btheta^*\}: P(\bDelta_{\overline\cM^\perp})\leq 3P(\bDelta_{\overline\cM})+4P(\btheta^*_{\cM^\perp})\},
\]
where $\bDelta_{\cN}$ represents the projection of $\bDelta$ to $\cN$ for arbitrary subspace $\cN$ of $\reals^d$. We assume, for all $\bDelta\in \cC(\cM,\overline\cM^\perp;\btheta^*)$, we have
\[
L_n(\btheta^*+\bDelta)-L_n(\btheta^*)-\langle \nabla L_n(\btheta^*),\bDelta\rangle \geq \kappa_L\norm{\bDelta}_2^2-\delta_L\norm{\bDelta}_2-\tau_L^2(\btheta^*),
\]
where $\kappa_L$, $\delta_L$, and $\tau_L^2(\btheta^*)$ are three constants.
\item {\bf (B5).} We assume
\[
\Psi(\overline \cM):=\sup_{\bv\in\overline \cM\setminus \zero}\frac{P(\bv)}{\norm{\bv}_2}<\infty~~{\rm and}~~\lambda_n\geq 2P^*(\nabla L_n(\btheta^*)).
\]
Here $P^*(\cdot)$ is the dual norm of $P(\cdot)$.
\end{itemize}

\hf{Assumptions {\bf (B1)}-{\bf (B5)} are posed for the purpose of involving estimators like $\hat\beta_{\alpha_n}$, and hence also (slightly) generalize the corresponding ones in \cite{negahban2012unified}.} Under the above assumptions, we have the following theorem hold, which is a straightforward extension to Theorem 1 in \cite{negahban2012unified}.
\begin{theorem}\label{thm:martin} Assume Assumptions {\bf (B1)-(B5)} hold. We then have
\[
\norm{\hat\btheta-\btheta^*}_2^2 \leq (2\lambda_n\Psi(\overline\cM)+\delta_L)^2/\kappa_L^2+2(\tau^2_L(\btheta^*)+2\lambda_nP(\btheta^*_{\cM^\perp}))/\kappa_L.
\]
\end{theorem}

\subsubsection{Stochastic error analysis}

For analyzing the high dimensional stochastic error, sparsity is commonly encouraged for model identifiability and efficient inference. We adopt this idea. In particular, we assume the following regularity condition:
\begin{itemize}
\item[] {\bf (A0).} The data $\{(Y_i, \bX_i), i=1,\ldots,n\}$ are generated in a triangular array setting as in \cite{greenshtein2004persistence}. We assume, for some pre-specified $\alpha_n$, $\norm{\bbeta^*_{r(\gamma),\alpha_n}}_0\leq s_n$, while the parameter $s_n$ changes with $n$. Note, in this setting, $\bbeta^*$ is not necessarily sparse. 
\end{itemize}  
This formulation of sparseness can be regarded as a working assumption, and intrinsically comes from the sieve idea \citep{grenander1981abstract, shen1999random}. Note a similar assumption is inexplicitly stated in \cite{fan2014robust}. 

For guaranteeing efficient inference, we still require three more assumptions, listed below.
\begin{itemize}
\item[] {\bf (A4).} Assume $F_{ii'}(\cdot)$ is twice-differentiable, and there exists an absolute constant $C_3>0$ such that $\sup_{u\in\reals}|\frac{d}{du}F_{ii'}(u)|\leq C_3$ for arbitrary pair $(i,i')$ with $1\leq i\ne i'\leq n$.
\item[] {\bf (A5).} We assume $\nabla^2\hat\cL_n(\bbeta)\succeq 0$ for arbitrary $\bbeta\in\R^{d}$ with $\beta_1=1$ such that $\|\bbeta-\bbeta^*_{r(\gamma),\alpha_n}\|_2\leq C_4\cdot r(\gamma)$ for some $C_4>1$.  
\item[] {\bf (A6).} There exists an absolute constant $C_5>0$ such that $C_5^{-1}\leq \lambda_{\min}(\bSigma)\leq \lambda_{\max}(\bSigma)\leq C_5$.  
\end{itemize}
Here Assumption {\bf (A4)} is easy to check. Actually, it is straightforward to verify sigmoid function, Gaussian, and double exponential CDFs as approximation functions introduced in Remark \ref{remark:example} all satisfy Assumption {\bf (A4)}. On the other hand, for Assumption {\bf (A6)}, we require a little bit more stringent assumption on $\lambda_{\min}(\bSigma)$ than what is required for the lasso. This is because of the additional effort on controlling the smoothness approximation error. Finally, noticing 
\[
\nabla^2\hat\cL_n(\bbeta)=-\frac{2}{n(n-1)}\sum_{i<i'}\alpha_n^2(\bX_{i}-\bX_{i'})(\bX_i-\bX_{i'})^\T F_{ii'}''(\tilde S_{ii'}\alpha_nZ_{ii'}(\bbeta))
\]
is very easy to calculate, we note Assumption {\bf (A5)} could be verified empirically.  As a matter of fact, in the appendix, we will show a lot of statistical models satisfy Assumption {\bf (A5)} via exhaustive simulation studies. 
\begin{remark}
Of note, \cite{sherman1993limiting} has shown that, under certain regularity conditions, $\nabla^2\cL(\bbeta^*)$ is positive definite. Using very similar arguments, we can show $\nabla^2\cL_n(\bbeta^*)=\E\nabla^2\hat\cL_n(\bbeta^*)$ is positive definite. Accordingly, by continuity, Assumption ${\bf (A5)}$ holds with high probability when $d/n$ is relatively small. However, empirical results in the appendix  Section \ref{sec:convex} show that, even when the population design matrix's condition number is very small and for $d/n$ very large (e.g., $n=50$ and $d=800$), the convexity property still holds with high probability.  
\end{remark}

Before presenting the main result in this section, let's define an additional parameter. We write the cone
\[
\cH:=\{\bv\in \reals^d: v_1=1, \norm{\bv-\bbeta^*}_2\leq r(\gamma), \norm{\bv_{S^c}}_1\leq 3\norm{\bv_S}_1\}
\]
with $S$ representing the index set of nonzero-entries in $\bbeta^*_{r,\alpha_n}$. We define a parameter $\kappa_n$ controlling the uniform convergence of $\hat\cL_n(\cdot)$ to $\cL_n(\cdot)$ within the cone $\cH$:
\[
\kappa_n:=\E\sup_{\cH}|\hat\cL_n(\bbeta)-\cL_n(\bbeta)|,
\]
where the expectation is in the outer probability sense. Of note, because both $\hat\cL_n(\bbeta)$ and $\cL_n(\bbeta)$ are bounded, we have $\kappa_n\lesssim 1$, but $\kappa_n$ could be much smaller. In particular, we have $\kappa_n=O(n^{-1/2})$ by standard U-statistics empirical process theory \citep{nolan1987u,arcones1993limit,sherman1994maximal} when we fix $d$ and choose $r(\gamma)=o(1)$. More refined theoretical evaluation on the order of $\kappa_n$ could be found in \cite{he2000parameters}.

With Assumptions {\bf (A0)-(A6)}, the following lemma then characterizes the stochastic error of the specified minimum $\hat\bbeta_{r(\gamma),\alpha_n}$ to its population counterpart $\bbeta^*_{r(\gamma),\alpha_n}$. 
\begin{lemma}\label{lemma:estimation_error}
If Assumptions {\bf (A0)-(A6)} all hold, we then have, when $\lambda_n\asymp \alpha_n\sqrt{\log d/n}$, 
for large enough $n$, $\norm{\hat\bbeta_{r(\gamma),\alpha_n}-\bbeta^*_{r(\gamma),\alpha_n}}_2^2 \stackrel{\P}{\lesssim}  \alpha_n^2s_n\log d/n + \alpha_n^{-1}+\kappa_n$.
\end{lemma}

\subsection{Main results}

Combining Lemmas \ref{lem:app-error} and \ref{lemma:estimation_error}, we are now ready to provide the main theorem. 
It characterizes the consistency property of the proposed smoothing estimators under a specified scaling condition.

\begin{theorem}\label{thm:error_bound}
Assume Assumptions {\bf (A0)-(A6)} hold for some pre-chosen 
\[
\alpha_n\asymp \{n/(s_n\log d)\}^{1/3}.
\]
Assume $\hat\bbeta_{\alpha_n}$ is a stationary point of optimization problem \eqref{eq:opt1}, satisfying $\norm{\hat\bbeta_{\alpha_n}-\bbeta^*}_2\leq r$ for some constant $r$ depending on $r(\gamma)$. Then $\hat\bbeta_{\alpha_n}$ exists and satisfies that, as long as $s_n\log d/n \rightarrow 0$ and $\kappa_n\rightarrow 0$, we have $\norm{\hat\bbeta_{\alpha_n}-\bbeta^*}_2 \stackrel{\P}{\rightarrow} 0$.  In particular, when $d$ is fixed, we have $\|\hat\bbeta_{\alpha_n}-\bbeta^*\|_2\stackrel{\P}{\rightarrow}  0$.
\end{theorem}

\hf{
In practice, it is very difficult to theoretically calculate exactly how large $d$ is allowed to be using Theorem \ref{thm:error_bound}. However, a rule of thumb in our case, mimicking the corresponding ones in robust statistics (cf. \cite{jurevckova2012methodology}), is $10\log d\leq n^{1/3}$.
}

\section{Discussions}\label{sec:discussion}

\hf{
In the manuscript, we are focused on studying the lasso-type penalty of formulation \eqref{eq:loss-han2}. It should be highlighted that  SCAD \citep{fan2001variable} and MCP \citep{zhang2010nearly} type penalties could be implemented and studied in the same manner, and similar theoretical and empirical performances can be expected. Since the extension from lasso-type penalty to non-convex ones is beyond the interest of this paper, 
we decide to leave them for future studies.  Another important issue which is only mildly touched is model diagnostic check. To our knowledge, there has not been much study on goodness-of-fit test of Han's model \eqref{eq:model} in high dimensions. In this manuscript we provided several heuristics to check monotonicity of $Y$ and a single index of $\bX$ (cf. Section \ref{sec:real} and Section \ref{sec:realdatadiagnostic} in the appendix). In the future, it will be interesting to develop a theoretically solid test for it.

This manuscript has shown the advantage of smoothed maximum rank correlation method both theoretically and empirically. We close this section with a discussion on some limitations. (i) Computationally, as shown in the appendix Section \ref{sec:runtime}, though better than several semiparametric competitors, RMRCE is significantly worse than the lasso in terms of demanding much more time in implementation. (ii) Theoretically, if the true generating model is linear and the noise is Gaussian, RMRCE loses efficiency compared to the lasso. 
Accordingly, if the practitioner has a strong belief that, for instance, a simple linear model of Gaussian noise applies to the studied data, then the lasso is recommended compared to RMRCE.  
}

\hf{
\section*{Acknowledgement}

We thank the Editor, AE, and two anonymous referees for their helpful comments. We are grateful to Dr. Peter Radchenko and Dr. Haileab Hilafu for providing codes to implement their methods. 
}


\section*{Appendix}

\begin{appendices}

\noindent This appendix provides all simulation results and the technical proofs.

\section{Synthetic data analysis}\label{sec:simulation}

This section empirically examines the finite sample performance of the proposed Regularized Maximum Rank Correlation Estimator (RMRCE) $\hat\bbeta_{\alpha_n}$ in (\ref{eq:opt1}) using the synthetic data. 
We demonstrate various empirical results to compare the proposed methods to \hf{the competing methods}. Our simulation highlights the distinctive attributes of the proposed method, that is, it has comparable performance to the lasso under the high dimensional linear model, but beats the lasso under non-linear models. \hf{The proposed methods outperform Hinge, SIM, and SDR under both linear and non-linear models.}


\subsection{Algorithm description}\label{sec:algorithm}
We exploit the coordinate descent algorithm \citep{fu1998penalized} without penalization for the first term to solve 
(\ref{eq:opt1}). 
This problem falls in the application regime of the coordinate descent algorithm theory \citep{nesterov2012efficiency}. 
For the comparison fairness, in the sequel we also do not penalize the first term in implementing the lasso. 

One issue in the implementation
is on choosing the tuning parameter $\lambda_n$ and the smoothing parameter $\alpha_n$. 
For tuning $\lambda_n$ and $\alpha_n$, we propose to use five-fold cross-validation. Using five randomly split subsets of equal size, we define the following loss function (\ref{cv}):
\begin{align}\label{cv}
CV(\lambda_n,\alpha_n):=\frac{1}{5}\sum_{k=1}^5(_2^{n_k})^{-1}\sum_{1\leq l<l'\leq n_k}\sign(Y_{i_l}-Y_{i_{l'}})\sign(\bX_{i_l}^\T\hat\bbeta^{(-k)}_{\alpha_n}(\lambda_n)-\bX_{i_{l'}}^\T\hat\bbeta^{(-k)}_{\alpha_n}(\lambda_n)),
\end{align}
where $n_k$ is the number of data points in the $k$-th part, and $\hat\bbeta^{(-k)}_{\alpha_n}(\lambda_n)$ is obtained from the other 4 parts of the training data with the tuning parameter $\lambda_n$ and smoothing parameter $\alpha_n$. We then select the $\lambda_n$ and $\alpha_n$ that maximize $CV(\lambda_n,\alpha_n)$ over a grid of possible values $(\lambda_n,\alpha_n)$. In addition, when $\alpha_n$ is pre-chosen, $\lambda_n$ is chosen to optimize \eqref{cv} with the chosen $\alpha_n$.

Another issue in the implementation is on choosing the starting point for the coordinate descent algorithm. For this, we consider the following strategy proposed in \cite{luo2015forward}. Specifically, let $\{\be_j\in\reals^d,1\leq j\leq d\}$ be the standard basis with the $j$-th entry equal to 1, and 0 at all the other entries. The algorithm starts with $\hat\bbeta^{(0)}=\sign(L_{j^*})\be_{j^*}$ by selecting the $j^*$-th coordinate with the index $j^*$ that maximizes the absolute value of $L_j$ with respect to $j\in\{2,\ldots,d\}$, i.e.,
\begin{align*}
j^*=\argmax_{j\in \{2,\ldots,d\}}\left\{|L_j|: L_j:=(_2^n)^{-1}\sum_{1\leq i<i'\leq n}\sign(Y_i-Y_{i'})\sign(X_{ij}-X_{i'j})\right\}.
\end{align*}
In practice, we could combine other starting point candidates like the lasso and sieve solutions, and select the one that maximizes the objective function in \eqref{eq:loss-han}. However, our numerical results indicate that the above simple strategy has worked very well in a variety of applications.

\subsection{Synthetic data analysis}
This section investigates the empirical performance of the proposed methods via synthetic data analysis. 
\hf{We compare the proposed methods to Hinge, the lasso, SIM, and SDR. The descriptions of these methods are in Section \ref{sec:real}}. 
First, we show, under the high dimensional linear model, the proposed methods perform reasonably well compared to the lasso. Secondly, we show the proposed methods beat the lasso under a monotonic transformation model \eqref{eq:model1}. \hf{Thirdly, the proposed methods beat Hinge, SIM, and SDR under both linear and non-linear models.}

\hf{We also explore the performance of the proposed methods with the tuning parameter $\alpha_n$ chosen by cross validation or set to be fixed values $1,3,5,7,9$. The results show that, as long as $\alpha_n$ is comparatively large ($\alpha_n \geq 3$), the estimates and selection results are not sensitive to $\alpha_n$. In addition, choosing $\alpha_n$ by cross validation only leads to marginal improvement of performance compared to fixed $\alpha_n$.}

We tried $d=50$ and $200$ with a series of sample sizes for variable selection and estimation. Due to the space limit, {below we mainly show} the results about variable selection and estimation performance with $d=200$. Similar patterns hold for $d=50$, \hf{some of whose results are put in later sections}. 


\subsubsection{High dimensional linear model}\label{sec:linear_model}

This section is focused on the high dimensional linear model\footnote{We have tried a lot of different combinations of $D(\cdot)$ and $\Lambda(\cdot,\cdot)$, confirming that the results are very similar to the linear case. Due to the space limit, we do not list them all in this paper.},
\begin{eqnarray}\label{model}
Y_i=\bX_i^\T\bbeta^0+\epsilon_i,i=1,2,\cdots, n,
\end{eqnarray}
with $\bbeta^0=(5,4,3,2,1,-1,-3,-5,0,\cdots,0)^\T$ and $\bX_i$ a $d$ dimensional random vector generated from a multivariate normal distribution $\text{N}_d(\mathbf{0},\bSigma=((\sigma_{jk})))$ with $\sigma_{jk}=0.5^{|j-k|}$ for $1\leq j,k\leq d$. 
Here the noise was generated from the standard normal and independent of the covariates. Note the difference between (\ref{model}) and (\ref{eq:model1}) in terms of $\bbeta^*=\bbeta^0/\beta^0_1$.  We also consider all three smoothing approximations. 



\begin{figure}[!htbp]
\centering
\includegraphics[width=0.9\textwidth]{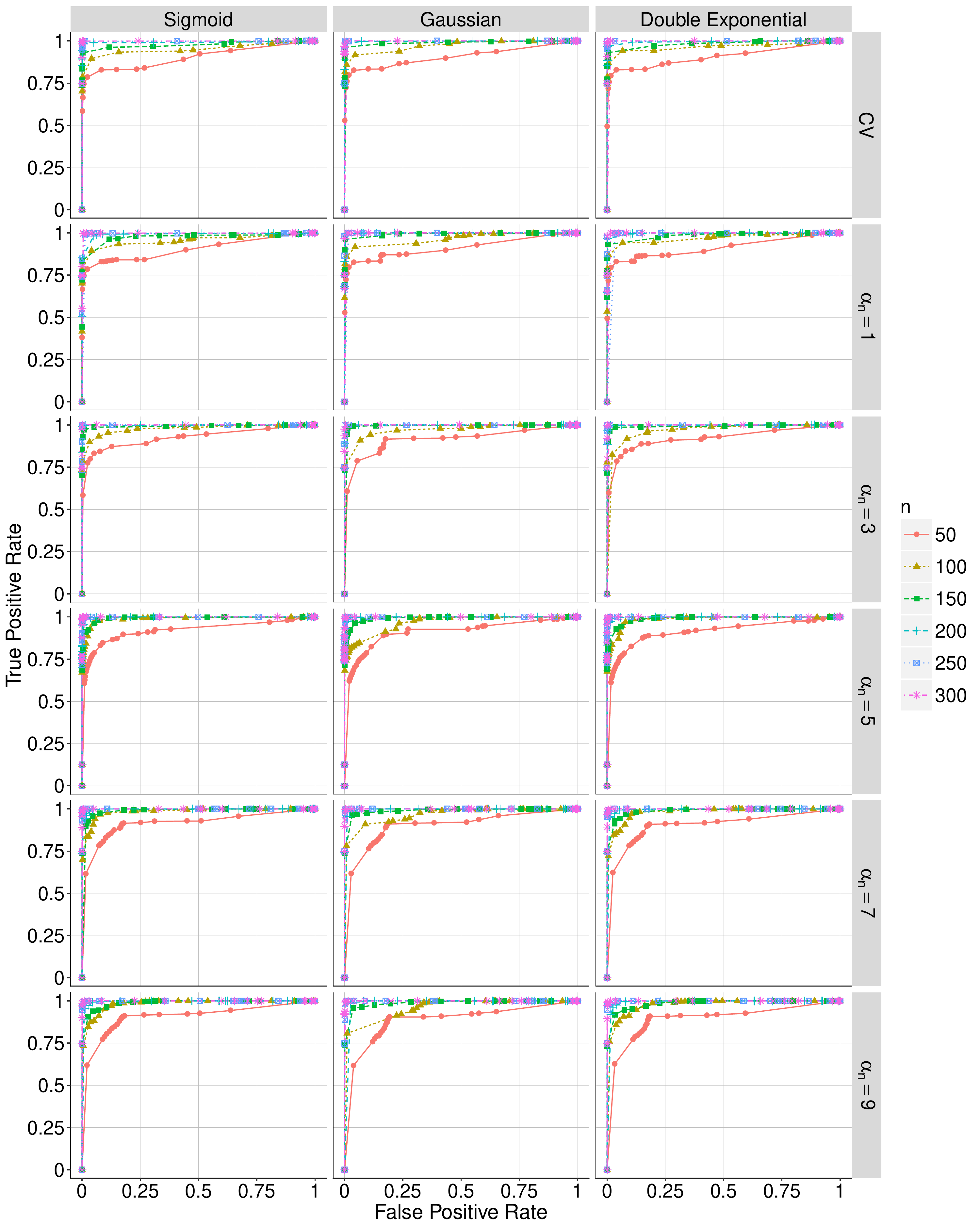}
\caption{\hf{\textbf{ROC curves for RMRCE under the high dimensional linear model.} $\alpha_n$ is selected by cross validation (CV) or set to be $1,3,5,7,9$. The results for Sigmoid, Gaussian and double exponential smoothing approximations are presented. The results are based on 200 replications with $d=200$.}}
\label{fig_s1_roc_RMRCE}
\end{figure}

\begin{figure}[!htbp]
\centering
\includegraphics[width=0.9\textwidth]{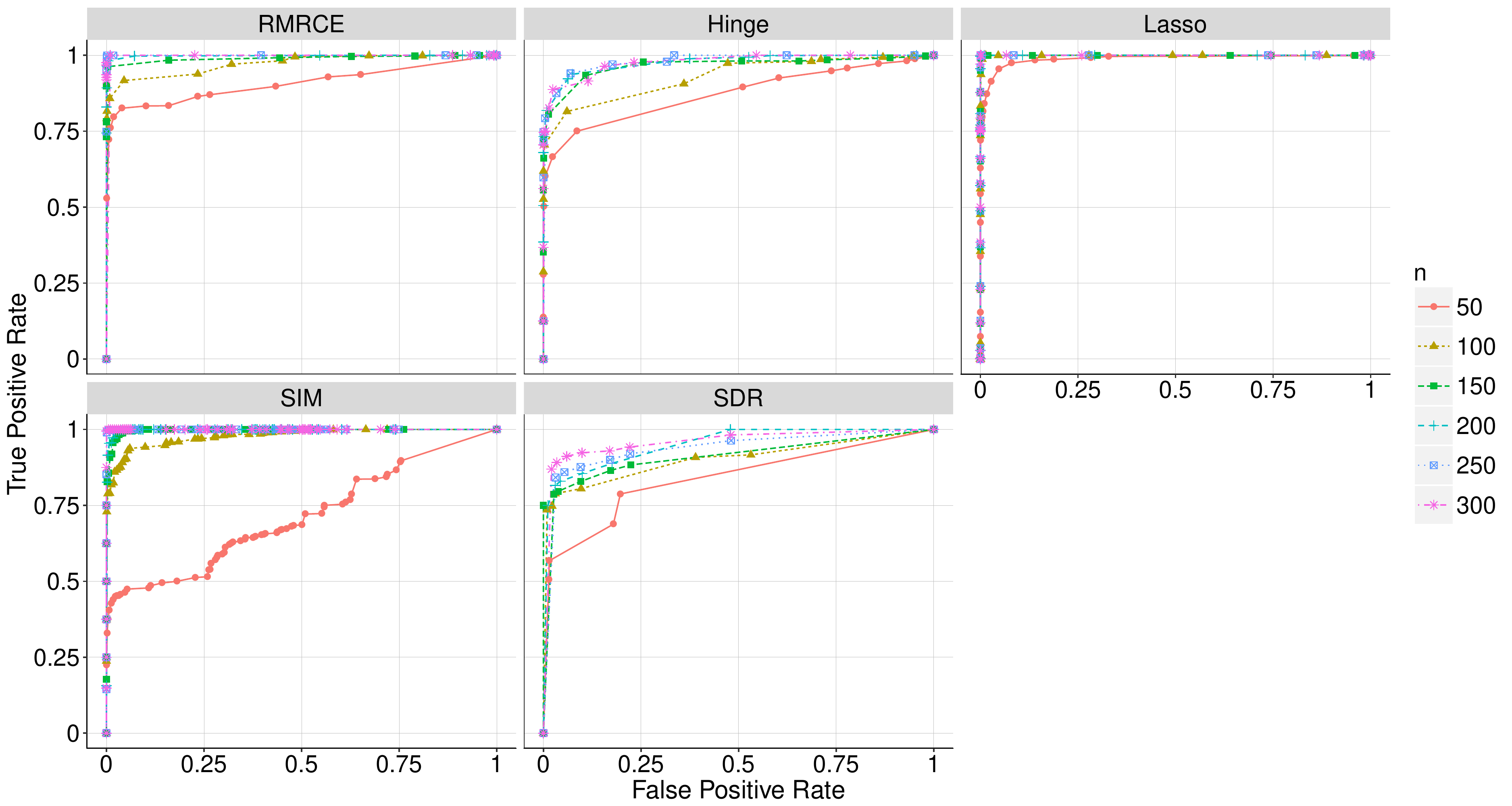}
\caption{\hf{\textbf{ROC curves for all competing methods  under the high dimensional linear model.} Methods included are: RMRCE ($\alpha_n$ selected by cross validation, Gaussian smoothing approximation), Hinge, the lasso, SIM, and SDR. The results are based on 200 replications with $d=200$.}}
\label{fig_s1_roc_allmethod}
\end{figure}

We first compare the variable selection performance with $d=200$. To this end, \hf{Figures \ref{fig_s1_roc_RMRCE} and \ref{fig_s1_roc_allmethod} plot the receiver operating characteristic (ROC) curves for RMRCE with different choices of tuning parameter $\alpha_n$, and all the competing methods. These results are calculated based on 200 replications.} It shows that the proposed methods have overall good variable selection performance compared to \hf{the competing methods}. Detailed comments are as follows.
(i) With sample size increasing, the ROC curves shift towards the upper left corner, as what we expect. (ii) The difference between different smoothing approximations \hf{and different choices of tuning parameter $\alpha_n$} is very slight via observing Figure \ref{fig_s1_roc_RMRCE}. \hf{The performance of RMRCE with $\alpha_n$ selected using cross validation is slightly better than those with fixed $\alpha_n$.} 
(iii) The lasso has slightly better variable selection performance as shown in Figure \ref{fig_s1_roc_allmethod}, especially regarding the small sample size. This is reasonable since the simulation setup in (\ref{model}) is the right model for the lasso.
\hf{(iv) RMRCE has better performance than Hinge. This shows that the smoothed rank correlation in RMRCE is better than the convex relaxation in Hinge. (v) RMRCE has better performance than the other two semiparametric regression methods: SIM and SDR.}
\hf{Table \ref{tab_s1_fptp_allmethod} gives the averaged false positive rates and the averaged true positive rates for all the competing methods.} Similar observations to Figure \ref{fig_s1_roc_allmethod} were observed. 

{
\renewcommand{\tabcolsep}{3pt}
\renewcommand{\arraystretch}{0.90}
\begin{table}[!htbp]
\caption{\label{tab_s1_fptp_allmethod}\textbf{Variable selection performance for \hf{all competing methods under the high dimensional linear model}.} \hf{Methods included are: RMRCE ($\alpha_n$ selected by cross validation, Gaussian smoothing approximation), Hinge, the lasso, SIM, and SDR.}  The results are based on 200 replications with $n=100$ and $d=200$. ``FPR" stands for averaged false positive rates with standard deviations in parentheses. ``TPR" stands for averaged true positive rates with standard deviations in parentheses. \hf{For SIM and SDR, $N$ stands for the number of selected variables.}}
\begin{center}
\subfigure[Subfigure 2 list of figures text][RMRCE]{
\footnotesize{\begin{tabular}{ccc}
\hline
$\lambda_n$         & FPR       & TPR     \\
\hline
0.005 & 0.321 (0.022) & 0.999 (0.012) \\
0.010 & 0.234 (0.025) & 0.999 (0.012) \\
0.030 & 0.009 (0.012) & 0.858 (0.111) \\
0.050 & 0.001 (0.002) & 0.792 (0.062) \\
0.100 & 0.001 (0.002) & 0.751 (0.110) \\
\hline
\end{tabular}}
\label{fptp_RMRCE_compete}}
\subfigure[Subfigure 2 list of figures text][Hinge]{
\footnotesize{\begin{tabular}{ccc}
\hline
$\lambda_n$        & FPR       & TPR     \\
\hline
0.040 & 0.870 (0.166) & 0.994 (0.027) \\
0.060 & 0.687 (0.188) & 0.986 (0.039) \\
0.080 & 0.472 (0.141) & 0.980 (0.049) \\
0.100 & 0.360 (0.307) & 0.906 (0.098) \\
0.150 & 0.060 (0.081) & 0.815 (0.117) \\
\hline
\end{tabular}}
\label{fptp_Hinge}}
\subfigure[Subfigure 2 list of figures text][Lasso]{
\footnotesize{\begin{tabular}{ccc}
\hline
$\lambda_n$        & FPR       & TPR     \\
\hline
0.100 & 0.157 (0.022) & 1.000 (0.000) \\
0.300 & 0.014 (0.009) & 0.999 (0.009) \\
0.500 & 0.001 (0.003) & 0.937 (0.080) \\
0.700 & 0.000 (0.001) & 0.833 (0.081) \\
1.000 & 0.000 (0.000) & 0.775 (0.050) \\
\hline
\end{tabular}}
\label{fptp_lasso}}
\subfigure[Subfigure 2 list of figures text][SIM]{
\footnotesize{\begin{tabular}{ccc}
\hline
$N$   & FPR       & TPR     \\
\hline
5.000 & 0.000 (0.001) & 0.623 (0.015) \\
10.000 & 0.018 (0.003) & 0.818 (0.076) \\
20.000 & 0.166 (0.262) & 0.948 (0.083) \\
30.000 & 0.309 (0.354) & 0.976 (0.054) \\
50.000 & 0.719 (0.352) & 0.996 (0.025) \\
\hline
\end{tabular}}
\label{fptp_SIM}}
\subfigure[Subfigure 2 list of figures text][SDR]{
\footnotesize{\begin{tabular}{ccc}
\hline
$N$    & FPR       & TPR     \\
\hline
2.000 & 0.009 (0.004) & 0.789 (0.059) \\
5.000 & 0.022 (0.011) & 0.736 (0.091) \\
10.000 & 0.031 (0.020) & 0.748 (0.086) \\
25.000 & 0.097 (0.004) & 0.805 (0.089) \\
50.000 & 0.390 (0.033) & 0.908 (0.070) \\
\hline
\end{tabular}}
\label{fptp_SDR}}
\end{center}
\end{table}
}


Secondly, we focus on comparing the estimation errors with $d=200$. 
\hf{Tables \ref{tab_s1_error_RMRCE} and \ref{tab_s1_error_allmethod}} present the estimation errors to $\bbeta^*$ compared to the tuning parameter $\lambda_n$ \hf{for RMRCE with different choices of tuning parameter $\alpha_n$, and all the competing methods across 200 replications}. Note, for lasso, the estimation error is calculated as $\|\hat{\bbeta}^{\text{lasso}}(\lambda_n)/\hat{\beta}_1^{\text{lasso}}(\lambda_n)-\bbeta^*\|_2$. \hf{The estimation errors are calculated in the same way for SIM and SDR. The results show that for RMRCE, Hinge, and the lasso,} the estimation error decreases first and then increases as tuning parameter increases. \hf{For SIM and SDR, the estimation error decreases first and then almost stays constant as the number of selected variables increases.} Besides, with increased sample size, the estimation error curves shift downward to 0. \hf{Table \ref{tab_s1_error_RMRCE}} also shows that the difference between the proposed methods with different approximation functions \hf{and different choices of $\alpha_n$} is very mild \hf{as long as $\alpha_n$ is large enough ($\alpha_n \geq 3$). The performance of RMRCE with $\alpha_n=1$ is comparatively worse than RMRCE with $\alpha_n \geq 3$}. \hf{Table \ref{tab_s1_error_RMRCE} and \ref{tab_s1_error_allmethod}} further show, under the high dimensional linear model (\ref{model}), the proposed methods perform reasonably well, although slightly worse than the lasso\footnote{Note that for fair comparison, the results for the lasso are based on the standardized estimation error $\|\hat{\bbeta}^{\text{lasso}}/\hat{\beta}^{\text{lasso}}_1-\bbeta^*\|_2$ with the lasso estimator $\hat{\bbeta}^{\text{lasso}}$ of $\bbeta^0$. \hf{Same for SIM and SDR.}}. \hf{In addition, the proposed methods have better performance than Hinge, SIM, and SDR.} 



{
\renewcommand{\tabcolsep}{6pt}
\renewcommand{\arraystretch}{0.88}
\begin{table}[!htbp]
\caption{\label{tab_s1_error_RMRCE}\hf{\textbf{Averaged estimation errors for RMRCE  under the high dimensional linear model.} Standard deviations are shown in parentheses. $\alpha_n$ is selected by cross validation or set to be $1,3,5,7,9$. The results for Sigmoid, Gaussian and double exponential (DE) smoothing approximations are presented. The results are based on 200 replications with $d=50$ or $d=200$. }}
\vspace{-20pt}
\begin{center}
\subfigure[Subfigure 2 list of figures text][$\alpha_n$ selected by cross validation]{
\footnotesize{\begin{tabular}{ccccccc}
\hline
$n$         & Sigmoid   (d=50)    & Sigmoid   (d=200)  & Gaussian (d=50) & Gaussian (d=200) & DE (d=50) & DE (d=200)    \\
\hline
50 & 0.068 (0.038) & 0.139 (0.089) & 0.099 (0.047) & 0.137 (0.104) & 0.081 (0.039) & 0.136 (0.101) \\
100 & 0.021 (0.013) & 0.033 (0.018) & 0.026 (0.014) & 0.035 (0.026) & 0.028 (0.017) & 0.046 (0.022) \\
150 & 0.013 (0.006) & 0.010 (0.008) & 0.012 (0.017) & 0.006 (0.010) & 0.014 (0.008) & 0.016 (0.014) \\
200 & 0.008 (0.004) & 0.008 (0.009) & 0.003 (0.004) & 0.003 (0.008) & 0.010 (0.005) & 0.006 (0.008) \\
250 & 0.006 (0.003) & 0.007 (0.005) & 0.002 (0.003) & 0.004 (0.002) & 0.006 (0.005) & 0.005 (0.004) \\
300 & 0.004 (0.003) & 0.002 (0.004) & 0.001 (0.007) & 0.004 (0.002) & 0.004 (0.003) & 0.003 (0.002) \\
\hline
\vspace{-17pt}
\end{tabular}}
\label{error_RMRCE_CV}}
\subfigure[Subfigure 2 list of figures text][$\alpha_n$ = 1]{
\footnotesize{\begin{tabular}{ccccccc}
\hline
$n$         & Sigmoid   (d=50)    & Sigmoid   (d=200)  & Gaussian (d=50) & Gaussian (d=200) & DE (d=50) & DE (d=200)    \\
\hline
50 & 0.239 (0.136) & 0.242 (0.197) & 0.113 (0.047) & 0.137 (0.104) & 0.109 (0.049) & 0.136 (0.101) \\
100 & 0.150 (0.061) & 0.156 (0.061) & 0.088 (0.022) & 0.077 (0.021) & 0.081 (0.022) & 0.088 (0.022) \\
150 & 0.133 (0.044) & 0.123 (0.038) & 0.084 (0.019) & 0.071 (0.011) & 0.082 (0.018) & 0.066 (0.015) \\
200 & 0.130 (0.037) & 0.114 (0.022) & 0.083 (0.013) & 0.071 (0.008) & 0.078 (0.015) & 0.066 (0.011) \\
250 & 0.125 (0.031) & 0.110 (0.017) & 0.078 (0.013) & 0.070 (0.008) & 0.078 (0.014) & 0.066 (0.009) \\
300 & 0.120 (0.025) & 0.112 (0.015) & 0.074 (0.011) & 0.080 (0.008) & 0.076 (0.012) & 0.070 (0.007) \\\hline
\vspace{-17pt}
\end{tabular}}
\label{error_RMRCE_alpha1}}
\subfigure[Subfigure 2 list of figures text][$\alpha_n$ = 3]{
\footnotesize{\begin{tabular}{ccccccc}
\hline
$n$         & Sigmoid   (d=50)    & Sigmoid   (d=200)  & Gaussian (d=50) & Gaussian (d=200) & DE (d=50) & DE (d=200)    \\
\hline
50 & 0.068 (0.038) & 0.139 (0.089) & 0.099 (0.047) & 0.169 (0.098) & 0.081 (0.039) & 0.170 (0.091) \\
100 & 0.021 (0.013) & 0.033 (0.018) & 0.026 (0.014) & 0.035 (0.026) & 0.028 (0.017) & 0.046 (0.022) \\
150 & 0.015 (0.009) & 0.010 (0.008) & 0.013 (0.006) & 0.006 (0.010) & 0.014 (0.008) & 0.016 (0.014) \\
200 & 0.011 (0.007) & 0.008 (0.004) & 0.003 (0.004) & 0.003 (0.008) & 0.010 (0.005) & 0.006 (0.008) \\
250 & 0.009 (0.005) & 0.007 (0.003) & 0.002 (0.003) & 0.004 (0.002) & 0.006 (0.005) & 0.005 (0.004) \\
300 & 0.007 (0.004) & 0.007 (0.002) & 0.001 (0.002) & 0.004 (0.002) & 0.004 (0.003) & 0.003 (0.002) \\\hline
\vspace{-17pt}
\end{tabular}}
\label{error_RMRCE_alpha3}}
\subfigure[Subfigure 2 list of figures text][$\alpha_n$ = 5]{
\footnotesize{\begin{tabular}{ccccccc}
\hline
$n$         & Sigmoid   (d=50)    & Sigmoid   (d=200)  & Gaussian (d=50) & Gaussian (d=200) & DE (d=50) & DE (d=200)    \\
\hline
50 & 0.095 (0.045) & 0.175 (0.099) & 0.119 (0.054) & 0.208 (0.098) & 0.111 (0.051) & 0.169 (0.090) \\
100 & 0.025 (0.014) & 0.051 (0.020) & 0.036 (0.073) & 0.072 (0.027) & 0.033 (0.015) & 0.061 (0.021) \\
150 & 0.013 (0.006) & 0.026 (0.008) & 0.012 (0.017) & 0.042 (0.016) & 0.017 (0.007) & 0.036 (0.011) \\
200 & 0.008 (0.004) & 0.017 (0.006) & 0.009 (0.011) & 0.013 (0.013) & 0.011 (0.005) & 0.023 (0.010) \\
250 & 0.006 (0.003) & 0.009 (0.006) & 0.008 (0.008) & 0.010 (0.007) & 0.008 (0.004) & 0.013 (0.008) \\
300 & 0.004 (0.002) & 0.002 (0.004) & 0.008 (0.005) & 0.008 (0.005) & 0.006 (0.003) & 0.005 (0.005) \\\hline
\vspace{-17pt}
\end{tabular}}
\label{error_RMRCE_alpha5}}
\subfigure[Subfigure 2 list of figures text][$\alpha_n$ = 7]{
\footnotesize{\begin{tabular}{ccccccc}
\hline
$n$         & Sigmoid   (d=50)    & Sigmoid   (d=200)  & Gaussian (d=50) & Gaussian (d=200) & DE (d=50) & DE (d=200)    \\
\hline
50 & 0.104 (0.050) & 0.171 (0.091) & 0.140 (0.068) & 0.304 (0.130) & 0.122 (0.056) & 0.232 (0.105) \\
100 & 0.030 (0.016) & 0.046 (0.027) & 0.044 (0.022) & 0.064 (0.037) & 0.040 (0.020) & 0.063 (0.033) \\
150 & 0.017 (0.007) & 0.017 (0.012) & 0.018 (0.012) & 0.031 (0.018) & 0.021 (0.009) & 0.028 (0.015) \\
200 & 0.010 (0.007) & 0.008 (0.009) & 0.012 (0.009) & 0.019 (0.009) & 0.012 (0.008) & 0.015 (0.011) \\
250 & 0.007 (0.005) & 0.007 (0.005) & 0.008 (0.006) & 0.016 (0.006) & 0.009 (0.006) & 0.012 (0.006) \\
300 & 0.004 (0.003) & 0.006 (0.003) & 0.001 (0.007) & 0.017 (0.005) & 0.006 (0.004) & 0.011 (0.004) \\
\hline
\vspace{-17pt}
\end{tabular}}
\label{error_RMRCE_alpha7}}
\subfigure[Subfigure 2 list of figures text][$\alpha_n$ = 9]{
\footnotesize{\begin{tabular}{ccccccc}
\hline
$n$         & Sigmoid   (d=50)    & Sigmoid   (d=200)  & Gaussian (d=50) & Gaussian (d=200) & DE (d=50) & DE (d=200)    \\
\hline
50 & 0.120 (0.055) & 0.211 (0.099) & 0.157 (0.073) & 0.395 (0.389) & 0.140 (0.065) & 0.313 (0.133) \\
100 & 0.037 (0.018) & 0.075 (0.027) & 0.049 (0.023) & 0.076 (0.041) & 0.044 (0.022) & 0.100 (0.035) \\
150 & 0.020 (0.009) & 0.021 (0.019) & 0.026 (0.012) & 0.021 (0.067) & 0.024 (0.011) & 0.029 (0.023) \\
200 & 0.014 (0.007) & 0.013 (0.011) & 0.013 (0.010) & 0.023 (0.010) & 0.017 (0.008) & 0.019 (0.013) \\
250 & 0.008 (0.005) & 0.011 (0.006) & 0.009 (0.007) & 0.019 (0.007) & 0.010 (0.007) & 0.015 (0.007) \\
300 & 0.006 (0.004) & 0.010 (0.004) & 0.001 (0.007) & 0.019 (0.005) & 0.008 (0.004) & 0.014 (0.005) \\\hline
\vspace{-17pt}
\end{tabular}}
\label{error_RMRCE_alpha9}}

\end{center}
\end{table}
}

{
\renewcommand{\tabcolsep}{4pt}
\renewcommand{\arraystretch}{0.90}
\begin{table}[!htbp]
\caption{\label{tab_s1_error_allmethod}\hf{\textbf{Averaged estimation errors for all competing methods  under the high dimensional linear model.} Standard deviations are shown in parentheses. Methods included are: RMRCE ($\alpha_n$ selected by cross validation, Gaussian smoothing approximation), Hinge, the lasso, SIM, and SDR. The results are based on 200 replications with $d=50$ or $d=200$. The tuning parameter $\lambda_n$ (RMRCE, Hinge, and the lasso) or number of selected variables (SIM and SDR) is determined through cross-validation.}}

\begin{center}
\subfigure[Subfigure 2 list of figures text][RMRCE]{
\footnotesize{\begin{tabular}{lcc}
\hline
&\multicolumn{2}{c}{$d$} \\
\cline{2-3}
$n$         & 50       & 200   \\
\hline
50 & 0.099 (0.047) & 0.137 (0.104) \\
100 & 0.026 (0.014) & 0.035 (0.026) \\
150 & 0.012 (0.017) & 0.006 (0.010) \\
200 & 0.003 (0.004) & 0.003 (0.008) \\
250 & 0.002 (0.003) & 0.004 (0.002) \\
300 & 0.001 (0.007) & 0.004 (0.002) \\
\hline
\end{tabular}}
\label{error_RMRCE_compete}}
\subfigure[Subfigure 2 list of figures text][Hinge]{
\footnotesize{\begin{tabular}{lcc}
\hline
&\multicolumn{2}{c}{$d$} \\
\cline{2-3}
$n$         & 50       & 200   \\
\hline
50 & 1.963 (0.248) & 2.143 (0.353) \\
100 & 2.580 (0.230) & 2.199 (0.271) \\
150 & 2.430 (0.203) & 2.220 (0.232) \\
200 & 2.260 (0.212) & 2.259 (0.155) \\
250 & 2.323 (0.191) & 2.591 (0.121) \\
300 & 2.362 (0.217) & 2.257 (0.109) \\
\hline
\end{tabular}}
\label{error_hinge}}
\subfigure[Subfigure 2 list of figures text][Lasso]{
\footnotesize{\begin{tabular}{lcc}
\hline
&\multicolumn{2}{c}{$d$} \\
\cline{2-3}
$n$         & 50       & 200   \\
\hline
50 & 0.034 (0.021) & 0.064 (0.035) \\
100 & 0.012 (0.007) & 0.020 (0.009) \\
150 & 0.008 (0.004) & 0.011 (0.005) \\
200 & 0.006 (0.003) & 0.008 (0.004) \\
250 & 0.004 (0.002) & 0.006 (0.003) \\
300 & 0.003 (0.002) & 0.005 (0.002) \\
\hline
\end{tabular}}
\label{error_lasso}}
\subfigure[Subfigure 2 list of figures text][SIM]{
\footnotesize{\begin{tabular}{lcc}
\hline
&\multicolumn{2}{c}{$d$} \\
\cline{2-3}
$n$         & 50       & 200   \\
\hline
50 & 1.109 (0.424) & 2.080 (0.937) \\
100 & 0.822 (0.535) & 0.902 (0.000) \\
150 & 0.815 (0.004) & 0.831 (0.014) \\
200 & 0.812 (0.003) & 0.824 (0.008) \\
250 & 0.809 (0.002) & 0.816 (0.000) \\
300 & 0.810 (0.002) & 0.814 (0.000) \\
\hline
\end{tabular}}
\label{error_SIM}}
\subfigure[Subfigure 2 list of figures text][SDR]{
\footnotesize{\begin{tabular}{lcc}
\hline
&\multicolumn{2}{c}{$d$} \\
\cline{2-3}
$n$         & 50       & 200   \\
\hline
50 & 1.952 (3.210) & 2.931 (2.246) \\
100 & 1.038 (3.691) & 1.046 (3.630) \\
150 & 0.950 (3.734) & 0.921 (0.015) \\
200 & 0.888 (3.746) & 1.025 (3.452) \\
250 & 0.861 (3.728) & 0.876 (3.676) \\
300 & 0.854 (3.764) & 0.854 (3.729) \\
\hline
\end{tabular}}
\label{error_SDR}}

\end{center}
\end{table}
}

\subsubsection{High dimensional generalized regression model}
This section considers the monotonic transformation model (\ref{eq:model1}) with $G(x)=x^3$:
\begin{eqnarray}\label{model_2}
Y_i=(\bX_i^\T\bbeta^0+\epsilon_i)^3,i=1,2,\cdots, n,
\end{eqnarray}
where $\bX_i$ and $\bbeta^0$ are the same as in Model (\ref{model}), and $\epsilon_i$ follows the standard normal distribution and is independent of $\bX_i$. This is the setting where the lasso could still possibly provide a consistent estimator \citep{li1989regression}. But the estimation/model selection efficiency is expected to be low, which will be demonstrated empirically here. 
\hf{Note that} under Model (\ref{model_2}), the proposed methods \hf{and Hinge} have exactly the same results as under Model (\ref{model}). \hf{For the convenience of comparison, we still include the results of Hinge and RMRCE ($\alpha_n$ chosen by cross validation, Gaussian approximation).} 

First, we reveal the variable selection performance for \hf{all competing methods} under Model (\ref{model_2}) by plotting the ROC curves in Figure \ref{fig_s2_roc_allmethod} with $d=200$. It confirms that  the lasso \hf{as well as other competing methods perform} much worse than the proposed methods . 
\hf{Table \ref{tab_s2_fptp_allmethod}} further illustrates the averaged TPRs and FPRs along with their standard deviations over 200 replications. As shown in Table \hf{\ref{tab_s2_fptp_allmethod}, Hinge, the lasso, SIM, and SDR have much worse performance than the proposed methods.} 
Secondly, 
\hf{Table \ref{tab_s2_error_allmethod}} shows the normalized $\ell_2$ estimation error. \hf{For the lasso, the estimation error is calculated as} $\|\hat{\bbeta}^{\text{lasso}}(\lambda_n)/\hat{\beta}_1^{\text{lasso}}(\lambda_n)-\bbeta^*\|_2$. \hf{The estimation errors are calculated in the same way for SIM and SDR}. With $d=200$, the comparison between the methods confirms the advantage of the proposed method. 

\begin{figure}[!htbp]
\centering
\includegraphics[width=\textwidth]{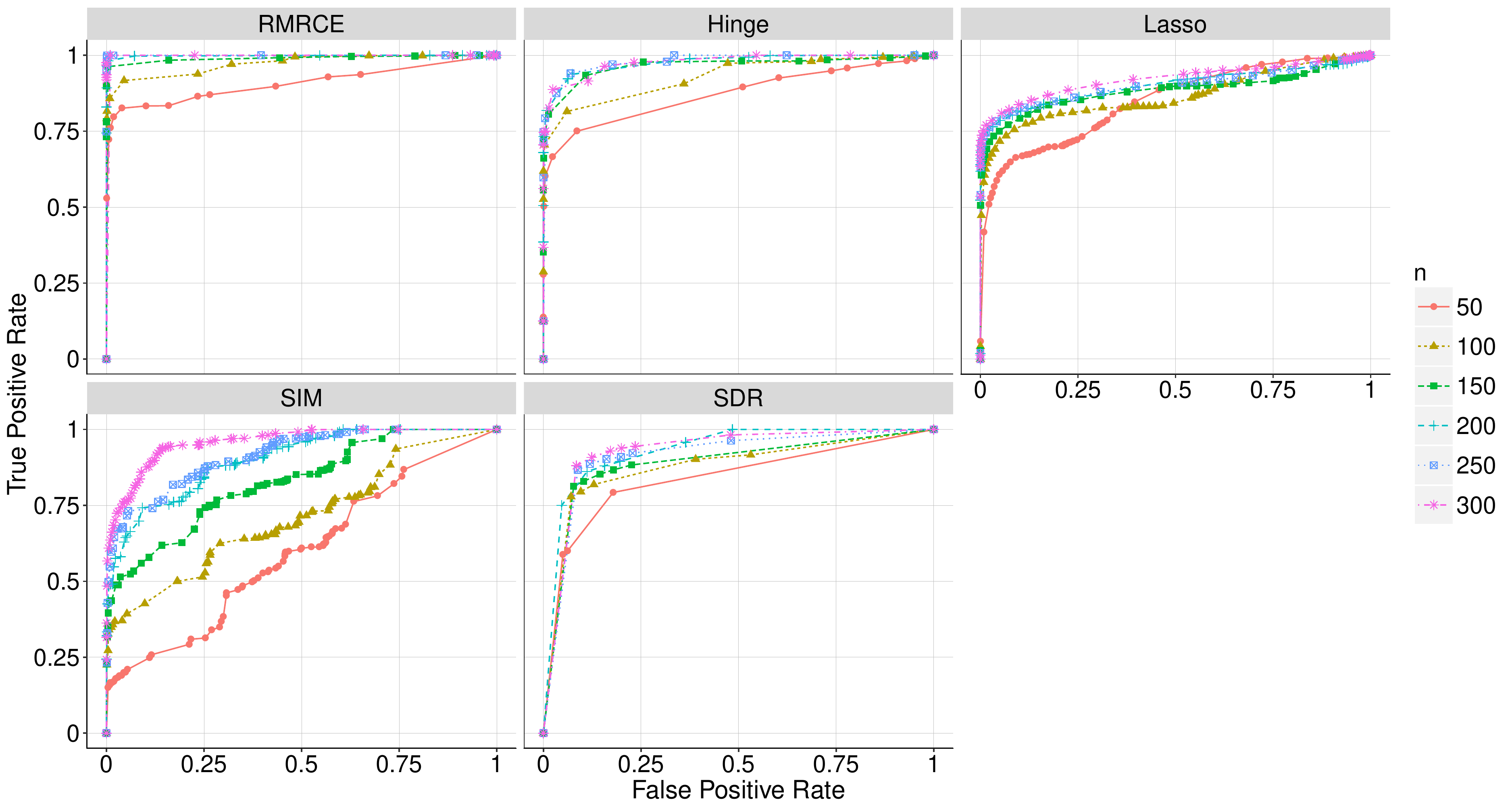}
\caption{\hf{\textbf{ROC curves for all competing methods with $G(x)=x^3$.} Methods included are: RMRCE ($\alpha_n$ selected by cross validation, Gaussian smoothing approximation), Hinge, the lasso, SIM, and SDR. The results are based on 200 replications with $d=200$.}}
\label{fig_s2_roc_allmethod}
\end{figure}


{
\renewcommand{\tabcolsep}{4pt}
\renewcommand{\arraystretch}{0.8}
\begin{table}[!htbp]
\caption{\label{tab_s2_fptp_allmethod}\hf{\textbf{Variable selection performance for all competing methods with $G(x)=x^3$.} Methods included are: RMRCE ($\alpha_n$ selected by cross validation, Gaussian smoothing approximation), Hinge, the lasso, SIM, and SDR. The results are based on 200 replications with $n=100$ and $d=200$. ``FPR" stands for averaged false positive rates with standard deviations in parentheses. ``TPR" stands for averaged true positive rates with standard deviations in parentheses. For SIM and SDR, $N$ stands for the number of selected variables.}}
\begin{center}
\subfigure[Subfigure 2 list of figures text][RMRCE]{
      \footnotesize{\begin{tabular}{ccc}
            \hline
            $\lambda_n$         & FPR       & TPR     \\
            \hline
0.005 & 0.321 (0.022) & 0.999 (0.012) \\
0.010 & 0.234 (0.025) & 0.999 (0.012) \\
0.030 & 0.009 (0.012) & 0.858 (0.111) \\
0.050 & 0.001 (0.002) & 0.792 (0.062) \\
0.100 & 0.001 (0.002) & 0.751 (0.110) \\
            \hline
            \end{tabular}}
      \label{s2_fptp_RMRCE_compete}}
\subfigure[Subfigure 2 list of figures text][Hinge]{
      \footnotesize{\begin{tabular}{ccc}
            \hline
            $\lambda_n$        & FPR       & TPR     \\
            \hline
0.040 & 0.870 (0.166) & 0.994 (0.027) \\
0.060 & 0.687 (0.188) & 0.986 (0.039) \\
0.080 & 0.472 (0.141) & 0.980 (0.049) \\
0.100 & 0.360 (0.307) & 0.906 (0.098) \\
0.150 & 0.060 (0.081) & 0.815 (0.117) \\
            \hline
            \end{tabular}}
      \label{s2_fptp_Hinge}}
\subfigure[Subfigure 2 list of figures text][Lasso]{
\footnotesize{\begin{tabular}{ccc}
\hline
$\lambda_n$        & FPR       & TPR     \\
\hline
1.000 & 0.881 (0.076) & 0.985 (0.044) \\
5.000 & 0.600 (0.042) & 0.889 (0.094) \\
10.000 & 0.542 (0.032) & 0.859 (0.095) \\
50.000 & 0.432 (0.032) & 0.832 (0.098) \\
100.000 & 0.365 (0.044) & 0.828 (0.093) \\
500.000 & 0.116 (0.061) & 0.774 (0.089) \\
\hline
\end{tabular}}
\label{s2_fptp_lasso}}
\subfigure[Subfigure 2 list of figures text][SIM]{
\footnotesize{\begin{tabular}{ccc}
\hline
$N$   & FPR       & TPR     \\
\hline
5 & 0.012 (0.005) & 0.327 (0.132) \\
10 & 0.052 (0.102) & 0.393 (0.191) \\
20 & 0.423 (0.379) & 0.652 (0.324) \\
30 & 0.578 (0.387) & 0.756 (0.304) \\
40 & 0.646 (0.380) & 0.796 (0.280) \\
50 & 0.727 (0.201) & 0.883 (0.204) \\
\hline
\end{tabular}}
\label{s2_fptp_SIM}}
\subfigure[Subfigure 2 list of figures text][SDR]{
\footnotesize{\begin{tabular}{ccc}
\hline
$N$    & FPR       & TPR     \\
\hline
5 & 0.071 (0.035) & 0.779 (0.098) \\
10 & 0.096 (0.050) & 0.795 (0.098) \\
20 & 0.126 (0.068) & 0.809 (0.084) \\
30 & 0.354 (0.217) & 0.851 (0.096) \\
40 & 0.531 (0.397) & 0.916 (0.099) \\
50 & 0.390 (0.033) & 0.902 (0.066) \\
\hline
\end{tabular}}
\label{s2_fptp_SDR}}
\end{center}
\end{table}
}

{
\renewcommand{\tabcolsep}{4pt}
\renewcommand{\arraystretch}{1.00}
\begin{table}[!htbp]
\caption{\label{tab_s2_error_allmethod}\hf{\textbf{Averaged estimation errors for all competing methods with $G(x)=x^3$.} Standard deviations are shown in parentheses. Methods included are: RMRCE ($\alpha_n$ selected by cross validation, Gaussian smoothing approximation), Hinge, the lasso, SIM, and SDR. The results are based on 200 replications with $d=50$ or $d=200$. The tuning parameter $\lambda_n$ (RMRCE, Hinge, and the lasso) or number of selected variables (SIM and SDR) is determined through cross-validation. }}
\begin{center}
\subfigure[Subfigure 2 list of figures text][RMRCE]{
      \footnotesize{\begin{tabular}{lcc}
            \hline
            &\multicolumn{2}{c}{$d$} \\
            \cline{2-3}
            $n$         & 50       & 200   \\
            \hline
50 & 0.099 (0.047) & 0.137 (0.104) \\
100 & 0.026 (0.014) & 0.035 (0.026) \\
150 & 0.012 (0.017) & 0.006 (0.010) \\
200 & 0.003 (0.004) & 0.003 (0.008) \\
250 & 0.002 (0.003) & 0.004 (0.002) \\
300 & 0.001 (0.007) & 0.004 (0.002) \\
            \hline
            \end{tabular}}
      \label{s2_error_RMRCE_compete}}
\subfigure[Subfigure 2 list of figures text][Hinge]{
      \footnotesize{\begin{tabular}{lcc}
            \hline
            &\multicolumn{2}{c}{$d$} \\
            \cline{2-3}
            $n$         & 50       & 200   \\
            \hline
50 & 1.963 (0.248) & 2.143 (0.353) \\
100 & 2.580 (0.230) & 2.199 (0.271) \\
150 & 2.430 (0.203) & 2.220 (0.232) \\
200 & 2.260 (0.212) & 2.259 (0.155) \\
250 & 2.323 (0.191) & 2.591 (0.121) \\
300 & 2.362 (0.217) & 2.257 (0.109) \\
            \hline
            \end{tabular}}
      \label{s2_error_hinge}}
\subfigure[Subfigure 2 list of figures text][Lasso]{
\footnotesize{\begin{tabular}{lcc}
\hline
&\multicolumn{2}{c}{$d$} \\
\cline{2-3}
$n$         & 50       & 200   \\
\hline
50 & 1.514 (0.347) & 2.163 (0.293) \\
100 & 0.629 (0.129) & 0.998 (0.213) \\
150 & 0.478 (0.092) & 0.614 (0.133) \\
200 & 0.407 (0.076) & 0.415 (0.084) \\
250 & 0.306 (0.058) & 0.382 (0.077) \\
300 & 0.260 (0.045) & 0.306 (0.054) \\
\hline
\end{tabular}}
\label{s2_error_lasso}}
\subfigure[Subfigure 2 list of figures text][SIM]{
\footnotesize{\begin{tabular}{lcc}
\hline
&\multicolumn{2}{c}{$d$} \\
\cline{2-3}
$n$         & 50       & 200   \\
\hline
50 & 2.589 (0.277) & 2.806 (0.926) \\
100 & 0.898 (1.301) & 2.061 (0.812) \\
150 & 0.860 (1.491) & 0.910 (0.243) \\
200 & 0.831 (0.023) & 0.897 (0.928) \\
250 & 0.836 (2.932) & 0.859 (3.495) \\
300 & 0.827 (1.480) & 0.850 (0.007) \\
\hline
\end{tabular}}
\label{s2_error_SIM}}
\subfigure[Subfigure 2 list of figures text][SDR]{
\footnotesize{\begin{tabular}{lcc}
\hline
&\multicolumn{2}{c}{$d$} \\
\cline{2-3}
$n$         & 50       & 200   \\
\hline
50 & 3.947 (1.824) & 2.504 (3.062) \\
100 & 1.096 (3.586) & 1.054 (3.636) \\
150 & 1.009 (3.636) & 0.970 (3.712) \\
200 & 0.889 (3.759) & 1.599 (3.606) \\
250 & 0.862 (3.746) & 0.883 (3.732) \\
300 & 0.857 (3.763) & 0.858 (3.702) \\
\hline
\end{tabular}}
\label{s2_error_SDR}}

\end{center}
\end{table}
}

\subsubsection{Comparison of computation time}\label{sec:runtime}

\hf{Table \ref{comptime} shows the median of the computation time of 200 replicated runs for RMRCE, Hinge, the lasso, SIM, and SDR with different $n$ and $d=50$. For demonstration purpose we only present the computation time with fixed tuning parameters. For RMRCE, the Gaussian approximation is used and the tuning parameters are set as $\alpha_n=5$ and $\lambda_n=1$. For Hinge and the lasso we set the tuning parameter as $\lambda_n=1$. SIM and SDR are run with the default parameters. The results are similar with other choices of the tuning parameters, as long as they are in a reasonable scale.

The lasso is the most efficient among all methods, and its computation time is close to zero. RMRCE takes significantly more time to finish compared to the lasso, but is still much faster than SIM and SDR. In addition, RMRCE is slightly slower than Hinge since in Hinge the non-convex rank correlation is replaced with a convex function, making the optimization faster. In summary RMRCE is much faster compared to SIM and SDR, but not as efficient as Hinge and the lasso.}

\begin{table}[ht]
\caption{\label{convex}\hf{\textbf{Median computation time (seconds) for all competing methods.} Methods included are: RMRCE, Hinge, the lasso, SIM, and SDR. The results are based on 200 replications with different $n$, $d=50$, and fixed tuning parameters $\alpha_n$ and $\lambda_n$.}}
\begin{center}
\footnotesize{\begin{tabular}{lccccccc}
\hline
&\multicolumn{6}{c}{$n$} \\
\cline{2-7}
Method & 50 & 100 & 150 & 200 & 250 & 300 \\
\hline
RMRCE & 1.350 & 5.177 & 14.872 & 24.83 & 30.638 & 33.771 \\
Hinge & 0.263 & 0.822 & 1.627 & 3.242 & 4.910 & 6.618 \\
Lasso & 0.002 & 0.011 & 0.157 & 0.002 & 0.002 & 0.002 \\
SIM & 72.202 & 144.404 & 168.132 & 190.23 & 218.331 & 316.347 \\
SDR & 59.849 & 122.514 & 132.736 & 157.172 & 198.831 & 237.852 \\
\hline
\end{tabular}}
\label{comptime}
\end{center}
\end{table}

\section{Empirical verification of the theory}

This section examines Theorem \ref{thm:error_bound} 
and Assumption {\bf (A5)} in Section \ref{sec:variance} using synthetic data. 

\subsection{Convexity verification}\label{sec:convex}
For investigating the stochastic error $\norm{\hat\bbeta_{\alpha_n}-\bbeta^*_{\alpha_n}}_2$, Assumption {\bf (A5)} is required to hold. In this section, we verify Assumption {\bf (A5)} via various empirical studies on the positive definiteness of the following Hessian matrix:
\[
\nabla^2\hat\cL_n(\bbeta^*)=-\frac{2}{n(n-1)}\sum_{i<i'}\alpha_n^2(\bX_{i}-\bX_{i'})(\bX_i-\bX_{i'})^\T F_{ii'}''(\tilde S_{ii'}\alpha_nZ_{ii'}(\bbeta^*)).
\]  

For presentation clearness, we focus on the high dimensional linear model \eqref{model} with $\bbeta^0=(5,4,3,2,1,-1,-3,-5,0,\cdots,0)^\T$ and $\bX_i$ a $d$ dimensional random vector generated from a multivariate normal distribution $\text{N}_d(\mathbf{0},\bSigma=((\sigma_{jk})))$ with $\sigma_{jk}=0.5^{|j-k|}$ for $1\leq j,k\leq d$. We further generate $\epsilon_i$ from a mixture of the standard normal and $\delta=$ 0\% or 20\% of the outliers following Cauchy($0, 0.01$). The noise is independent of $\bX_i$. 

We demonstrate the results under different situations with different noise distributions as well as different smoothing approximations. Table \ref{convex} gives the results, considering all three examples of smoothing approximations in Remark \ref{remark:example}, with pure Gaussian noise and the mixture of Gaussian noise with $20$\% Cauchy outliers. Here, via exhaustive simulation studies,  $\alpha_n$ is recommended to be $5$ for sigmoid, Gaussian, and double exponential CDF approximations. The results are calculated based on 200 replications.

There are several noteworthy discoveries. First, the performances of all cases are similar. Specifically, as sample size $n$ increases, the proportion of positive definite Hessian matrices increases to 1. In addition, small dimension $d$ leads to higher proportion of convexity. Secondly, for comparison between different noise terms, we find pure Gaussian noise enjoys higher proportion of convexity compared to the mixture noise with large $n$. Thirdly, for comparison between different smoothing approximations, they perform similarly, though the Gaussian CDF approximation enjoys slightly higher proportions of positive definite Hessian matrices. 


\begin{table}[ht]
\caption{\label{convex}\textbf{Proportion of positive definite Hessian matrices.} Results are based on 200 replications. The first row is for $\delta=0$ and the second row is for $\delta=0.2$. ``DE" stands for double exponential CDF approximation.}
\begin{center}
\subfigure[Subfigure 2 list of figures text][Sigmoid, $\delta=0$]{
\footnotesize{\begin{tabular}{lccc}
\hline
&\multicolumn{3}{c}{$d$} \\
\cline{2-4}
n         & 50       & 200  & 800  \\
\hline
50        & 0.830 & 0.815 & 0.850 \\
100       & 1.000 & 0.925 &  0.910 \\
150       & 1.000 & 0.985 & 0.935 \\
200       & 1.000 & 1.000 & 0.955 \\
250       & 1.000 & 1.000 & 0.940 \\
300       & 1.000 & 1.000 & 0.970 \\
\hline
\end{tabular}}}
\hspace{0.5cm}
\subfigure[Subfigure 2 list of figures text][Gaussian, $ \delta=0$]{
\footnotesize{\begin{tabular}{lccc}
\hline
&\multicolumn{3}{c}{$d$} \\
\cline{2-4}
n         & 50       & 200  & 800 \\
\hline
50        & 0.865 & 0.840 & 0.875 \\
100       & 1.000 & 0.955 &  0.930\\
150       & 1.000 & 0.990 &  0.975\\
200       & 1.000 & 1.000 &  0.965\\
250       & 1.000 & 1.000 &  0.970\\
300       & 1.000 & 1.000 &  0.980\\
\hline
\end{tabular}}}
\hspace{0.5cm}
\subfigure[Subfigure 2 list of figures text][DE, $\delta=0$]{
\footnotesize{\begin{tabular}{lccc}
\hline
&\multicolumn{2}{c}{$d$} \\
\cline{2-4}
n         & 50       & 200  & 800 \\
\hline
50        & 0.815 & 0.845 & 0.870  \\
100       & 1.000 & 0.920 &  0.925\\
150       & 1.000 & 0.975 &  0.950\\
200       & 1.000 & 0.995 &  0.945\\
250       & 1.000 & 1.000 &  0.980\\
300       & 1.000 & 1.000 &  0.975\\
\hline
\end{tabular}}}

\subfigure[Subfigure 2 list of figures text][Sigmoid, $ \delta=0.2$]{
\footnotesize{\begin{tabular}{lccc}
\hline
&\multicolumn{2}{c}{$d$} \\
\cline{2-4}
n         & 50       & 200   & 800 \\
\hline
50        & 0.890 & 0.845 &  0.880\\
100       & 1.000 & 0.925 &  0.925\\
150       & 1.000 & 0.945 &  0.955\\
200       & 1.000 & 0.980 &  0.950\\
250       & 1.000 & 0.990 &  0.960\\
300       & 1.000 & 0.990 &  0.965\\
\hline
\end{tabular}}}
\hspace{0.5cm}
\subfigure[Subfigure 2 list of figures text][Gaussian, $ \delta=0.2$]{
\footnotesize{\begin{tabular}{lccc}
\hline
&\multicolumn{3}{c}{$d$} \\
\cline{2-4}
n         & 50       & 200  & 800 \\
\hline
50        & 0.900 & 0.870 &  0.910\\
100       & 1.000 & 0.930 &  0.930\\
150       & 1.000 & 0.975 &  0.945\\
200       & 1.000 & 0.990 &  0.955\\
250       & 1.000 & 0.990 &  0.950\\
300       & 1.000 & 0.995 &  0.960\\
\hline
\end{tabular}}}
\hspace{0.5cm}
\subfigure[Subfigure 2 list of figures text][DE, $ \delta=0.2$]{
\footnotesize{\begin{tabular}{lccc}
\hline
&\multicolumn{3}{c}{$d$} \\
\cline{2-4}
n         & 50       & 200   & 800 \\
\hline
50        & 0.845 & 0.865 & 0.865 \\
100       & 1.000 & 0.945 &  0.925 \\
150       & 1.000 & 0.965 &  0.935 \\
200       & 1.000 & 0.970 &  0.930 \\
250       & 1.000 & 0.985 &  0.955 \\
300       & 1.000 & 0.995 &  0.950 \\
\hline
\end{tabular}}}
\end{center}
\end{table}

\subsection{Verification of the theorem}

This section verifies the property given in Theorem \ref{thm:error_bound}. Simulations with different generating models have shown that the proposed methods 
are robust to the monotonic functions $D(\cdot)$ and $\Lambda(\cdot,\cdot)$. Hence to demonstrate the scaling property of the estimation errors $\norm{\hat\bbeta_{\alpha_n}-\bbeta^*}_2$, 
we only focus on the high dimensional linear model (\ref{model}) with $\epsilon_i$ generated from the standard normal and independent of $\bX_i$. \hf{For simplicity, we only show the results for RMRCE with $\alpha=5$. The results are similar for other choices of $\alpha_n$.}

\begin{figure}[ht]
\centering
\subfigure[Subfigure 2 list of figures text][{\sf\small Sigmoid}]{
\includegraphics[width=0.245\textwidth]{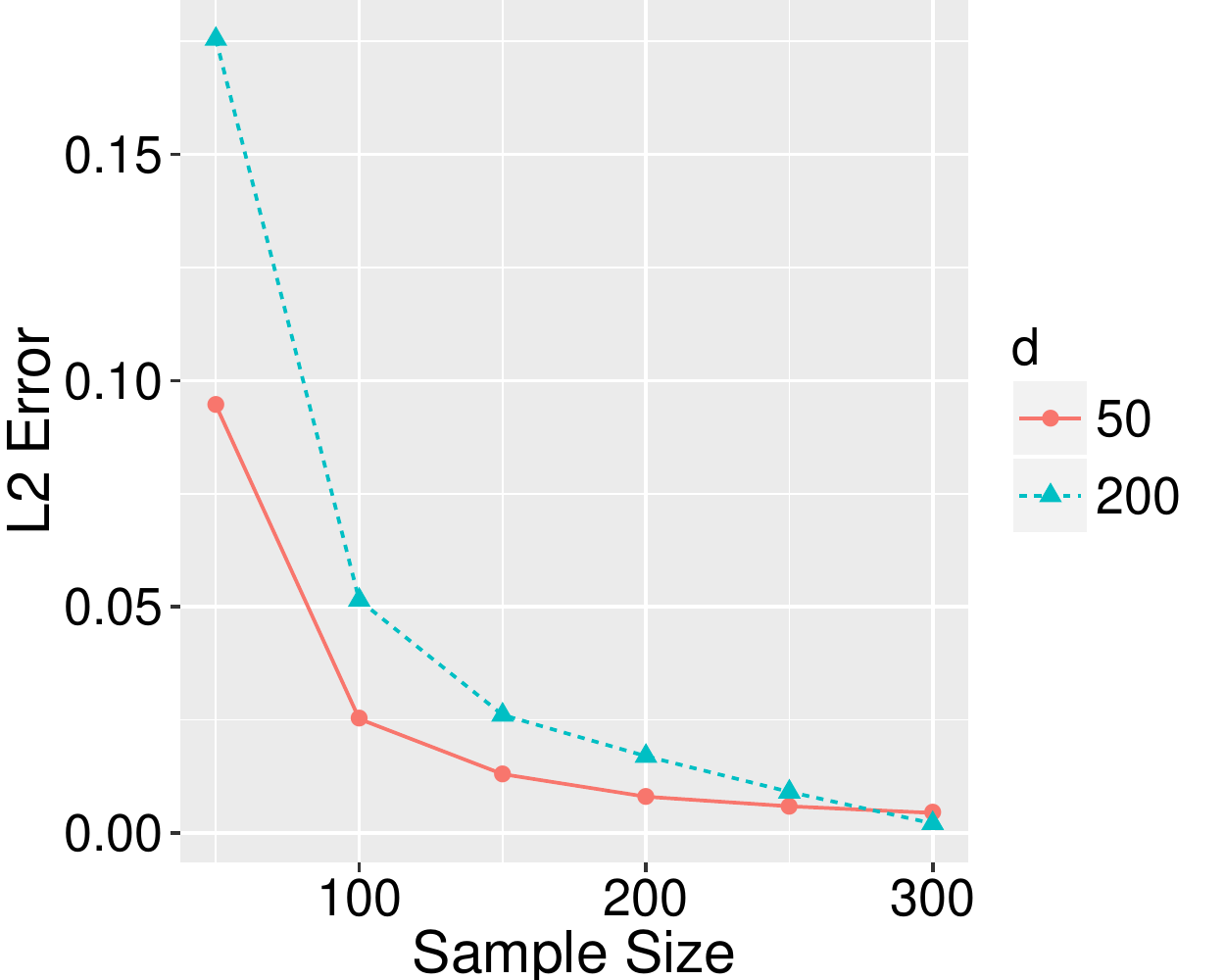}
\label{stack_logistic}}
\hspace{-0.5cm}
\subfigure[Subfigure 2 list of figures text][{\sf\small Gaussian}]{
\includegraphics[width=0.245\textwidth]{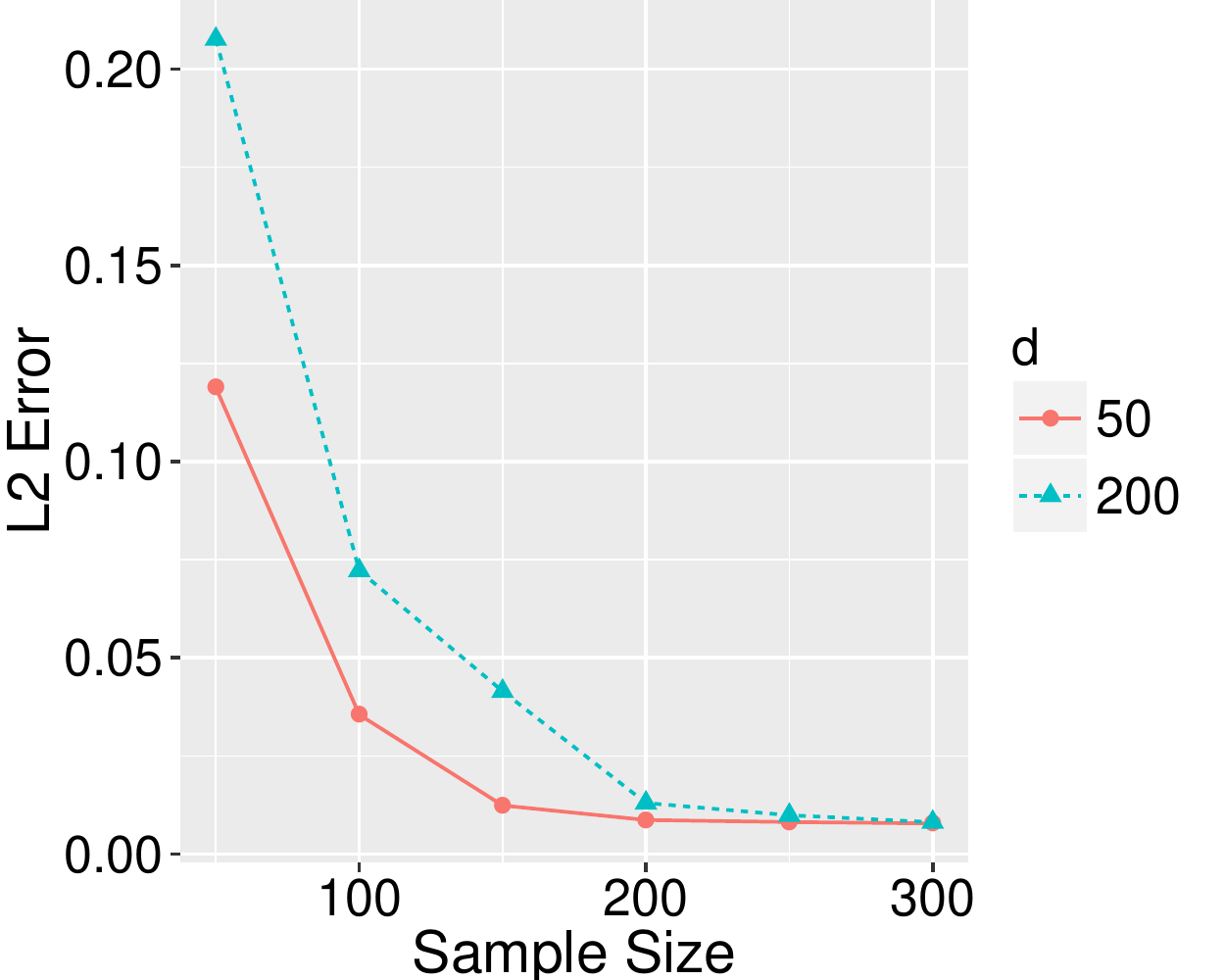}
\label{stack_gaussian}}
\hspace{-0.5cm}
\subfigure[Subfigure 2 list of figures text][{\sf\small Double Exponential}]{
\includegraphics[width=0.245\textwidth]{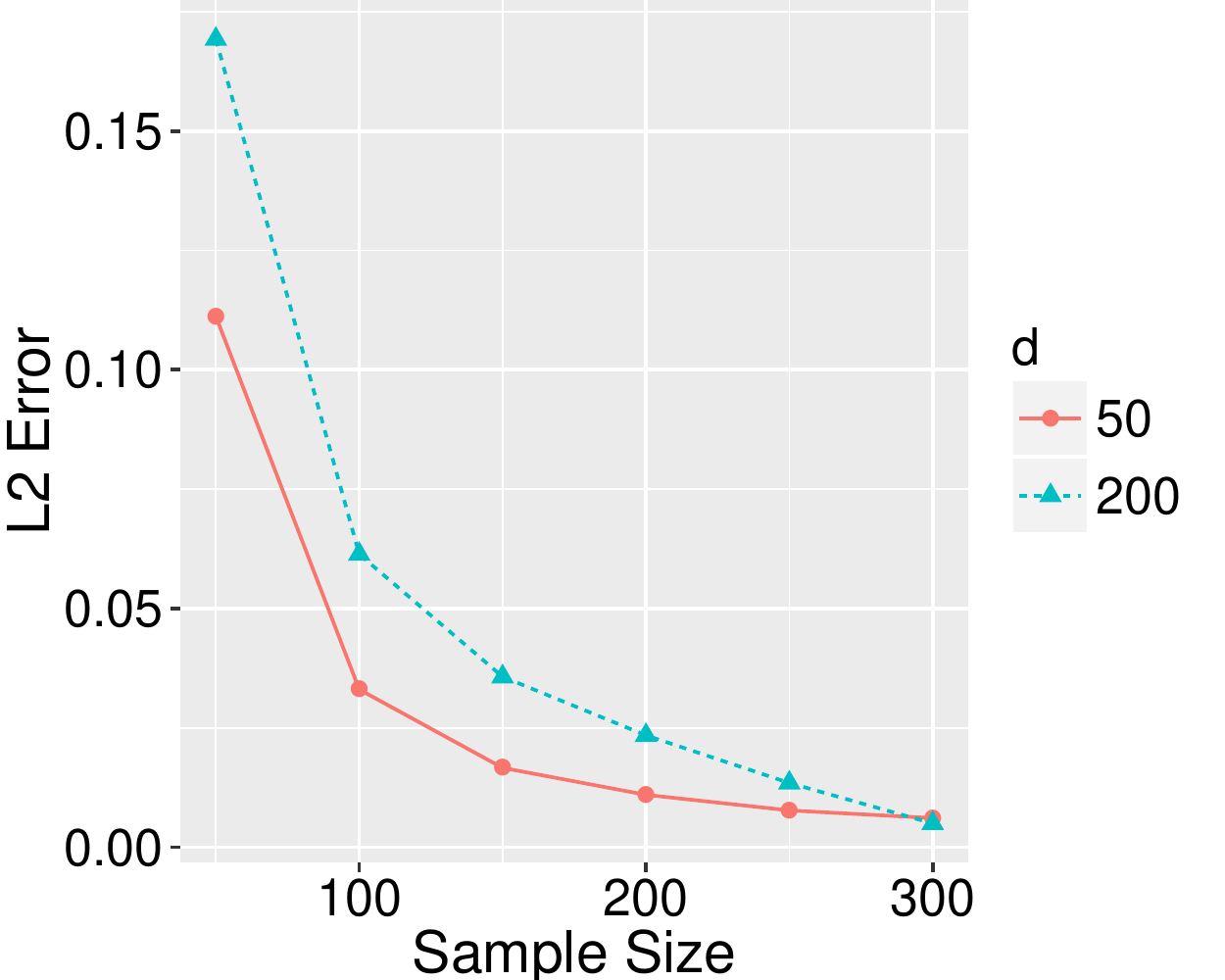}
\label{stack_double_exp}}
\hspace{-0.5cm}
\subfigure[Subfigure 2 list of figures text][{\sf\small lasso}]{
\includegraphics[width=0.245\textwidth]{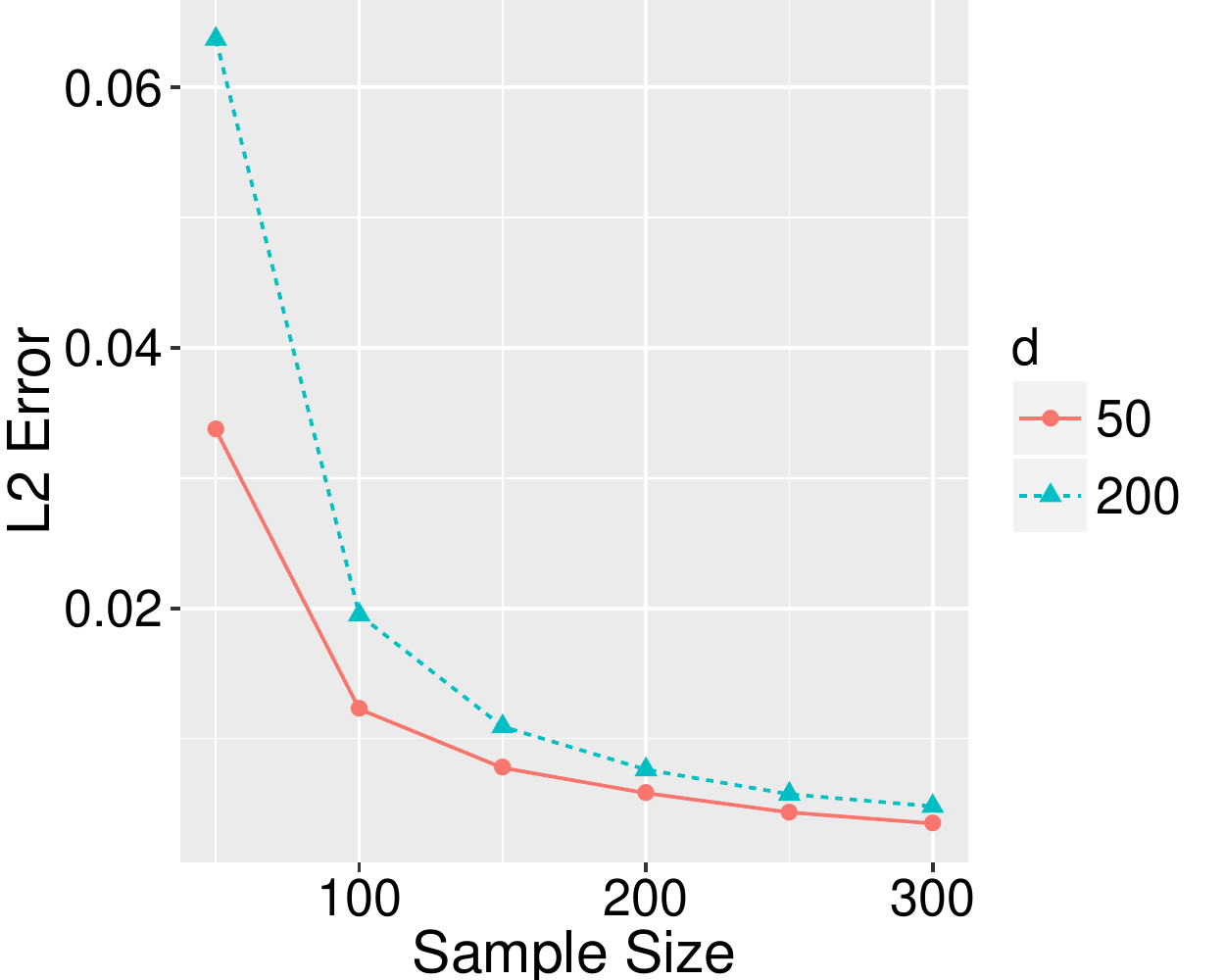}
\label{stack_lasso}}
\subfigure[Subfigure 2 list of figures text][{\sf\small Sigmoid}]{
\includegraphics[width=0.245\textwidth]{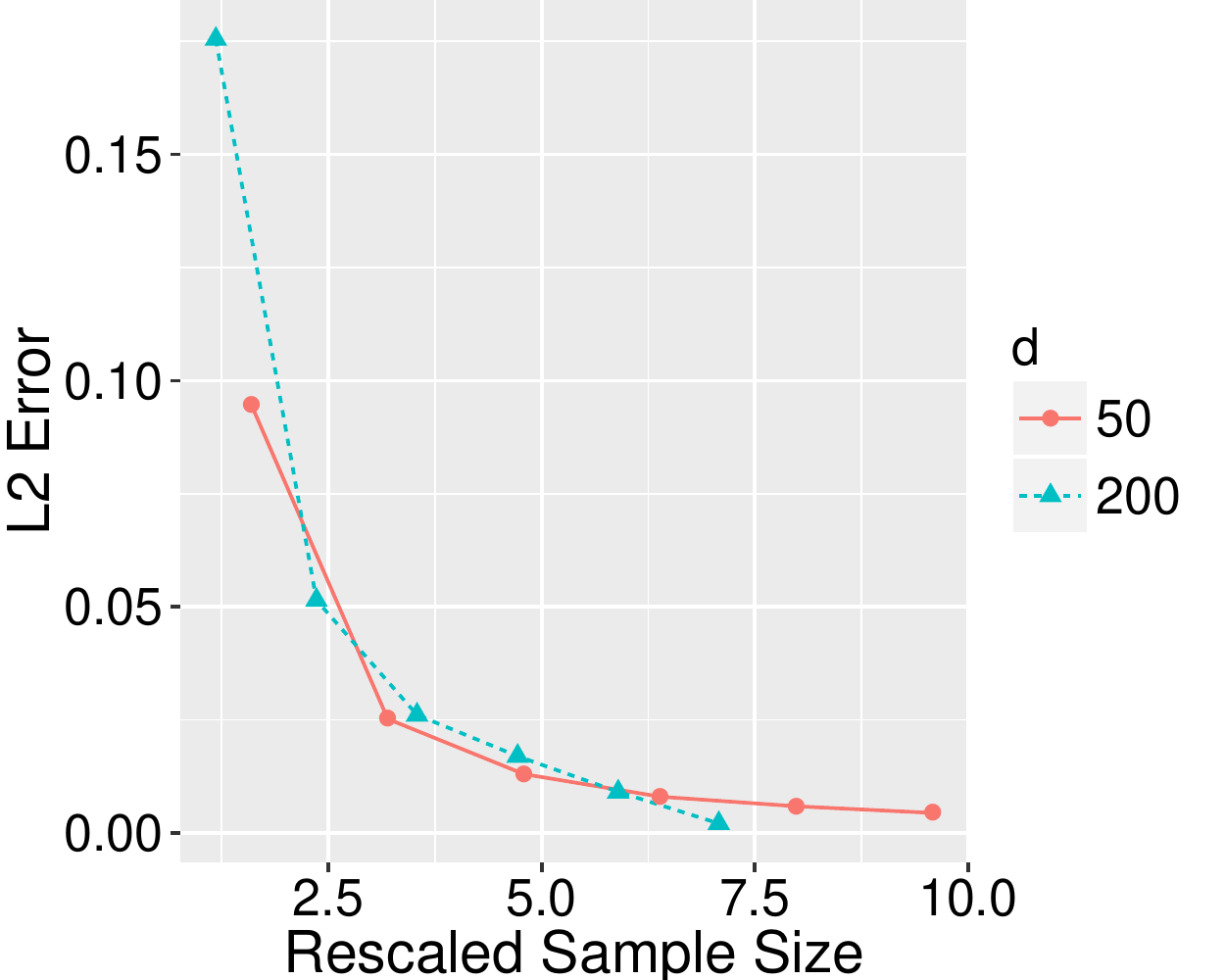}
\label{stack_logistic_re}}
\hspace{-0.5cm}
\subfigure[Subfigure 2 list of figures text][{\sf\small Gaussian}]{
\includegraphics[width=0.245\textwidth]{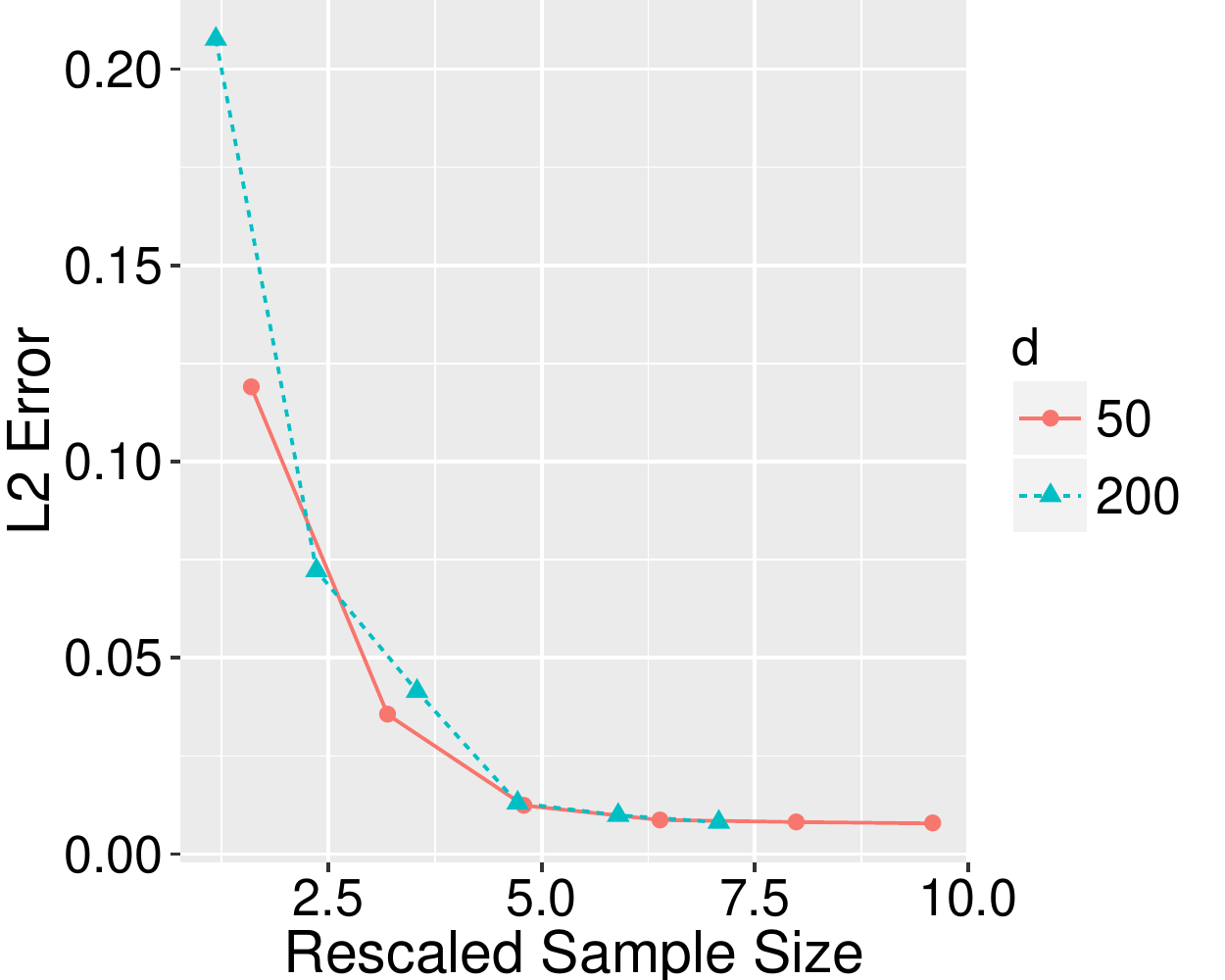}
\label{stack_gaussian_re}}
\hspace{-0.5cm}
\subfigure[Subfigure 2 list of figures text][{\sf\small Double Exponential}]{
\includegraphics[width=0.245\textwidth]{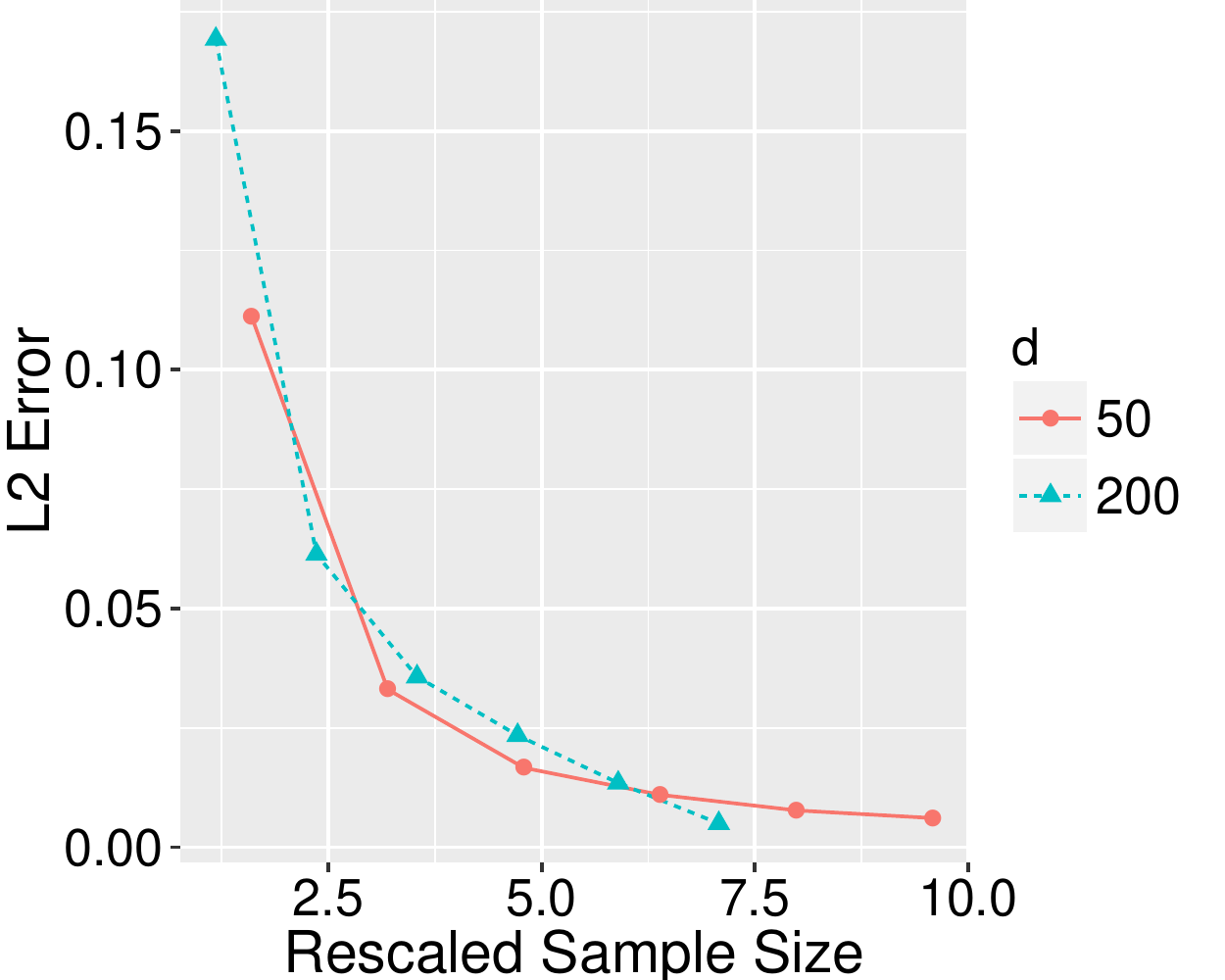}
\label{stack_double_exp_re}}
\hspace{-0.5cm}
\subfigure[Subfigure 2 list of figures text][{\sf\small lasso}]{
\includegraphics[width=0.245\textwidth]{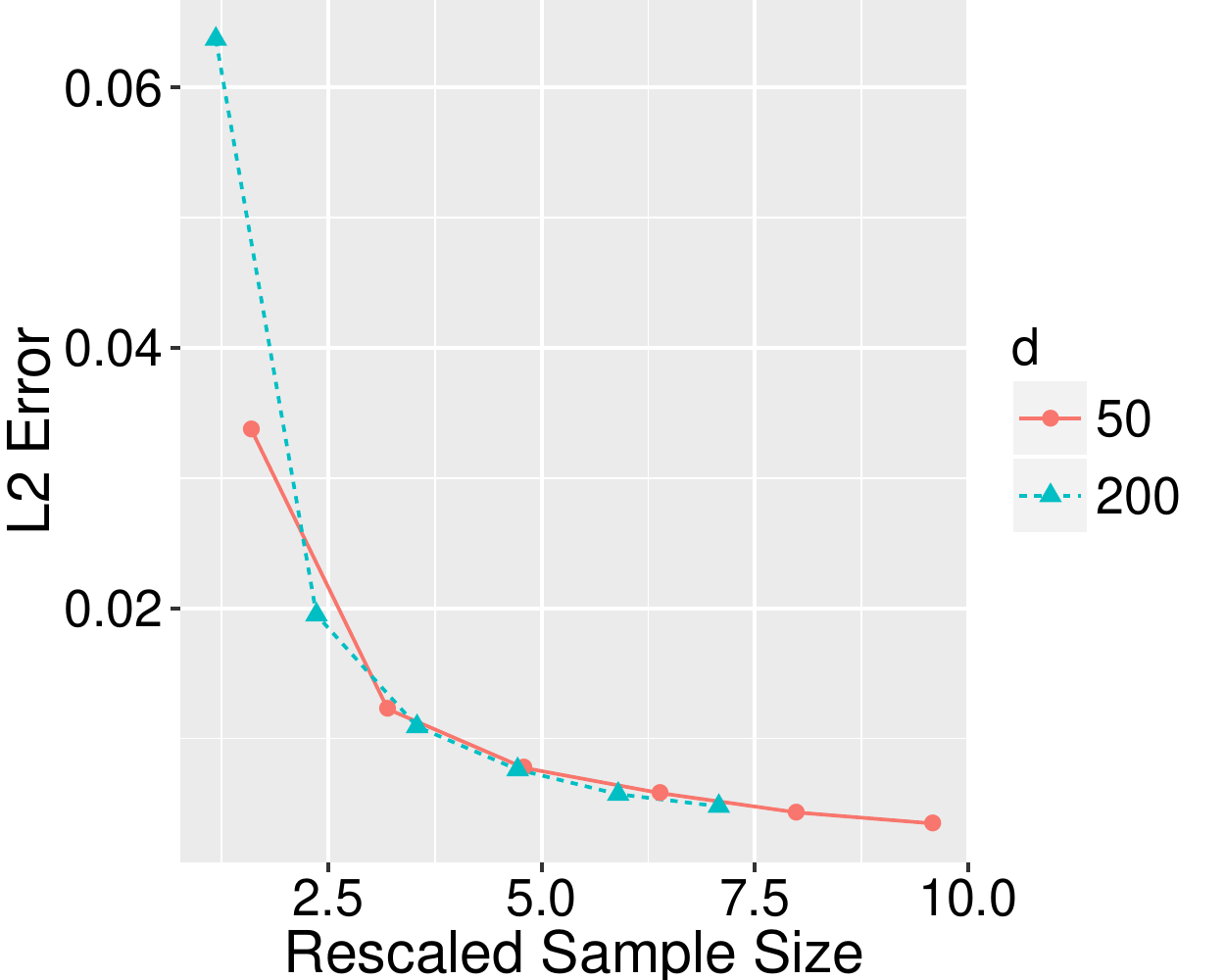}
\label{stack_lasso_re}}
\caption{\textbf{Estimation error curves for different methods.}  The first row is for estimation error curves compared to the sample size $n$, and the second row is for estimation error curves compared to the rescaled sample size $n/(s\log d)$. }
\label{stack}
\end{figure}

To this end, Figure \ref{stack} illustrates the scaling property of the averaged estimation errors $\norm{\hat\bbeta_{\alpha_n}-\bbeta^*}_2$ compared to $n$ and $n/(s\log d)$ across 200 replications for different approximation functions. The performance is also compared to the lasso. 

Figure \ref{stack} clearly shows a ``stacking curve" phenomena. 
Specifically, regarding the rescaled sample size $n/(s\log d)$, the error curves corresponding to different $d$'s are all aligned together. In addition, the error decays to zero as sample size $n$ increases, and increases as $d$ increases. Hence Figure \ref{stack} confirms the theoretical discovery in Theorem \ref{thm:error_bound}.  
For comparison, we also include the lasso's stacking curves in Figure \ref{stack_lasso} and \ref{stack_lasso_re}. They show similar ``stacking curve" phenomena. 

\section{Additional materials for real data example}\label{sec:realdata}

\subsection{Data processing and normalization for TF prediction}\label{supp:pp}

We downloaded DNase-seq and gene expression exon array data from 57 different cell types generated by ENCODE at the University of Washington \citep{thurman2012accessible}.
Exon array samples were consistently normalized using the GeneBASE software \citep{kapur2007exon}. The output of GeneBASE was gene-level expression values for all genes. From these values, we extracted expression values for all human TFs documented in the Animal Transcription Factor Database (AminalTFDB) \cite{zhang2012animaltfdb}. We then filtered out TFs whose expression values were nearly constant across all samples (defined as coefficient of variation $<0.2$) because one would expect them to behave like an intercept term in a regression model. These nearly-constant TFs were not expected to provide much information to explain variation of the response variable across different cell types. After filtering, 169 TFs were retained and they were used as predictors $\bX$ in our regression models. We obtained the observed values of $\bX$ for all samples. Replicate samples from the same cell type were averaged. This resulted in a gene expression matrix $\mathbb{X}$ with 57 rows and 169 columns. Here rows correspond to 57 cell types ($n=57$). Columns correspond to 169 TFs ($d=169$). Each row is a realization of $\bX$.

For the responses, we processed the DNase-seq data from the same 57 cell types as follows. First, the human genome was divided into 200 base pair (bp) non-overlapping windows, yielding approximately $1.65\times10^7$ windows. For each window and each DNase-seq sample, we calculated the DNase-seq signal by counting the number of DNase-seq reads overlapping with the window. The read count was then normalized by the library size. To do so, the window read count was divided by the total read count of the sample and then multiplied with a constant 17,002,867, which is the minimum sample read count of all samples. The normalized counts were $\log_2$ transformed after adding a pseudocount of 1. Since windows without any DNase-seq signal are unlikely to be cis-elements,  we only retained windows that had non-zero read count in at least 10 cell types and had coefficient of variation no less than 1. From these retained windows, we randomly sampled 1,000 windows to serve as our testing cis-elements. For each cis-element, we extracted the normalized and log-transformed read count from all DNase-seq samples to serve as the measurements of their DNase I hypersensitivity. Replicate samples from the same cell type were averaged. This produced a matrix $\mathbb{Y}$ with 57 rows and 1,000 columns. The rows represent 57 cell types, and the columns represent 1,000 cis-elements. Each column corresponds to a response $Y$, and it contains the observed values of $Y$ in all 57 cell types.

Lastly, we used $\mathbb{X}$ and $\mathbb{Y}$ to perform the analyses. Our regression model was fitted for each cis-element (i.e., each column of $\mathbb{Y}$) separately. Each regression has sample size $n=57$ and dimension {$d=169$}. An analysis of the whole dataset involves fitting 1,000 regression models.

\subsection{RMRCE performance with difference choices of $\alpha_n$}\label{sec:realdataalpha}

\hf{With regard to the real data experiment in Section \ref{sec:real}, Figure \ref{TFacc_RMRCE} further compares the accuracy of RMRCE with $\alpha_n$ chosen by cross validation or set to be fixed values $1,3,5,7,9$. The performances are similar for RMRCE with difference choices of $\alpha_n$. $\alpha_n$ chosen by cross validation only leads to marginal performance gain compared to fixed $\alpha_n$. This agrees with the conclusion from the synthetic data analysis.}

\hf{In real data applications, since different choices of $\alpha_n$ (as long as $\alpha_n$ is large enough) lead to fairly robust results, we recommend using a fixed $\alpha_n=5$, although choosing the optimal $\alpha_n$ via a cross validation procedure is encouraged if enough time and resource for computation are available.}

\begin{figure}[!htbp]
\centering
\includegraphics[width=0.8\textwidth]{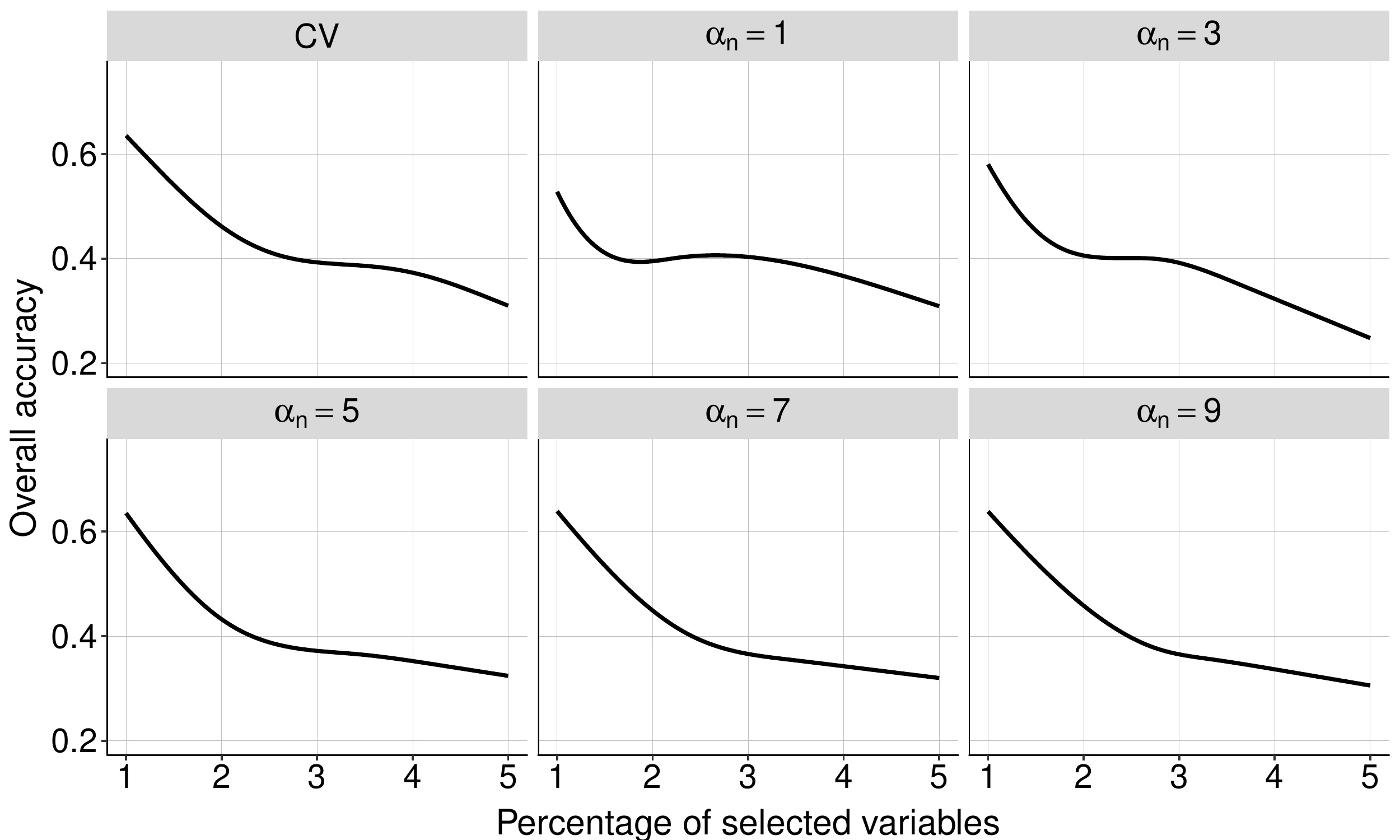}
\caption{\hf{Overall accuracy of RMRCE with $\alpha_n$ chosen by cross validation (CV) or set to be fixed values $1,3,5,7,9$. X-axis shows the averaged percentage of selected TFs out of all 169 TFs. Y-axis shows the overall accuracy.}}
\label{TFacc_RMRCE}
\end{figure}

\subsection{A model diagnostic heuristic}\label{sec:realdatadiagnostic}

\hf{The monotonicity assumption is the most important and intrinsic feature of the proposed generalized regression model \eqref{eq:model}, and this section provides another heuristic to examine this assumption in real data applications. Figures \ref{cislassoexample} and \ref{cisRMRCEexample} provide some empirical illustrations that $Y$ has a monotonically increasing relationship with $\bX^\T \widehat {\bm \beta}^{\rm RMRCE}$ for two cis-elements. In this section we further discuss a model diagnostic tool to check the monotonicity assumption of our model, and we apply this tool to the real data example.

For model \eqref{eq:model}, our goal is to verify the assumption that the response $Y$ has a monotonically increasing relationship with the linear term $\bX^\T\bbeta^*$. However, this assumption cannot be directly verified in reality since $\bbeta^*$ is unknown. Instead we develop a model diagnostic heuristic by replacing $\bbeta^*$ with $\widehat {\bm \beta}^{\rm RMRCE}$ and checking whether $Y$ has a monotonically increasing relationship with $\bX^\T \widehat {\bm \beta}^{\rm RMRCE}$ in a cross-validation procedure. 
Specifically, we split the response and explanatory variables $(Y,\bX)$ into two parts with equal number of observations: the first part $(Y_1,\bX_1)$ has the first half of all observations and the second part $(Y_2,\bX_2)$ has the second half of all observations. We fit RMRCE and obtain $\widehat {\bm \beta}^{\rm RMRCE}$ on $(Y_1,\bX_1)$ and calculate the Spearman's rank correlation between $Y_2$ and $\bX_2^\T \widehat {\bm \beta}^{\rm RMRCE}$. We use the one-sided Spearman's rank correlation test \citep{hollander2013nonparametric} to test whether the Spearman's rank correlation is significantly positive. 

As a demonstration, we apply the model diagnostic heuristic to the real data example of TF prediction. For the protein-binding activity $Y$ and gene expression level $\bX$ of each of the 1,000 cis-elements, we split $(Y,\bX)$ into $(Y_1,\bX_1)$ and $(Y_2,\bX_2)$, fit RMRCE on $(Y_1,\bX_1)$, and lastly perform the one-sided Spearman's rank correlation test on $Y_2$ and $\bX_2^\T \widehat {\bm \beta}^{\rm RMRCE}$ to examine the monotonicity assumption. 1,000 $p$-values are obtained and adjusted for multiple testing using Bonferroni method. With an adjusted $p$-value smaller than 0.05, we reject the null hypothesis that there is none or a negative association. 
The whole procedure is repeated for different choices of RMRCE tuning parameters $\alpha_n$ and $\lambda_n$. Table \ref{diagnostic} shows the percentage of the adjusted $p$-values that are smaller than 0.05. Most of the adjusted $p$-values are smaller than 0.05.}

{
\renewcommand{\tabcolsep}{20pt}
\renewcommand{\arraystretch}{1.00}
\begin{table}[ht]
\caption{\label{convex}\hf{\textbf{Model diagnostic results for RMRCE in 1,000 studied real datasets.} The percentage of 1,000 adjusted $p$-values that are smaller than 0.05 for Spearman's rank correlation tests, or equivalently the percentage of 1,000 cis-elements that pass the model diagnostic heuristic tests. Results for different $\alpha_n$ and $\lambda_n$ are shown.}}
\begin{center}
\footnotesize{\begin{tabular}{lcccccc}
\hline
&\multicolumn{5}{c}{$\alpha_n$} \\
\cline{2-6}
$\lambda_n$ & 1 & 3 & 5 & 7 & 9 \\
\hline
0.01 & 0.796 & 0.761 & 0.747 & 0.732 & 0.725 \\
0.02 & 0.759 & 0.756 & 0.722 & 0.709 & 0.701 \\
0.03 & 0.775 & 0.752 & 0.725 & 0.716 & 0.703 \\
0.04 & 0.772 & 0.722 & 0.707 & 0.727 & 0.709 \\
0.05 & 0.775 & 0.747 & 0.711 & 0.696 & 0.693 \\
\hline
\end{tabular}}
\label{diagnostic}
\end{center}
\end{table}
}

\section{Proofs}\label{sec:proofs}

This section collects the proofs of the results in the paper. Recall  $Z_{ii'}(\bbeta):=(\bX_i-\bX_{i'})^\T\bbeta$, and $S_{ii'}:=\mathds{1}(Y_i>Y_{i'})$. In the sequel, 
we define $p_0(\bX_i,\bX_{i'};\bbeta^*):=\P(S_{ii'}=0|\bX_i,\bX_{i'})=\P(Y_i<Y_{i'}|\bX_i,\bX_{i'})$, and $p_1(\bX_i,\bX_{i'};\bbeta^*):=\P(S_{ii'}=1|\bX_i,\bX_{i'})=\P(Y_i>Y_{i'}|\bX_i,\bX_{i'})$.

\subsection{Proof of Proposition \ref{prop:1} }
\begin{proof}

(i) Given the link function $D\circ \Lambda(u,v)=u+v$, we have $Y=\bX^\T\bbeta^*+\epsilon$ but  without requiring $\bX$ and $\epsilon$ to be centered.
First, by simple calculation, using the fact that $\bX$ is independent with $\epsilon$, it immediately follows
\[
\frac{\E [(Y_1-Y_2) (\bX_1^\T\bbeta-\bX_2^\T\bbeta)]}{\sqrt{\cov(Y_1-Y_2)}\sqrt{\cov(\bX_1^\T\bbeta-\bX_2^\T\bbeta)}}=\frac{\E [\bbeta^{*\T}(\bX_1-\bX_2)(\bX_1-\bX_2)^\T\bbeta]}{\sqrt{\cov(Y_1-Y_2)}\sqrt{\cov(\bX_1^\T\bbeta-\bX_2^\T\bbeta)}}.
\]
Noticing that $\bX_1-\bX_2$ is a random vector with mean $\mathbf{0}$ and covariance $2\bSigma$, we have
\[
\argmax_{\bbeta\in\reals^d}\Big\{\frac{\E [(Y_1-Y_2) (\bX_1^\T\bbeta-\bX_2^\T\bbeta)]}{\sqrt{\cov(Y_1-Y_2)}\sqrt{\cov(\bX_1^\T\bbeta-\bX_2^\T\bbeta)}}\Big\}=\argmax_{\bbeta\in\reals^d}\frac{\bbeta^{*\T}\bSigma\bbeta}{\sqrt{\bbeta^\T\bSigma\bbeta}},
\]
since $\cov(Y_1-Y_2)$ is a constant as a function of $\bbeta$. Let $\mathbf{a}:=\bSigma^{1/2}\bbeta^*$ and $\mathbf{b}:=\bSigma^{1/2}\bbeta/\|\bSigma^{1/2}\bbeta\|_2$. Using Cauchy-Schwarz inequality yields that $\max_{\mathbf{b}:\|\mathbf{b}\|_2=1}\mathbf{a}^\T\mathbf{b}$ achieves at $\mathbf{b}=\mathbf{a}/\|\mathbf{a}\|_2$. From the inversibility of $\bSigma$, it then follows $\argmax_{\bbeta\in\reals^d}\bbeta^{*\T}\bSigma\bbeta/\sqrt{\bbeta^\T\bSigma\bbeta}=\bbeta^*$ up to a scaling. 

(ii) This follows directly from the proof of Theorem 1 in \cite{han1987non}. 

Hence we complete the proof of the proposition.
\end{proof}

\subsection{Proof of Lemma \ref{lem:approx}}
\begin{proof}
First of all, recall $\cL(\bbeta)=-\E\big(S_{ii'}\mathds{1}(Z_{ii'}(\bbeta)>0)+(1-S_{ii'})\mathds{1}(Z_{ii'}(\bbeta)<0)\big)$ and $\cL_n(\bbeta)=-{2}\sum_{i<i'}\E\Big\{S_{ii'}F_{ii'}(\alpha_nZ_{ii'}(\bbeta))+(1-S_{ii'})(1-F_{ii'}(\alpha_nZ_{ii'}(\bbeta)))\Big\}/\{n(n-1)\}$.
By the definition of $p_0(\bX_i,\bX_{i'};\bbeta^*)$, and $p_1(\bX_i,\bX_{i'};\bbeta^*)$, after taking conditional expectation of the response given the covariates, we have
\begin{align*}
\cL(\bbeta)-\cL_n(\bbeta)
&=\frac{2}{n(n-1)}\sum_{i<i'}\E\Big\{p_1(\bX_i,\bX_{i'};\bbeta^*)\big[F_{ii'}(\alpha_nZ_{ii'}(\bbeta))-\mathds{1}(Z_{ii'}(\bbeta)>0)\big]\\
&~~~~~~~~~~~~~~~~~~~~~+p_0(\bX_i,\bX_{i'};\bbeta^*)\big[1-F_{ii'}(\alpha_nZ_{ii'}(\bbeta))-\mathds{1}(Z_{ii'}(\bbeta)<0)\big]\Big\}.
\end{align*}
According to whether $Z_{ii'}(\bbeta)>0$ or not, we rewrite the above expression as follows,
\begin{align*}
&\cL(\bbeta)-\cL_n(\bbeta)\\
=&\frac{2}{n(n-1)}\sum_{i<i'}\E\Big\{\big[p_1(\bX_i,\bX_{i'};\bbeta^*)-p_0(\bX_i,\bX_{i'};\bbeta^*)\big]F_{ii'}(\alpha_nZ_{ii'}(\bbeta))\mathds{1}(Z_{ii'}(\bbeta)<0)\\
&+\big[p_0(\bX_i,\bX_{i'};\bbeta^*)-p_1(\bX_i,\bX_{i'};\bbeta^*)\big]\big[1-F_{ii'}(\alpha_nZ_{ii'}(\bbeta))\big]\mathds{1}(Z_{ii'}(\bbeta)>0)\Big\}.
\end{align*}
Because of $|p_1(\bX_i,\bX_{i'};\bbeta^*)-p_0(\bX_i,\bX_{i'};\bbeta^*)|\leq 1$, it yields
\begin{align}\label{lem_approx_temp1}
\cL(\bbeta)-\cL_n(\bbeta)
&\leq\frac{2}{n(n-1)}\sum_{i<i'}\E\Big\{F_{ii'}(\alpha_nZ_{ii'}(\bbeta))\mathds{1}(Z_{ii'}(\bbeta)<0)\notag \\
&~~~~~~~~~~~~~~~~~~~~~+\big[1-F_{ii'}(\alpha_nZ_{ii'}(\bbeta))\big]\mathds{1}(Z_{ii'}(\bbeta)>0)\Big\}.
\end{align}
Following Assumption {\bf (A1)}, we have $Z_{ii'}(\bbeta)\sim N_1(0,\sigma^2_{\bbeta})$ with $\sigma^2_{\bbeta}:=2\bbeta^{\T}\bSigma\bbeta$ for all $1\leq i\neq i'\leq n$. Hence with the expectation $\E$ in (\ref{lem_approx_temp1}) taken with respect to $\P_{\bbeta}(z)\sim N_1(0,\sigma^2_{\bbeta})$, it follows
\begin{align*}
\cL(\bbeta)-\cL_n(\bbeta)
&\leq \frac{2}{n(n-1)}\sum_{i<i'}\Big\{\int_0^{\infty}\{1-F_{ii'}(\alpha_nz)\}d\P_{\bbeta}(z)+\int^0_{-\infty}F_{ii'}(\alpha_nz)d\P_{\bbeta}(z)\Big\}.
\end{align*}
Notice that with Assumption {\bf (A2)}, we have $F_{ii'}(-u)=1-F_{ii'}(u)$ for arbitrary $u\geq 0$, together with $\P_{\bbeta}(z)\sim N_1(0,\sigma^2_{\bbeta})$, simple calculation immediately yields
\begin{align*}
\cL(\bbeta)-\cL_n(\bbeta)
&\leq \frac{4}{n(n-1)}\sum_{i<i'}\int_0^{\infty}\frac{1}{\sqrt{2\pi}\sigma_{\bbeta}}\{1-F_{ii'}(\alpha_nz)\}\exp\big(-\frac{z^2}{2\sigma^2_{\bbeta}}\big)dz.
\end{align*}
Assumption {\bf (A2)} also assumes the existence of some absolute constants $C_1, C_2>0$ such that $1-F_{ii'}(u)\leq C_1\exp(-C_2u)$ for any $u>0$. Thus we further have
\begin{align*}
\cL(\bbeta)-\cL_n(\bbeta)
&\leq 2\int_0^{\infty}\frac{C_1}{\sqrt{2\pi}\sigma_{\bbeta}}\exp\big(-\frac{z^2}{2\sigma^2_{\bbeta}}-C_2\alpha_n z\big)dz=2C_1e^{\sigma_{\bbeta}^2C_2^2\alpha_n^2}(1-\Phi(\sigma_{\bbeta}C_2\alpha_n))\\
\leq\frac{2C_1}{C_2\sigma_{\bbeta}\alpha_n}.
\end{align*}
And similarly we can prove $-(\cL(\bbeta)-\cL_n(\bbeta))\leq\frac{2C_1}{C_2\sigma_{\bbeta}\alpha_n}$. So we finally obtain
\[
\sup_{\bbeta:\bbeta_1=1}|\cL_n(\bbeta)-\cL(\bbeta)|\leq\frac{2C_1}{C_2\alpha_n}\sup_{\bbeta:\bbeta_1=1}\frac{1}{\sqrt{2\bbeta^\T\bSigma\bbeta}}.
\]
This ends the proof.
\end{proof}

\subsection{Proof of Lemma \ref{lem:app-error}}
\begin{proof}
We begin by introducing some additional notation. For any function $f:\R^d\to\R$, define $\inf f:=\inf_{\bbeta:\bbeta_1=1} f(\bbeta)$. Given $r>0$ and $\bbeta^*\in\R^d$, we denote $\inf_r f:=\inf_{\bbeta:\bbeta_1=1,\|\bbeta-\bbeta^*\|_2\leq r}f(\bbeta)$. And similarly we define the corresponding versions for ``$\sup f$" and ``$\sup_r f$".
With 
\[
\bbeta_{r,\alpha_n}^*=\argmin_{\bbeta_1=1,\norm{\bbeta-\bbeta^*}_2\leq r} \cL_n(\bbeta)~~~{\rm and} ~~~\bbeta^*=\argmin_{\bbeta_1=1} \cL(\bbeta), 
\]
we immediately have
\begin{align*}
\cL(\bbeta_{r,\alpha_n}^*)-\cL(\bbeta^*)=|-\inf\cL+\cL(\bbeta_{r,\alpha_n}^*)|=|-\inf\cL+\inf_r\cL_n-\inf_r\cL_n+\cL(\bbeta_{r,\alpha_n}^*)|.
\end{align*}
By the triangular inequality, $\cL(\bbeta_{r,\alpha_n}^*)-\cL(\bbeta^*)\leq|-\inf\cL+\inf_r\cL_n|+|\cL(\bbeta_{r,\alpha_n}^*)-\inf_r\cL_n|$. Since $\bbeta_{r,\alpha_n}^*$ is the minimizer of $\cL_n(\bbeta)$ under the restrictions $\bbeta_1=1$ and $\norm{\bbeta-\bbeta^*}_2\leq r$, we further have
\begin{align*}
\cL(\bbeta_{r,\alpha_n}^*)-\cL(\bbeta^*)&\leq|-\inf_r\cL_n+\inf\cL|+|\cL(\bbeta_{r,\alpha_n}^*)-\cL_n(\bbeta_{r,\alpha_n}^*)|\\
&\leq |-\inf_r\cL_n+\inf\cL|+\sup|\cL-\cL_n|.
\end{align*}
Notice that 
\begin{align*}
-\inf_r\cL_n+\inf\cL=\sup_r(-\cL_n)-\sup(-\cL)=\sup_r(-\cL_n+\cL-\cL)-\sup(-\cL)\\
\leq\sup_r(|\cL_n-\cL|-\cL)-\sup(-\cL)\leq\sup_r|\cL_n-\cL|+\sup_r(-\cL)-\sup(-\cL)\leq\sup|\cL_n-\cL|,
\end{align*}
and similarly, $\inf_r\cL_n-\inf\cL\leq\sup|\cL_n-\cL|$. They lead to $|-\inf_r\cL_n+\inf\cL|\leq\sup|\cL_n-\cL|$. Hence we have $\cL(\bbeta_{r,\alpha_n}^*)-\cL(\bbeta^*)\leq 2\sup|\cL-\cL_n|$.

Now using Assumptions {\bf (A1)} and {\bf (A2)}, it follows from Lemma \ref{lem:approx} that
\begin{align}\label{lem_app_error_temp1}
\cL(\bbeta_{r,\alpha_n}^*)-\cL(\bbeta^*)\leq 2\sup_{\bbeta:\bbeta_1=1}|\cL_n(\bbeta)-\cL(\bbeta)|\leq\frac{4C_1}{C_2\alpha_n}\sup_{\bbeta:\bbeta_1=1}\frac{1}{\sqrt{2\bbeta^\T\bSigma\bbeta}}.
\end{align}
Given Assumptions {\bf (A1)} and {\bf (A3)}, Proposition \ref{prop:convexity} guarantees that for any given positive constants $\gamma:=\{\gamma_1,\gamma_2\}$ with $\gamma_2/\gamma_1-1$ arbitrarily close to 0, such that for some small enough $r(\gamma)>0$ only depending on $\gamma$, as long as  $ \|\bbeta-\bbeta^*\|_2\leq r(\gamma)$, we have
\begin{align}\label{lem_app_error_temp2}
\cL(\bbeta)-\cL(\bbeta^*)\asymp\norm{\bbeta-\bbeta^*}_2^2.
\end{align}
Hence when $ \|\bbeta_{r(\gamma),\alpha_n}^*-\bbeta^*\|_2\leq r(\gamma)$ holds, combining (\ref{lem_app_error_temp1}) and (\ref{lem_app_error_temp2}) implies
\[
\|\bbeta_{r(\gamma),\alpha_n}^*-\bbeta^*\|_2^2\lesssim \alpha_n^{-1}\cdot\sup_{\bbeta:\bbeta_1=1}\frac{1}{\sqrt{2\bbeta^\T\bSigma\bbeta}}.
\]
This completes the proof.
\end{proof}

\subsection{Proof of Lemma \ref{lem:approx2}}
\begin{proof}
Without loss of generality, assume $M=1$. First of all, let us recall 
\begin{align*}
\mathcal{L}_n(\bbeta)-\mathcal{L}(\bbeta)=&\E\Big\{\big[p_1(\bX_i,\bX_{i'};\bbeta^*)-p_0(\bX_i,\bX_{i'};\bbeta^*)\big]F_{ii'}(\alpha_nZ_{ii'}(\bbeta))\mathds{1}(Z_{ii'}(\bbeta)<0)\\
&+\big[p_0(\bX_i,\bX_{i'};\bbeta^*)-p_1(\bX_i,\bX_{i'};\bbeta^*)\big]\big[1-F_{ii'}(\alpha_nZ_{ii'}(\bbeta))\big]\mathds{1}(Z_{ii'}(\bbeta)>0)\Big\}.
\end{align*}
Under the monotonic transformation model \eqref{eq:model1}, we have further derivation,
\begin{align*}
\mathcal{L}_n(\bbeta)-\mathcal{L}(\bbeta)
&=\E\Big\{\big[2F_\epsilon(Z_{ii'}(\bbeta^*))-1\big]F_{ii'}(\alpha_nZ_{ii'}(\bbeta))\mathds{1}(Z_{ii'}(\bbeta)<0)\Big\}\\
&~~~+\E\Big\{\big[1-2F_\epsilon(Z_{ii'}(\bbeta^*))\big]\big[1-F_{ii'}(\alpha_nZ_{ii'}(\bbeta))\big]\mathds{1}(Z_{ii'}(\bbeta)>0)\Big\}.
\end{align*}
Due to the fact $F_{\epsilon}(-u)=1-F_{\epsilon}(u)$ and the property $F_{ii'}(-u)=1-F_{ii'}(u)$, it follows
\begin{align}\label{lem_approx2_temp1}
\begin{split}
\mathcal{L}_n(\bbeta)-\mathcal{L}(\bbeta)
&=\E\Big\{\big[2F_\epsilon(Z_{ii'}(\bbeta^*))-1\big]F_{ii'}(\alpha_nZ_{ii'}(\bbeta))\mathds{1}(Z_{ii'}(\bbeta)<0)\Big\}\\
&~~~+\E\Big\{\big[2F_\epsilon(-Z_{ii'}(\bbeta^*))-1\big]\big[F_{ii'}(-\alpha_nZ_{ii'}(\bbeta))\big]\mathds{1}(-Z_{ii'}(\bbeta)<0)\Big\}.
\end{split}
\end{align}
Under Assumption {\bf (A1)}, it yields immediately
\begin{align*}
\begin{pmatrix}
Z_{ii'}(\bbeta^*)\\
Z_{ii'}(\bbeta)
\end{pmatrix}\sim-\begin{pmatrix}
Z_{ii'}(\bbeta^*)\\
Z_{ii'}(\bbeta)
\end{pmatrix}\sim\text{N}_2\left(\mathbf{0},\begin{pmatrix}
2\bbeta^{*\intercal}\bSigma\bbeta^* & 2\bbeta^{*\intercal}\bSigma\bbeta\\
2\bbeta^{\intercal}\bSigma\bbeta^* & 2\bbeta^{\intercal}\bSigma\bbeta
\end{pmatrix}\right),
\end{align*}
which straightforwardly implies 
\begin{align*}
\E\Big\{\big[2F_\epsilon(-Z_{ii'}(\bbeta^*))-1\big]\big[F_{ii'}(-\alpha_nZ_{ii'}(\bbeta))\big]\mathds{1}(-Z_{ii'}(\bbeta)<0)\Big\}\\
=\E\Big\{\big[2F_\epsilon(Z_{ii'}(\bbeta^*))-1\big]\big[F_{ii'}(\alpha_nZ_{ii'}(\bbeta))\big]\mathds{1}(Z_{ii'}(\bbeta)<0)\Big\}.
\end{align*}
Hence (\ref{lem_approx2_temp1}) can be further simplified as
\begin{align}\label{lem_approx2_temp2}
\mathcal{L}_n(\bbeta)-\mathcal{L}(\bbeta)=2\E\Big\{\big[2F_\epsilon(Z_{ii'}(\bbeta^*))-1\big]F_{ii'}(\alpha_nZ_{ii'}(\bbeta))\mathds{1}(Z_{ii'}(\bbeta)<0)\Big\}.
\end{align}

(1) If we have $\bbeta^{*\intercal}\bSigma\bbeta=0$, that is $Z_{ii'}(\bbeta^*)$ is independent with $Z_{ii'}(\bbeta)$, we then have
\begin{align*}
\mathcal{L}_n(\bbeta)-\mathcal{L}(\bbeta)=2\E\Big\{\big[2F_\epsilon(Z_{ii'}(\bbeta^*))-1\big]\Big\}\E\Big\{F_{ii'}(\alpha_nZ_{ii'}(\bbeta))\mathds{1}(Z_{ii'}(\bbeta)<0)\Big\}.
\end{align*}
Since $Z_{ii'}(\bbeta^*)$ has the same distribution as $-Z_{ii'}(\bbeta^*)$, with 
\[
2F_\epsilon(Z_{ii'}(\bbeta^*))-1=1-2F_\epsilon(-Z_{ii'}(\bbeta^*)), 
\]
it follows 
$$\E\Big\{\big[2F_\epsilon(Z_{ii'}(\bbeta^*))-1\big]\Big\}=\E\Big\{\big[2F_\epsilon(-Z_{ii'}(\bbeta^*))-1\big]\Big\}=-\E\Big\{\big[2F_\epsilon(Z_{ii'}(\bbeta^*))-1\big]\Big\},$$
which implies $\E\Big\{\big[2F_\epsilon(Z_{ii'}(\bbeta^*))-1\big]\Big\}=0$.
Hence we obtain $\mathcal{L}_n(\bbeta)-\mathcal{L}(\bbeta)=0$.

(2) With $\bSigma=\mathbf{I}$, we have 
\begin{align*}
\begin{pmatrix}
Z_{ii'}(\bbeta^*)\\
Z_{ii'}(\bbeta)
\end{pmatrix}\sim\text{N}_2\left(\mathbf{0},2\begin{pmatrix}
1 & \rho\\
\rho & 1
\end{pmatrix}\right), \text{with }  \rho=\bbeta^{*\intercal}\bbeta.
\end{align*}
Using \eqref{lem_approx2_temp2}, with the above joint normal distribution, it follows 
\begin{align*}
W(\rho):=\mathcal{L}_n(\bbeta)-\mathcal{L}(\bbeta)
&=2\int_{-\infty}^{\infty}\int_{-\infty}^0\big[2F_\epsilon(u)-1\big]F_{ii'}(\alpha_nv)\frac{1}{4\pi\sqrt{1-\rho^2}}e^{-\frac{u^2+v^2-2\rho uv}{4(1-\rho^2)}}dv du.
\end{align*}
After further simplification, we deduce
$$W(\rho)=2\int_{0}^{\infty}\int_0^{\infty}\big[2F_\epsilon(u)-1\big]\big[1-F_{ii'}(\alpha_nv)\big]\frac{1}{4\pi\sqrt{1-\rho^2}}\Big\{e^{-\frac{u^2+v^2+2\rho uv}{4(1-\rho^2)}}-e^{-\frac{u^2+v^2-2\rho uv}{4(1-\rho^2)}}\Big\}dv du.$$
%
Now suppose that the noise term $\epsilon$ follows a normal distribution with standard deviation $\sigma$. It follows that we have $F_\epsilon(u)=\Phi(u/(\sqrt{2}\sigma))$, which renders
$$W(\rho)=2\int_{0}^{\infty}\int_0^{\infty}\big[2\Phi(\frac{u}{\sqrt{2}\sigma})-1\big]\big[1-F_{ii'}(\alpha_nv)\big]\frac{1}{4\pi\sqrt{1-\rho^2}}\Big\{e^{-\frac{u^2+v^2+2\rho uv}{4(1-\rho^2)}}-e^{-\frac{u^2+v^2-2\rho uv}{4(1-\rho^2)}}\Big\}dv du.$$
For sigmoid, Gaussian CDF, and double exponential CDF approximations, we conclude that the corresponding $W(\rho)$ is an increasing function of $|\rho|$ by employing numeric integrations.

(3) With $F_\epsilon(u)=\Phi(u/(\sqrt{2}\sigma))$ and $F_{ii'}(\alpha_nv)=1/(1+e^{-\alpha_nv})$, we have
\begin{align*}
&\mathcal{L}_n(\bbeta^*)-\mathcal{L}(\bbeta^*)=2\E\Big\{\big[2F_\epsilon(Z_{ii'}(\bbeta^*))-1\big]F_{ii'}(\alpha_nZ_{ii'}(\bbeta^*))\mathds{1}(Z_{ii'}(\bbeta^*)<0)\Big\}\\
=&2\int_{-\infty}^{0}\big[2F_\epsilon(u)-1\big]F_{ii'}(\alpha_nu)\frac{1}{2\sqrt{\pi}}e^{-\frac{u^2}{4}}du
=2\int_{0}^{\infty}\big[1-2\Phi(\frac{u}{\sqrt{2}\sigma})\big]\frac{1}{1+e^{\alpha_nu}}\frac{1}{2\sqrt{\pi}}e^{-\frac{u^2}{4}}du.
\end{align*}
Based on this, numerical integrations show $\mathcal{L}_n(\bbeta^*)-\mathcal{L}(\bbeta^*)\asymp\alpha_n^{-2}$. And together with the result in Item (2), we have for any $\bbeta\in\S^{d-1}$, $|\cL_n(\bbeta)-\cL(\bbeta)| \lesssim \alpha_n^{-2}$. This completes the proof.
\end{proof}

\subsection{Proof of Proposition \ref{prop:gaussian_cauchy}}
\begin{proof}
We aim to validate the following results for the i.i.d. noise terms $\{\epsilon_i, i=1,2,\cdots, n\}$ with Gaussian or Cauchy distribution, i.e.,
\[
\int_{-\infty}^{0}f_{\epsilon}(x)\exp(-x^2/(2b_n^2))dx= Cb_n(1+o(1))~~ \rm{as}~~b_n\rightarrow 0,
\]
where $f_{\epsilon}(\cdot)$ represents the PDF of $\epsilon_2-\epsilon_1$.

For the noise term $\epsilon_i\sim\text{N}(\mu,\sigma^2)$, we have $f_{\epsilon}(x)=e^{-x^2/(4\sigma^2)}/(2\sqrt{\pi}\sigma)$. With bounded $\sigma^2$, it yields
\begin{align*}
\int_{-\infty}^{0}f_{\epsilon}(x)e^{-\frac{x^2}{2b_n^2}}dx
=\frac{1}{\sqrt{2}\sigma\sqrt{\frac{1}{2\sigma^2}+\frac{1}{b_n^2}}}\Phi(0)=\frac{b_n}{\sqrt{b_n^2+2\sigma^2}}= \frac{b_n(1+o(1))}{\sqrt{2\sigma^2}},
\end{align*}
as $b_n\to 0$, where $\Phi(\cdot)$ is the CDF of standard normal distribution. For the noise term $\epsilon_i\sim\text{Cauchy}(\mu,\gamma)$, it is easy to see $f_{\epsilon}(x)=1/\{2\pi\gamma(1+(x/(2\gamma))^2)\}$. By simple calculation, we deduce
\begin{align}\label{prop_gaussian_cauchy_temp1}
\int_{-\infty}^{0}f_{\epsilon}(x)e^{-\frac{x^2}{2b_n^2}}dx&=\frac{1}{\pi}\int_{-\infty}^{0}\frac{1}{(1+x^2)} e^{-\frac{x^2}{2(b_n/(2\gamma))^2}}dx\notag\\
&=\frac{1}{\pi}\frac{\pi}{2}e^{\frac{1}{2(b_n/(2\gamma))^2}}[1-\Phi(\sqrt{2}/\sqrt{2(b_n/(2\gamma))^2})]\notag\\
&=\frac{1}{2}e^{2\gamma^2/b_n^2}\left[1-\Phi(2\gamma/b_n)\right].
\end{align}
With $\phi(\cdot)$ the PDF of standard normal distribution, using the fact 
\[
\phi(x)/(x+1/x)<1-\Phi(x)<\phi(x)/x ~{\rm for~any}~ x>0, 
\]
we have
\begin{align}\label{prop_gaussian_cauchy_temp2}
\frac{1}{\sqrt{2\pi}}e^{-2\gamma^2/b_n^2}/[2\gamma/b_n+b_n/(2\gamma)]<1-\Phi(2\gamma/b_n)<\frac{1}{\sqrt{2\pi}}e^{-2\gamma^2/b_n^2}/[2\gamma/b_n].
\end{align}
Combining (\ref{prop_gaussian_cauchy_temp1}) and (\ref{prop_gaussian_cauchy_temp2}) leads to
$$\frac{\gamma b_n}{\sqrt{2\pi}(4\gamma^2+b_n^2)}<\int_{-\infty}^{0}f_{\epsilon}(x)e^{-\frac{x^2}{2b_n^2}}dx<\frac{b_n}{4\sqrt{2\pi}\gamma}.$$
Having bounded $\gamma$, we immediately have $\int_{-\infty}^{0}f_{\epsilon}(x)e^{-\frac{x^2}{2b_n^2}}dx= b_n(1+o(1))/(4\sqrt{2\pi}\gamma)$ as $b_n\to 0$. Hence we complete the proof of the proposition for both Gaussian and Cauchy distributed noises.
\end{proof}

\subsection{Proof of Lemma \ref{lem:convexity}}
\begin{proof}
Recall $\cL(\bbeta)=-\E\big(S_{ii'}\mathds{1}(Z_{ii'}(\bbeta)>0)+(1-S_{ii'})\mathds{1}(Z_{ii'}(\bbeta)<0)\big)$. Under the monotonic transformation model (\ref{eq:model1}), i.e.,\ $Y=G(\bX^\T\bbeta^*+\epsilon)$, we have 
\[
p_0(\bX_i,\bX_{i'};\bbeta^*)=\P(Y_i<Y_{i'}|\bX_i,\bX_{i'})=\P\big\{G(\bX_{i}^\T\bbeta^*+\epsilon_{i})<G(\bX_{i'}^\T\bbeta^*+\epsilon_{i'})\big|\bX_{i},\bX_{i'}\big\}, 
\]
and 
\[
p_1(\bX_i,\bX_{i'};\bbeta^*)=\P(Y_i>Y_{i'}|\bX_i,\bX_{i'})=\P\big\{G(\bX_{i}^\T\bbeta^*+\epsilon_{i})>G(\bX_{i'}^\T\bbeta^*+\epsilon_{i'})\big|\bX_{i},\bX_{i'}\big\}. 
\]
By the monotonicity of $G(\cdot)$, we can write
\begin{align}
p_0\{Z_{ii'}(\bbeta^*)\}&:=p_0(\bX_i,\bX_{i'};\bbeta^*)=\P\big\{Z_{ii'}(\bbeta^*)<\epsilon_{i'}-\epsilon_{i}\big|\bX_{i},\bX_{i'}\big\},\label{lem_convexity_tempn1}\\
p_1\{Z_{ii'}(\bbeta^*)\}&:=p_1(\bX_i,\bX_{i'};\bbeta^*)=\P\big\{Z_{ii'}(\bbeta^*)>\epsilon_{i'}-\epsilon_{i}\big|\bX_{i},\bX_{i'}\big\}.\label{lem_convexity_tempn2}
\end{align}
Notice in $\cL(\bbeta)$ we have both of the two terms $Z_{ii'}(\bbeta^*)$ and $Z_{ii'}(\bbeta)$ involved. Using Assumption {\bf (A1)},  
we immediately have the following joint distribution for $(Z_{ii'}(\bbeta^*),Z_{ii'}(\bbeta))^\T$:
\begin{align}\label{lem_convexity_tempn3}
\begin{pmatrix}
Z_{ii'}(\bbeta^*)\\
Z_{ii'}(\bbeta)
\end{pmatrix}\sim\text{N}_2\left(\mathbf{0},\begin{pmatrix}
2\bbeta^{*\T}\bSigma\bbeta^* & 2\bbeta^{*\T}\bSigma\bbeta\\
2\bbeta^{\T}\bSigma\bbeta^* & 2\bbeta^{\T}\bSigma\bbeta
\end{pmatrix}\right),
\end{align}
which is independent of the indices $i$ and $i'$. For convenience, let's define
\begin{align}\label{def:rho}
\rho:=\frac{\bbeta^{\T}\bSigma\bbeta^*}{\sqrt{\bbeta^\T\bSigma\bbeta}\sqrt{\bbeta^{*\T}\bSigma\bbeta^*}}.
\end{align}

After taking conditional expectations with respect to the response given the covariates, we have
\begin{align*}
\cL(\bbeta)-\cL(\bbeta^*)
&=\E\left\{p_1\{Z_{ii'}(\bbeta^*)\}\big[\mathds{1}\{Z_{ii'}(\bbeta^*)>0\}-\mathds{1}\{Z_{ii'}(\bbeta)>0\}\big]\right\}\\
&~~~+\E\left\{p_0\{Z_{ii'}(\bbeta^*)\}\big[\mathds{1}\{Z_{ii'}(\bbeta^*)<0\}-\mathds{1}\{Z_{ii'}(\bbeta)<0\}\big]\right\}.
\end{align*}
In order to further simplify the above expression, we define the following regions
\begin{align*}
\mathcal{A}_1=\left\{Z_{ii'}(\bbeta^*)>0; Z_{ii'}(\bbeta)>0\right\};\ & \mathcal{A}_2=\left\{Z_{ii'}(\bbeta^*)>0; Z_{ii'}(\bbeta)<0\right\};\\
\mathcal{A}_3=\left\{Z_{ii'}(\bbeta^*)<0; Z_{ii'}(\bbeta)>0\right\};\ & \mathcal{A}_4=\left\{Z_{ii'}(\bbeta^*)<0; Z_{ii'}(\bbeta)<0\right\}.
\end{align*}
Simple calculation then leads to
\begin{align*}
&\cL(\bbeta)-\cL(\bbeta^*)\\
=&\E\left\{\big[p_1\{Z_{ii'}(\bbeta^*)\}-p_0\{Z_{ii'}(\bbeta^*)\}\big]\mathds{1}(\mathcal{A}_2)+\big[p_0\{Z_{ii'}(\bbeta^*)\}-p_1\{Z_{ii'}(\bbeta^*)\}\big]\mathds{1}(\mathcal{A}_3)\right\}\\
=&\E\left\{\big[p_1\{Z_{ii'}(\bbeta^*)\}-p_0\{Z_{ii'}(\bbeta^*)\}\big]\mathds{1}(\mathcal{A}_2)\right\}+\E\left\{\big[p_0\{Z_{ii'}(\bbeta^*)\}-p_1\{Z_{ii'}(\bbeta^*)\}\big]\mathds{1}(\mathcal{A}_3)\right\}.
\end{align*}
With the monotonic transformation model (\ref{eq:model1}), using monotonicity and the assumption of i.i.d. noise terms $\{\epsilon_i,i=1,2,\cdots,n\}$ assumed in Assumption {\bf (A1)}, we have
\begin{align}\label{expectation:positive}
\big[p_1\{Z_{ii'}(\bbeta^*)\}-p_0\{Z_{ii'}(\bbeta^*)\}\big]\mathds{1}(\mathcal{A}_2)\geq  0 \notag\\
~~\text{and}~~
\big[p_0\{Z_{ii'}(\bbeta^*)\}-p_1\{Z_{ii'}(\bbeta^*)\}\big]\mathds{1}(\mathcal{A}_3)\geq 0.
\end{align}
According to the simplified notation in (\ref{lem_convexity_tempn1}) and (\ref{lem_convexity_tempn2}) under Model (\ref{eq:model1}), with the monotonicity of $G$, it follows
\begin{align*}
p_1\{Z_{ii'}(\bbeta^*)\}-p_0\{Z_{ii'}(\bbeta^*)\}
&=\P\big\{Z_{ii'}(\bbeta^*)>\epsilon_{i'}-\epsilon_{i}\big|\bX_{i},\bX_{i'}\big\}-\P\big\{Z_{ii'}(\bbeta^*)<\epsilon_{i'}-\epsilon_{i}\big|\bX_{i},\bX_{i'}\big\}\\
&=1-2\P\big\{Z_{ii'}(\bbeta^*)<\epsilon_{i'}-\epsilon_{i}\big|\bX_{i},\bX_{i'}\big\}=1-2F_{\epsilon}\big\{-Z_{ii'}(\bbeta^*)\big\},
\end{align*}
where $F_{\epsilon}$ is the CDF for $\epsilon_i-\epsilon_{i'}$ with $i\neq i'$. Recall for any $p>0$ and any random variable $U\geq 0$, 
\[
\E U^p=\int_0^{\infty} pu^{p-1}\P(U>u)du. 
\]
With  \eqref{expectation:positive}, applying the above formula to the random variable 
\[
U=\big[p_1\{Z_{ii'}(\bbeta^*)\}-p_0\{Z_{ii'}(\bbeta^*)\}\big]\mathds{1}(\mathcal{A}_2)
\]
gives us that
\begin{align}\label{lem_convexity_temp1}
&\E\left\{\big[p_1\{Z_{ii'}(\bbeta^*)\}-p_0\{Z_{ii'}(\bbeta^*)\}\big]\mathds{1}(\mathcal{A}_2)\right\}=\int_0^{\infty}\P\left\{1-2F_{\epsilon}\big\{-Z_{ii'}(\bbeta^*)\big\}>a, \mathcal{A}_2\right\}da\notag\\
=&\int_0^{\infty}\P\left\{-Z_{ii'}(\bbeta^*)<F^{-1}_{\epsilon}\big(\frac{1-a}{2}\big), -Z_{ii'}(\bbeta)>0\right\}da.
\end{align}
By simple calculation, with $\sigma_1^2:=\sigma_{\bbeta^*}^2=2\bbeta^{*\T}\bSigma\bbeta^*$ and $\sigma_2^2:=\sigma_{\bbeta}^2=2\bbeta^{\T}\bSigma\bbeta$, according to the joint distribution specified in (\ref{lem_convexity_tempn3}), we have
\begin{align}\label{lem_convexity_temp2}
H(x;\rho)&:=\P\left\{-Z_{ii'}(\bbeta^*)<x, -Z_{ii'}(\bbeta)>0\right\} \notag\\
&=\int_{-\infty}^{x}\int_{-\infty}^{0}\frac{1}{2\pi\sigma_1\sigma_2\sqrt{1-\rho^2}}\exp\left\{-\frac{1}{2(1-\rho^2)}\left[\frac{u^2}{\sigma_1^2}+\frac{v^2}{\sigma_2^2}+2\rho\frac{uv}{\sigma_1\sigma_2}\right]\right\}du dv \notag\\
&=\frac{1}{2\pi\sigma_1\sqrt{1-\rho^2}}\int_{-\infty}^{x}e^{-\frac{u^2}{2\sigma_1^2}}\left[\int_{-\infty}^{0}\exp\left\{-\frac{(v+\rho u/\sigma_1)^2}{2(1-\rho^2)}\right\}dv\right] du \notag\\
&=\int_{-\infty}^{x}\frac{1}{\sqrt{2\pi}\sigma_1}e^{-\frac{u^2}{2\sigma_1^2}}\Phi(\frac{\rho u}{\sigma_1\sqrt{1-\rho^2}})du,
\end{align}
where $\Phi(\cdot)$ is the CDF of the standard normal distribution.

Combining (\ref{lem_convexity_temp1}) and (\ref{lem_convexity_temp2}), we deduce
\begin{align}
K(\rho):=\E\left\{\big[p_1\{Z_{ii'}(\bbeta^*)\}-p_0\{Z_{ii'}(\bbeta^*)\}\big]\mathds{1}(\mathcal{A}_2)\right\}&=\int_0^{\infty}H\left\{F^{-1}_{\epsilon}\big(\frac{1-a}{2}\big);\rho\right\}da\notag\\
&=2\int_{-\infty}^{F^{-1}_{\epsilon}(1/2)}H(x;\rho)f_{\epsilon}(x)dx.
\end{align}
Via exchange of taking derivative and integral, we have the first derivative of the above function $K(\rho)$ is of the form
\begin{align*}
K'(\rho)=2\int_{-\infty}^{F^{-1}_{\epsilon}(1/2)}f_{\epsilon}(x)\int_{-\infty}^{x}\frac{1}{\sqrt{2\pi}\sigma_1}e^{-\frac{u^2}{2\sigma_1^2}}\phi(\frac{\rho u}{\sigma_1\sqrt{1-\rho^2}})\frac{u}{\sigma_1(1-\rho^2)^{3/2}}du dx,
\end{align*}
where $\phi(\cdot)$ represents the PDF for the standard normal distribution. With simple calculation, we can simplify $K'(\rho)$ further as follows:
\begin{align}\label{lem_convexity_tempn4}
K'(\rho)&=\frac{1}{\pi\sigma_1^2(1-\rho^2)^{3/2}}\int_{-\infty}^{F^{-1}_{\epsilon}(1/2)}f_{\epsilon}(x)\int_{-\infty}^{x}ue^{-\frac{u^2}{2\sigma_1^2(1-\rho^2)}}du dx\notag\\
&=\frac{1}{\pi\sigma_1^2(1-\rho^2)^{3/2}}\int_{-\infty}^{F^{-1}_{\epsilon}(1/2)}f_{\epsilon}(x)\left\{-\sigma_1^2(1-\rho^2)e^{-\frac{x^2}{2\sigma_1^2(1-\rho^2)}}\right\} dx\notag\\
&=\frac{-1}{\pi\sqrt{1-\rho^2}}\int_{-\infty}^{F^{-1}_{\epsilon}(1/2)}f_{\epsilon}(x)e^{-\frac{x^2}{2\sigma_1^2(1-\rho^2)}}dx.
\end{align}
Due to $F^{-1}_{\epsilon}(1/2)=0$ since $F_{\epsilon}$ is the CDF for $\epsilon_2-\epsilon_1$, it yields immediately from (\ref{lem_convexity_tempn4}) that
\begin{align*}
K'(\rho)=\frac{-1}{\pi\sqrt{1-\rho^2}}\int_{-\infty}^{0}f_{\epsilon}(x)e^{-\frac{x^2}{2\sigma_1^2(1-\rho^2)}}dx.
\end{align*}
By Assumption {\bf (A3')}, it follows
\[
\int_{-\infty}^{0}f_{\epsilon}(x)e^{-\frac{x^2}{2\sigma_1^2(1-\rho^2)}}dx=C\sigma_1\sqrt{1-\rho^2}(1+o(1))~~ {\rm as} ~~\rho\to 1.
\]
It yields
\[
K'(\rho)= -C(\sigma_1/\pi)(1+o(1)) ~~\text{as} ~~\rho\to 1.
\]
When $\rho=1$, we have $\bbeta^*=\bbeta$, which leads to $K(1)=0$. Applying Taylor expansion at $\rho=1$ gives us
\begin{align*}
K(\rho)=K(1)+K'(\rho)(\rho-1)+o(\rho-1)=K'(\rho)(\rho-1)+o(\rho-1), \text{as } {\rho}\to 1.
\end{align*}
Hence by the definition of $K(\rho)$, we have
\[
\E\left\{\big[p_1\{Z_{ii'}(\bbeta^*)\}-p_0\{Z_{ii'}(\bbeta^*)\}\big]\mathds{1}(\mathcal{A}_2)\right\}=K(\rho)=C(\sigma_1/\pi)(1-\rho)(1+o(1)), \text{as } \rho\to 1.
\]
Similarly, by symmetry, we conclude
\[
\E\left\{\big[p_0\{Z_{ii'}(\bbeta^*)\}-p_1\{Z_{ii'}(\bbeta^*)\}\big]\mathds{1}(\mathcal{A}_3)\right\}= C(\sigma_1/\pi)(1-\rho)(1+o(1)), \text{as } \rho\to 1.
\]
In summary, we have
\begin{align*}
&\cL(\bbeta)-\cL(\bbeta^*)\\
=&\E\left\{\big[p_1\{Z_{ii'}(\bbeta^*)\}-p_0\{Z_{ii'}(\bbeta^*)\}\big]\mathds{1}(\mathcal{A}_2)\right\}+\E\left\{\big[p_0\{Z_{ii'}(\bbeta^*)\}-p_1\{Z_{ii'}(\bbeta^*)\}\big]\mathds{1}(\mathcal{A}_3)\right\}\\
=& c(1-\rho)(1+o(1)), \text{as } \|\bbeta-\bbeta^*\|_2\to 0,
\end{align*}
for some absolute constant $c>0$. We can then pick positive constant set $\gamma:=\{\gamma_1,\gamma_2\}$ with $\gamma_2/\gamma_1-1$ arbitrarily close to 0, such that for some small enough $r(\gamma)>0$ only depending on $\gamma$, as long as  $ \|\bbeta-\bbeta^*\|_2\leq r(\gamma)$, we have
\[
\gamma_1 \cdot \Big(1-\frac{\bbeta^\T\bSigma\bbeta^*}{\sqrt{\bbeta^\T\bSigma\bbeta}\sqrt{\bbeta^{*\T}\bSigma\bbeta^*}}\Big) \leq \cL(\bbeta)-\cL(\bbeta^*)\leq \gamma_2  \cdot \Big(1-\frac{\bbeta^\T\bSigma\bbeta^*}{\sqrt{\bbeta^\T\bSigma\bbeta}\sqrt{\bbeta^{*\T}\bSigma\bbeta^*}}\Big),
\]
due to the definition of $\rho$ in (\ref{def:rho}).
\end{proof}

\subsection{Proof of Theorem \ref{thm:martin}}

\begin{proof}
The proof is split to three steps.

(1) In the first step, we show 
\begin{align}\label{eq:0}
\hat\bDelta:=\hat\btheta-\btheta^*\in \cC(\cM,\overline\cM^\perp;\btheta^*). 
\end{align}
By Assumption {\bf (B3)} and simple algebra, for any $\bDelta\in\reals^d$, we have 
\begin{align}\label{eq:1}
P(\btheta^*+\bDelta)-P(\btheta^*) \geq P(\bDelta_{\overline\cM^{\perp}})-P(\bDelta_{\overline\cM})-2P(\btheta^*_{\cM^\perp}).
\end{align}
In addition, due to the convex differentiability of $L_n(\cdot)$ (Assumption {\bf (B2)}), we have 
\begin{align*}
L_n(\btheta^*+\hat\bDelta)-L_n(\btheta^*)\geq -|\langle \nabla L_n(\btheta^*),\hat\bDelta\rangle|.
\end{align*}
Holder's inequality then yields 
\[
L_n(\btheta^*+\hat\bDelta)-L_n(\btheta^*) \geq -P^*(\nabla L_n(\btheta^*))P(\hat\bDelta). 
\]
Using Assumption {\bf (B5)}, we further have
\begin{align}\label{eq:2}
L_n(\btheta^*+\hat\bDelta)-L_n(\btheta^*) \geq -\frac{\lambda_n}{2}(P(\hat\bDelta_{\overline\cM})+P(\hat\bDelta_{\overline\cM^\perp})).
\end{align}
Combining \eqref{eq:1} and \eqref{eq:2}, and using the fact $\hat\btheta$ minimizes $L_n(\btheta)+\lambda_nP(\btheta)$, we have
\[
0\geq \frac{\lambda_n}{2}\Big(P(\hat\bDelta_{\overline\cM^\perp})-3P(\hat\bDelta_{\overline\cM})-4P(\btheta^*_{\cM^\perp})\Big).
\]
This then proves \eqref{eq:0}. 

(2). Let's define 
\[
\cF(\bDelta):=L_n(\btheta^*+\bDelta)+\lambda_nP(\btheta^*+\bDelta)-L_n(\btheta^*)-\lambda_nP(\btheta^*).
\]
In the second step, we proceed to prove the following assertion: if for all $\bDelta\in\cC(\cM,\overline\cM^\perp;\btheta^*)\cap\{\norm{\bDelta}_2=\gamma\}$ we have $\cF(\bDelta)>0$, then $\norm{\hat\bDelta}_2\leq \gamma$. To this end, let's assume $\norm{\hat\bDelta}_2>\gamma$. Then because $\cC(\cM,\overline\cM^\perp;\btheta^*)$ is star-shaped (by Assumption {\bf (B1)}), we can always find $t^*\in(0,1)$ such that 
\[
t^*\hat\bDelta\in \cC(\cM,\overline\cM^\perp;\btheta^*)\cap\{\norm{\bDelta}_2=\gamma\}.
\]
However, by Assumption {\bf (B2)} we have
\begin{align*}
\cF(t^*\hat\bDelta)\leq t^*\cF(\hat\bDelta)\leq 0.
\end{align*}
Therefore, we have a contradiction, and accordingly $\norm{\hat\bDelta}_2\leq \gamma$. 

(3). In the end, let's prove the main theorem. For all $\bDelta\in \cC(\cM,\overline\cM^\perp;\btheta^*)\cap\{\norm{\bDelta}_2=\gamma\}$, using Assumption {\bf (B4)}, we have
\begin{align*}
&\cF(\bDelta)=L_n(\btheta^*+\bDelta)-L_n(\btheta^*)+\lambda_n(P(\btheta^*+\bDelta)-P(\btheta^*))\\
\geq& -|\langle \nabla L_n(\btheta^*),\bDelta\rangle| + \kappa_L\norm{\bDelta}_2^2-\delta_L\norm{\bDelta}_2-\tau_L^2(\btheta^*)+\lambda_n(P(\btheta^*+\bDelta)-P(\btheta^*))\\
\geq& -\frac{\lambda_n}{2}P(\bDelta)+\kappa_L\norm{\bDelta}_2^2-\delta_L\norm{\bDelta}_2-\tau_L^2(\btheta^*)+\lambda_n\Big(P(\bDelta_{\overline\cM^\perp})-P(\bDelta_{\overline\cM})-2P(\btheta^*_{\cM^\perp})\Big)\\
\geq& -\frac{3}{2}\lambda_nP(\bDelta_{\overline\cM})-2\lambda_nP(\btheta^*_{\cM^\perp})+\kappa_L\norm{\bDelta}_2^2-\delta_L\norm{\bDelta}_2-\tau_L^2(\btheta^*).
\end{align*}
Finally, using Assumption {\bf (B5)}, we derive
\[
\cF(\bDelta) \geq -(\frac{3}{2}\lambda_n\Psi(\overline\cM)+\delta_L)\norm{\bDelta}_2+\kappa_L\norm{\bDelta}_2^2-\tau_L^2(\btheta^*)-2\lambda_nP(\btheta^*_{\cM^\perp}).
\]
Hence, by picking 
\[
\gamma^2= (2\lambda_n\Psi(\overline\cM)+\delta_L)^2/\kappa_L^2+2(\tau^2_L(\btheta^*)+2\lambda_nP(\btheta^*_{\cM^\perp}))/\kappa_L,
\]
we have, for all $\bDelta\in \cC(\cM,\overline\cM^\perp;\btheta^*)\cap\{\norm{\bDelta}_2=\gamma\}$,
\[
\cF(\bDelta) \geq 0.
\]
This completes the proof.
\end{proof}

\subsection{Proof of Lemma \ref{lemma:estimation_error}}
\begin{proof}
In the following we only consider the constrainted version of $\hat\cL_n$ that takes value infinity outside of a small ball of $\bbeta^*_{r(\gamma),\alpha_n}$. First, with Assumptions {\bf (A1)} and {\bf (A2)}, Lemma \ref{lem:approx} gives us
\begin{align}\label{eq:key1}
\sup_{\bbeta:\bbeta_1=1}|\cL_n(\bbeta)-\cL(\bbeta)|\leq\frac{2C_1}{C_2\alpha_n}\sup_{\bbeta:\bbeta_1=1}\frac{1}{\sqrt{2\bbeta^\T\bSigma\bbeta}}.
\end{align}


Secondly, using Assumptions {\bf (A1)} and {\bf (A3)}, Equation (9) in \cite{sherman1993limiting} and Proposition \ref{prop:convexity} implies that for $\bDelta$ small enough with $\gamma_2/\gamma_1-1$ close to 0,
\begin{align}\label{eq:key2}
&\gamma_1\lambda_{\min}(\bGamma)\norm{\bDelta}_2^2\leq\cL(\bbeta^*+\bDelta)-\cL(\bbeta^*)\leq \gamma_2\lambda_{\max}(\bGamma)\norm{\bDelta}_2^2 \notag\\
&{\rm and}~~~\cL(\bbeta^*+\bDelta)-\cL(\bbeta^*)=\bDelta^\T\bGamma\bDelta(1/4+o(1)).
\end{align}
Thirdly, given Assumptions {\bf (A1)}-{\bf (A3)}, Lemma \ref{lem:app-error} shows, under further Assumption {\bf (A6)},
\begin{align}\label{eq:key3}
\norm{\bbeta^*_{r(\gamma),\alpha_n}-\bbeta^*}_2\lesssim \alpha_n^{-1/2}.
\end{align}
Combining \eqref{eq:key1}, \eqref{eq:key2}, and \eqref{eq:key3} as well as Assumption {\bf (A6)} yields
\[
\cL_n(\bbeta^*+\bDelta)-\cL_n(\bbeta^*)\geq C'_1\norm{\bDelta}_2^2-C'_2/\alpha_n,
\]
and
\[
\cL_n(\bbeta^*_{r(\gamma),\alpha_n}+\bDelta)-\cL_n(\bbeta^*_{r(\gamma),\alpha_n}) \geq C'_3\norm{\bDelta}_2^2-C'_4/\alpha_n-C_5'\alpha_n^{-1/2}\norm{\bDelta}_2.
\]
In fact, the first one is trivial and the second one can be derived in detail as follows. First note
\begin{align*}
&\cL_n(\bbeta^*_{r(\gamma),\alpha_n}+\bDelta)-\cL_n(\bbeta^*_{r(\gamma),\alpha_n})\\
=&\cL_n(\bbeta^*_{r(\gamma),\alpha_n}+\bDelta)-\cL(\bbeta^*_{r(\gamma),\alpha_n}+\bDelta)+\cL(\bbeta^*_{r(\gamma),\alpha_n}+\bDelta)-\cL(\bbeta^*_{r(\gamma),\alpha_n})\\
&+\cL(\bbeta^*_{r(\gamma),\alpha_n})-\cL_n(\bbeta^*_{r(\gamma),\alpha_n}),
\end{align*}
where the first two terms and the last two terms can be lower bounded through (\ref{eq:key1}),
\[
\cL_n(\bbeta^*_{r(\gamma),\alpha_n}+\bDelta)-\cL(\bbeta^*_{r(\gamma),\alpha_n})+\cL(\bbeta^*_{r(\gamma),\alpha_n})-\cL_n(\bbeta^*_{r(\gamma),\alpha_n})\geq -C'_4/\alpha_n.
\]
Hence it immediately follows
\begin{align}\label{lem_estimation_temp0}
\cL_n(\bbeta^*_{r(\gamma),\alpha_n}+\bDelta)-\cL_n(\bbeta^*_{r(\gamma),\alpha_n})\geq -C'_4/\alpha_n+\cL(\bbeta^*_{r(\gamma),\alpha_n}+\bDelta)-\cL(\bbeta^*_{r(\gamma),\alpha_n}).
\end{align}
Notice that by adding and subtracting same terms, we have
\begin{align}\label{lem_estimation_temp1}
\cL(\bbeta^*_{r(\gamma),\alpha_n}+\bDelta)-\cL(\bbeta^*_{r(\gamma),\alpha_n})
=\cL(\bbeta^*_{r(\gamma),\alpha_n}+\bDelta)-\cL(\bbeta^*)-(\cL(\bbeta^*_{r(\gamma),\alpha_n})-\cL(\bbeta^*)).
\end{align}
In \eqref{lem_estimation_temp1}, following from \eqref{eq:key2}, we have
\begin{align}\label{lem_estimation_temp2}
\cL(\bbeta^*_{r(\gamma),\alpha_n}+\bDelta)-\cL(\bbeta^*)&=0.25(\bbeta^*_{r(\gamma),\alpha_n}+\bDelta-\bbeta^*)^\T\bGamma(\bbeta^*_{r(\gamma),\alpha_n}+\bDelta-\bbeta^*)(1+o(1)),\notag\\
\cL(\bbeta^*_{r(\gamma),\alpha_n})-\cL(\bbeta^*)&=0.25(\bbeta^*_{r(\gamma),\alpha_n}-\bbeta^*)^\T\bGamma(\bbeta^*_{r(\gamma),\alpha_n}-\bbeta^*)(1+o(1)).
\end{align}
Combining \eqref{lem_estimation_temp0}, \eqref{lem_estimation_temp1}, and \eqref{lem_estimation_temp2} implies
\begin{align*}
\cL_n(\bbeta^*_{r(\gamma),\alpha_n}+\bDelta)-\cL_n(\bbeta^*_{r(\gamma),\alpha_n})
\geq&-C'_4/\alpha_n+0.25(\bDelta^\T\bGamma\bDelta+2\bDelta^\T\bGamma(\bbeta^*_{r(\gamma),\alpha_n}-\bbeta^*))(1+o(1)).
\end{align*}
Further with Cauchy inequality, (\ref{eq:key3}), and Assumption {\bf(A3)}, it follows
\begin{align*}
\cL_n(\bbeta^*_{r(\gamma),\alpha_n}+\bDelta)-\cL_n(\bbeta^*_{r(\gamma),\alpha_n})\geq& -C'_4/\alpha_n-C_5'\alpha_n^{-1/2}\norm{\bDelta}_2+C_3'\norm{\bDelta}_2^2.
\end{align*}
By the definition of $\kappa_n$, it then follows
\[
\hat \cL_n(\bbeta^*_{r(\gamma),\alpha_n}+\bDelta)-\hat \cL_n(\bbeta^*_{r(\gamma),\alpha_n})\geq C'_3\norm{\bDelta}_2^2-C'_4/\alpha_n-C_5'\alpha_n^{-1/2}\norm{\bDelta}_2-O_P(\kappa_n),
\]
and accordingly
\begin{align*}
\delta \hat \cL_n(\bDelta,\bbeta^*_{r(\gamma),\alpha_n})&:= \hat \cL_n(\bbeta^*_{r(\gamma),\alpha_n}+\bDelta)-\hat \cL_n(\bbeta^*_{r(\gamma),\alpha_n})-\langle\nabla \hat \cL_n(\bbeta^*_{r(\gamma),\alpha_n}), \bDelta\rangle\\
&\geq C'_3\norm{\bDelta}_2^2-C'_4/\alpha_n-O_P(\kappa_n)-C_5'\alpha_n^{-1/2}\norm{\bDelta}_2-\langle\nabla \hat \cL_n(\bbeta^*_{r(\gamma),\alpha_n}), \bDelta\rangle.
\end{align*}
We then determine the value of $\langle\nabla \hat \cL_n(\bbeta^*_{r(\gamma),\alpha_n}), \bDelta\rangle$. By simple algebra, we have
\[
\nabla\hat{\mathcal{\cL}}_n(\bbeta)=-\frac{2}{n(n-1)}\sum_{i<i'}\tilde{S}_{ii'}\alpha_n\tilde{\bX}_{ii'}F'_{ii'}\{\tilde{S}_{ii'}\alpha_nZ_{ii'}(\bbeta)\}.
\]
Now note $\nabla\hat\cL_n(\bbeta)$ is a U-statistic of order two, written as
\[
\nabla\hat\cL_n(\bbeta):=-\frac{2}{n(n-1)}\sum_{i<i'}h(\{\bX_i,\epsilon_i\},\{\bX_{i'},\epsilon_{i'}\}),
\]
with $h(\{\bX_i,\epsilon_i\},\{\bX_{i'},\epsilon_{i'}\})=\tilde{S}_{ii'}\alpha_n\tilde{\bX}_{ii'}F'_{ii'}\{\tilde{S}_{ii'}\alpha_nZ_{ii'}(\bbeta)\}$.
In addition, let $\norm{\cdot}_{\psi_2}$ be the subgaussian norm defined in \cite{vershynin2010introduction}:
\[
\norm{X}_{\psi_2}:=\sup_{p\geq 1}\frac{1}{\sqrt{p}}\Big(\E|X|^p\Big)^{1/p}.
\]
Combined with Assumption {\bf (A4)} that $\sup_{u\in\reals}|F_{ii'}'(u)|\leq C_3$, we have, for arbitrary $j\in\{1,\ldots,d\}$,
\[
\Big\|\tilde{S}_{ii'}\alpha_n[\tilde{\bX}_{ii'}]_{j}F'_{ii'}\{\tilde{S}_{ii'}\alpha_nZ_{ii'}(\bbeta)\}\Big\|_{\psi_2}\leq C_3\alpha_n\Big\|[\tilde{\bX}_{ii'}]_{j}\Big\|_{\psi_2}
\]
is upper bounded by an absolute constant. Therefore, for any pair $(i,i')$, we have 
\[
h(\{\bX_i,\epsilon_i\},\{\bX_{i'},\epsilon_{i'}\})
\] 
is subgaussian. Moreover, it is easy to see that
\[
\E\nabla\hat\cL_n(\bbeta^*_{r(\gamma),\alpha_n})=\nabla\E\hat\cL_n(\bbeta^*_{r(\gamma),\alpha_n})=\nabla\cL_n(\bbeta^*_{r(\gamma),\alpha_n})=0.
\]
Therefore, $\nabla\hat\cL_n(\bbeta^*_{r(\gamma),\alpha_n})$ is a U-statistic of centered subgaussian distributed elements. Employing the standard Hoeffding's decoupling technique and Bonferroni's adjustment then yields
\[
\norm{\nabla\hat{\mathcal{L}}_n(\bbeta^*_{r(\gamma),\alpha_n})}_{\infty}\stackrel{\P}{\lesssim} \alpha_n\sqrt{\log d/n}.
\]
Cauchy inequality then gives us
\[
|\langle\nabla \hat \cL_n(\bbeta^*_{r(\gamma),\alpha_n}), \bDelta\rangle| \leq \norm{\nabla\hat{\mathcal{L}}_n(\bbeta^*_{r(\gamma),\alpha_n})}_{\infty}\norm{\bDelta}_1,
\]
which yields
\[
\delta \hat \cL_n(\bDelta,\bbeta^*_{r(\gamma),\alpha_n})\stackrel{\P}{\gtrsim} C'_3\norm{\bDelta}_2^2-C'_4/\alpha_n-C_5'\alpha_n^{-1/2}\norm{\bDelta}_2-O_P(\kappa_n)-C'_6\alpha_n\sqrt{\log d/n}\norm{\bDelta}_1.
\]

Then under Assumption {\bf (A5)} and Assumption {\bf (A0)} that for some $\alpha_n$ large enough,
\[
\norm{\bbeta^*_{r(\gamma),\alpha_n}}_0\leq s_n,
\]
we have, letting $\hat\bDelta:=\hat\bbeta_{r(\gamma),\alpha_n}-\bbeta^*_{r(\gamma),\alpha_n}$,
\[
\norm{\hat\bDelta}_1\leq 4\norm{\hat\bDelta_S}_1\leq 4\sqrt{s_n}\norm{\hat\bDelta}_2.
\]
And by Theorem \ref{thm:martin}, setting $P(\btheta)=\sum_{j=2}^d|\btheta_j|$ and $\cA:=\{\btheta\in\reals^d:\btheta_1=1,\norm{\btheta-\bbeta^*}_2\leq r\}$, we have, when $\lambda_n\gtrsim \alpha_n\sqrt{\log d/n}$,
\[
\norm{\hat\bDelta}_2^2\stackrel{\P}{\lesssim} s_n\lambda_n^2 + \alpha_n^{-1}+\alpha_n^2s_n\log d/n + \kappa_n,
\]
which implies, when $\lambda_n\asymp \alpha_n\sqrt{\log d/n}$,
\[
\norm{\hat\bDelta}_2^2\stackrel{\P}{\lesssim} \alpha_n^2s_n\log d/n+ \alpha_n^{-1} + \kappa_n.
\]
This completes the proof.
\end{proof}

\subsection{Proof of Theorem \ref{thm:error_bound}}
\begin{proof}
Picking $\alpha_n\asymp ( n/(s_n\log d))^{1/3}$,  Lemma \ref{lemma:estimation_error} then yields
\begin{align}\label{thm_error_bound_temp1}
\norm{\hat\bbeta_{r(\gamma),\alpha_n}-\bbeta^*_{r(\gamma),\alpha_n}}_2\stackrel{\P}{\lesssim}  \Big( \frac{s_n\log d}{n} \Big)^{1/6}+\kappa_n^{1/2}.
\end{align}
Due to Lemma \ref{lem:app-error} , we have $\norm{\bbeta^*_{r(\gamma),\alpha_n}-\bbeta^*}_2^2\stackrel{\P}{\lesssim} \alpha_n^{-1}\asymp ( n/(s_n\log d))^{-1/3}$, which together with (\ref{thm_error_bound_temp1}) leads to
\begin{align*}
\norm{\hat\bbeta_{r(\gamma),\alpha_n}-\bbeta^*}_2 &\leq \norm{\hat\bbeta_{r(\gamma),\alpha_n}-\bbeta^*_{r(\gamma),\alpha_n}}_2  + \norm{\bbeta^*_{r(\gamma),\alpha_n}-\bbeta^*}_2 \stackrel{\P}{\lesssim} \Big( \frac{s_n\log d}{n} \Big)^{1/6}+\kappa_n^{1/2}.
\end{align*}
Thus as long as $s_n\log d/n \rightarrow 0$ and $\kappa_n\rightarrow 0$, we have $\norm{\hat\bbeta_{r(\gamma),\alpha_n}-\bbeta^*}_2 \stackrel{\P}{\rightarrow} 0$.

Finally, noticing that for $n$ large enough, by Lemma \ref{lem:app-error} and the above result, $\bbeta^*_{r(\gamma),\alpha_n}$ and $\hat\bbeta_{r(\gamma),\alpha_n}$ are both within the ball of $\{\bbeta: \bbeta_1=1,\norm{\bbeta-\bbeta^*}_2\leq r(\gamma)\}$. Since $r(\gamma)$ only depends on $\gamma$, it could be picked as an absolute constant by fixing $\gamma_1,\gamma_2$, with $\gamma_2/\gamma_1=1.01$ for example. Then we have $\bbeta^*_{r(\gamma),\alpha_n}$ and $\hat\bbeta_{r(\gamma),\alpha_n}$ are indeed local minima for $n$ large enough. This completes the proof.
\end{proof}

\end{appendices}

\bibliographystyle{apalike}
\bibliography{MSIM}

\end{document}